\documentclass[superscriptaddress,nofootinbib,11pt,preprintnumbers,floatfix,longbibliography]{revtex4-2}
\usepackage[utf8]{inputenc}

\usepackage{booktabs}
\usepackage{graphicx}
\usepackage{amssymb}
\usepackage[dvipsnames]{xcolor}
\definecolor{red}{rgb}{0.9, 0,0}
\definecolor{cerulean}{rgb}{0., 0.42,0.9}
\definecolor{navy}{rgb}{0.05, 0.05,0.8}

\usepackage{float}

\usepackage[colorlinks]{hyperref}
\hypersetup{
    colorlinks = true,
    citecolor  = red,
	linkcolor  = navy
}
\usepackage{subcaption}
\usepackage{xfrac}
\usepackage{amsthm}
\usepackage{mathtools}
\usepackage{multirow}

\usepackage{enumitem}

\usepackage{xcolor}
\usepackage{etoolbox}
\usepackage{environ}

\definecolor{draftcol}
{HTML}{4A90D9}

\newtoggle{showdraft}
\toggletrue{showdraft}                  %

\usepackage{bm}
\makeatletter
\renewcommand{\vec}[1]{\ifcat a\noexpand#1\mathbf{#1}\else\bm{#1}\fi}
\newcommand{\unit}[1]{\ifcat a\noexpand#1\mathbf{\hat{#1}}\else\bm{\hat{#1}}\fi}
\makeatother

 \definecolor{lightblue}{rgb}{0.2,0.5,1}
 \hypersetup{colorlinks=true,
 linkcolor=lightblue,
 citecolor=lightblue,
 urlcolor=lightblue,
 linktocpage=lightblue,
 pdfproducer=lightblue}
\begin{document}

\title{Pulsar Timing Sensitivity to Dark Matter Substructure in the Presence of a Stochastic Gravitational-Wave Background}

\author{Abhiram Cherukupalli}
\email{abhiram@caltech.edu}
\affiliation{California Institute of Technology,
1200 E. California Boulevard, Pasadena, CA 91125, USA}

\author{Vincent S. H. Lee}
\email{vincentszehimlee@berkeley.edu}
\affiliation{Department of Physics, University of California, Berkeley,
Berkeley, CA 94720, USA}
\affiliation{Department of Physics, University of California, San Diego, La Jolla, CA 92093-0319, USA}

\author{Kim Berghaus}
\email{kim.berghaus@kit.edu}
\affiliation{
Institute for Theoretical Physics (ITP), Karlsruhe Institute of Technology (KIT), Wolfgang-Gaede-Str. 1,
76131 Karlsruhe, Germany
}
\affiliation{
Institute for Astroparticle Physics (IAP), Karlsruhe Institute of Technology (KIT),
Hermann-von-Helmholtz-Platz 1, 76344 Eggenstein-Leopoldshafen, Germany}
\affiliation{Walter Burke Institute for Theoretical Physics, California Institute of Technology,
1200 E. California Boulevard, Pasadena, CA 91125, USA}

\author{Kathryn M. Zurek}
\email{kzurek@caltech.edu}
\affiliation{Walter Burke Institute for Theoretical Physics, California Institute of Technology,
1200 E. California Boulevard, Pasadena, CA 91125, USA}

\date{\today}

\preprint{CALT-TH-2026-025}
\preprint{KA-TP-14-2026}
\preprint{N3AS-26-014}

\begin{abstract}

Pulsar timing arrays (PTAs) can detect dark matter (DM) substructure through the small shifts a transiting DM subhalo imprints on pulse arrival times. Recently found evidence for a stochastic gravitational-wave background (GWB) acts as red noise and competes with the substructure signal. Here we provide an analytic understanding of how this background degrades PTA sensitivity to DM substructure.
We derive the full gauge-invariant proper-time observable induced by a transiting DM subhalo and develop a framework for the expected signal-to-noise ratio in the presence of red noise, accounting for the degeneracy with the pulsar timing model. From this we obtain simple scaling relations for the reach across the static, dynamic, and stochastic regimes, and numerically compute the reach for a Square Kilometre Array benchmark at three representative points of the NANOGrav 15-year posterior.  The GWB background suppresses the sensitivity to DM substructure by one to three orders of magnitude compared to forecasts in the presence of only white noise, and the suppression depends on the amplitude and spectral index of the background within a factor of three. The dynamic Shapiro signal suffers the smallest suppression and gives the best sensitivity to DM substructure near $10^{-2}\, M_\odot$. %
Probing the regime where subhalos make up all or part of the DM %
remains a challenge even for surveys with more pulsars and longer observing time. Despite this, PTA measurements remain a 
competitive probe of DM substructure, and future surveys will increase in sensitivity by up to two orders of magnitude from existing NANOGrav limits.

\end{abstract}
\maketitle
\newpage
\makeatletter

\renewcommand{\l@subsubsection}[2]{}
\def\l@f@section{\addpenalty{\@secpenalty}\addvspace{0.6em plus\p@}}
\let\oldl@section\l@section
\renewcommand{\l@section}[2]{\oldl@section{\textbf{#1}}{\textbf{#2}}}

\makeatother
\tableofcontents
\newpage

\section{Introduction}\label{sec:intro}
Dark matter (DM) dominates the Universe, but little is known about its fundamental nature.  The way that dark matter clumps through the pull of gravity on %
scales smaller than a galaxy will tell us something about its fundamental nature.  For example, axions form miniclusters~\cite{Hogan:1988mp, Kolb:1993zz, Zurek:2006sy, Buschmann:2019icd, Eggemeier:2019khm, Xiao:2021nkb} if the Pecci-Quinn (PQ) symmetry~\cite{Peccei:1977hh} is broken after inflation, while dark matter interacting through the weak force has a characteristic earth mass scale ($\sim 10^{-6}\,M_{\odot}$) below which we expect no substructure~\cite{Green:2005fa}. While a map of the dark matter power spectrum on small scales would be a powerful tool for understanding dark matter, little is known observationally about dark matter structure on such small scales.

One particularly promising avenue for detecting DM substructure in the galaxy is pulsar timing arrays (PTAs)~\cite{Siegel:2007fz, Baghram:2011is, Clark:2015sha, Schutz:2016khr, Kashiyama:2018gsh, Dror:2019twh, Ramani:2020hdo, Lee:2020wfn, Lee:2021zqw, Gresham:2022biw, NANOGrav:2023hvm, Berghaus:2025kvn}. Pulsars are extremely stable clocks~\cite{Goldreich:1969sb, Sturrock:1971zc}. By measuring the times of arrival of pulses from an array of pulsars across the sky, one can look for timing deviations sourced by changes in the gravitational potential. PTAs were originally motivated as a way to detect a stochastic gravitational-wave background (GWB)~\cite{Taylor:2021yjx} 
as expected from a population of merging supermassive black hole binaries~\cite{Phinney:2001di}, but in recent years they have also been proposed as a tool for DM substructure detection. Conventional gravitational probes lose sensitivity in particular to  diffuse DM substructure, potentially leaving PTAs as the only viable probe. 
In this work we model the substructure as DM subhalos with point masses of mass $M$. This is the most optimistic case, but previous PTA forecast have found a modest reach suppression for diffuse profiles in comparison~\cite{Ramani:2020hdo, Lee:2020wfn}, and we expect the same to hold under the inferred GWB.

In the past few years, there has been strong evidence indicating the existence of this stochastic GWB in the data as observed by the North American Nanohertz Observatory for Gravitational Waves (NANOGrav), whose origin has not yet been pinpointed~\cite{NANOGrav:2023gor}. 
The theoretical framework for using PTAs to detect DM substructure has been developed in a series of works over the past two decades ~\cite{Siegel:2007fz, Baghram:2011is, Clark:2015sha, Schutz:2016khr, Kashiyama:2018gsh, Dror:2019twh, Ramani:2020hdo, Lee:2020wfn}. The analyses are conceptually challenging because the signal originates from pulsar acceleration, photon propagation and clock effects, only fully developed until Ref.~\cite{Ramani:2020hdo}.  In addition, the analyses are numerically challenging because pulsar timing measurements have multiple sources of uncertainty, including pulsar-intrinsic white and red noise, and the contributions from a GWB. 
A Bayesian statistics based analysis pipeline was developed in Ref.~\cite{Lee:2021zqw} to address these issues numerically in a realistic dataset, and has been applied to search for substructure in NANOGrav's 15-year dataset~\cite{NANOGrav:2023gor}, yielding an upper limit on the substructure abundance in the galaxy~\cite{NANOGrav:2023hvm}. However, an analytic understanding of how the different noise sources affect the search is still lacking, %
which prevents answering questions such as 
how sensitivity to substructure improves with future datasets that include more pulsars and longer observation times, and how the sensitivity %
scales with the values of the amplitude and spectral index of a GWB.

In this work, we directly address this issue by studying how the potential presence of a GWB affects PTA sensitivity to signals from DM substructure. We first develop a framework for estimating the expected sensitivity to a generic signal in pulsar timing data given a particular noise source. A key ingredient is the \textit{timing model} of the pulsar, which describes the intrinsic evolution of the pulsar phase in the absence of any signal~\cite{Taylor:2021yjx}, since degeneracy with the signal of interest can degrade sensitivity. We then specialize to DM substructure. We derive the full gauge-invariant observable due to a transiting DM for pulsar timing, which is identified as the proper time elapsed between consecutive pulses as measured by an observer on Earth as shown in Ref.~\cite{Lee_2026}. We show that the individual contributions reduce to the expressions used in prior analyses~\cite{Dror:2019twh, Ramani:2020hdo, Lee:2020wfn}. Finally, we apply this signal to our framework and estimate the sensitivity for both deterministic (a single DM encounter with a fixed waveform) and stochastic (an ensemble of DM encounters described by a power spectrum) signals. We find that the projected sensitivity depends strongly on the spectral index of the stochastic GWB. While the theoretically predicted value of the spectral index is 13/3~\cite{Phinney:2001di}, this value lies more than $2\sigma$ away from the posterior distribution inferred from NANOGrav's observation~\cite{NANOGrav:2023gor}, indicating a large uncertainty in its value should the GWB exist. Moreover, we find that the stochastic GWB substantially diminishes PTA sensitivity to DM substructure. Hints of this issue already appeared in earlier works using Bayesian frameworks to search for DM substructure with both mock~\cite{Lee:2021zqw} and real PTA data~\cite{NANOGrav:2023hvm}, but we quantify its extent analytically for the first time in this work, in a manner that can be generalized to future PTA surveys. %

Our paper is organized as follows. In Sec.~\ref{sec:SNR}, we describe our timing and noise model and develop the framework for computing the expected signal-to-noise ratio (SNR) for any signal in the presence of red noise. In Sec.~\ref{sec:dark_matter_residual}, we compute the gauge-invariant observable, given by the proper time between pulse arrivals at Earth, in the presence of a transiting DM subhalo. In Sec.~\ref{sec:analytical_estimates},
we use analytic estimates to compute the expected SNR for signals localized to a single pulsar across the static, dynamic, and stochastic regimes.
In Sec.~\ref{sec:results_and_discussion}, we present our main results and show the sensitivity of PTA experiments to DM substructure in the presence of a stochastic GWB. We conclude in Sec.~\ref{sec:conclusion}.

The appendices are a collection of several technical results. App.~\ref{app:pulsar_SNRs} derives the SNR framework of Sec.~\ref{sec:SNR}, App.~\ref{app:gauge_invariance} explicitly demonstrates the gauge invariance of the observable in Sec.~\ref{sec:dark_matter_residual}, App.~\ref{sec:appendix_integral} computes the red-noise suppression of the sensitivity integrals, %
App.~\ref{app:noise_weighted_norm} evaluates the projected signal norms, %
and App~\ref{app:numerics} gives details of the numerics.

\section{Detecting a Signal in the Presence of Red Noise}\label{sec:SNR}

Pulsar timing arrays are sensitive to a wide range of new-physics signals beyond the stochastic GWB~\cite{Mitridate:2023oar, NANOGrav:2023hvm}, such as DM substructure (see references in Sec.~\ref{sec:intro}), cosmic strings~\cite{Ellis:2020ena, Blasi:2020mfx, Ellis:2023tsl}, and ultralight bosonic fields~\cite{Graham:2015ifn, Porayko:2018sfa, Kaplan:2022lmz,  Kim:2023kyy, EPTA:2023xxk, EuropeanPulsarTimingArray:2023egv,  Smarra:2024kvv, Gan:2025icr,  Boddy:2025oxn, Dror:2025nvg, Hu:2026yop}. However, the detection of any such signal must contend with noise in the timing data, which includes pulsar-intrinsic processes and potentially the GWB
itself, which can obscure the signals of interest. In this section, we develop the statistical framework for assessing the detectability of a generic signal in the presence of such noise. In this section we will consider two types of signals --- deterministic signals (those with a known waveform that can be modeled as a specific time-dependent function in each pulsar's timing residuals), and stochastic signals (those characterized statistically by their power spectrum rather than by a specific waveform).

We derive the SNR for a generic signal after marginalizing over the pulsar timing model --- a quadratic polynomial 
to the pulse arrival times that accounts for the pulsar's spin frequency and spin-down rate. Because these parameters can only be inferred from the same timing data used to search for new physics, any low-frequency signal component that  is degenerate %
with the timing model is unavoidably absorbed into the fit, which weakens signal sensitivity. After defining the contributions to the timing residuals (Sec.~\ref{sec:pulsar_model}) and the projection procedure (Sec.~\ref{sec:timing_model_marginalization}), we derive the SNR for deterministic (Sec.~\ref{sec:deterministic_signal_statistics}) and stochastic (Sec.~\ref{sec:snr_stochastic}) signals. 

Throughout this section, we first restrict to signals localized to a single pulsar, so that each pulsar provides an independent measurement of the signal. For a signal in pulsar $a$, the timing-model projection and SNR are built using only the noise autocorrelation of that pulsar. This is the statistic we later apply to the DM pulsar-term and line-of-sight signals in Sec.~\ref{sec:analytical_estimates}. This statistic deliberately ignores the possibility of using the other pulsars to condition the inter-pulsar-correlated part of the noise in pulsar $a$. For a GWB-like covariance and well-separated pulsars, such conditioning can reduce only the correlated component; the uncorrelated GWB contribution remains as a floor, so the resulting improvement in sensitivity is only order unity for these single-pulsar estimates. %
We extend this framework to coherent signals and correlated noise across the array, using the full noise covariance, in a companion paper \cite{in_prep}.

\subsection{Contributions to Pulsar Timing Residuals}
\label{sec:pulsar_model}
Despite their exceptionally stable rotational periods, the measured pulse times-of-arrival %
from millisecond pulsars can be affected by several astrophysical effects that must be accounted for when estimating sensitivity to new-physics signals. We therefore model the timing residuals of the $a$-th pulsar as (see, \textit{e.g.}, Ref.~\cite{Taylor:2021yjx})
\begin{equation}\label{eqn:time_shift_a}
    \delta t_a(t) = \delta t_{m}(t;\vec{\beta}_a)+\delta t_{\mathrm{sig},a}(t)+n_a(t) \, ,
\end{equation}
where $\delta t_{\mathrm{sig},a}(t)$ is the signal of interest, and the remaining contributions include 
\begin{itemize}[leftmargin=*]

    \item a timing model, $\delta t_m(t;\vec{\beta}_a)$,
    which describes the deterministic spin-down contribution to the pulse arrival times. For millisecond pulsars %
    , we approximate this by a quadratic polynomial in $t$,
    \begin{equation}\label{eqn:timing_model}
        \delta t_{m}(t;\vec{\beta}_a) = \beta_{a}^{(0)} + \beta_{a}^{(1)} t + \frac{1}{2} \beta_{a}^{(2)} t^2,
    \end{equation}
    where $\vec{\beta}_a$ denotes the timing-model coefficients. These parameters are not known {\em a priori} and must be marginalized over when constructing the detection statistic. 
    In a full timing analysis the model generally contains additional deterministic contributions~\cite{Taylor:2021yjx}, %
     such as dispersion-measure variations, but for the analytic estimates in this work we restrict to the quadratic form above for tractability.
        
    \item stochastic fluctuations, $n_a(t)$,
    which represent random contributions to the pulse arrival times %
    from noise processes in the pulsar timing data. These can be modeled as a stationary Gaussian process with one-sided power spectral density (PSD), $S_{ab}(f)$, defined by
    \begin{equation}\label{eqn:noise_PSD_def}
        \langle \tilde{n}_a(f)\tilde{n}_b(f') \rangle \equiv \frac{1}{2}S_{ab}(f)\delta(f-f') \, .
    \end{equation}
    These stochastic fluctuations can arise from three different sources: %
    
    \begin{enumerate}[leftmargin=*]
        \item \textbf{white noise}: noise with a flat PSD across all sampling frequencies that is uncorrelated across pulsars. This includes effects such as radiometer noise, instrumental noise and pulse phase jitter~\cite{Lentati:2016ygu}. The amplitude of the white noise PSD is given by %
        \begin{equation}\label{eqn:white_noise_PSD}
            N_a(f) = 2\Delta t_a\, t_{\mathrm{rms}, a}^2 \, , 
        \end{equation}
        where $\Delta t_a$ is the observational cadence, and $t_{\mathrm{rms}, a}$ is the root-mean-squared timing residual of the $a$-th pulsar.
        \item \textbf{pulsar-intrinsic red noise}: long-timescale noise processes with the noise power predominantly at low sampling frequencies that is uncorrelated across pulsars. This noise is modeled empirically as a power law in the PSD and physically arises from processes such as spin noise and rotational irregularities. %
        \item \textbf{stochastic GWB}:  a potential background that affects timing measurements in a correlated manner across pulsars. This correlation is known as the Hellings-Downs (HD) correlation~\cite{Hellings:1983fr}. The PSD of the stochastic GWB can be modeled as a power law with amplitude $A_{\mathrm{GWB}}$ and spectral index $\gamma_{\mathrm{GWB}}$, given by
        \begin{equation}\label{eqn:GWB_PSD}
            S_{ab}^{(\mathrm{GWB})}(f) = \left[\frac{1}{2}\delta_{ab}+\Gamma^\text{HD}_{ab}\right]\frac{A^2_\text{GWB}}{12\pi^2} \left(\frac{1\text{yr}^{-1}}{f}\right)^{\!\gamma_\text{GWB}}\text{yr}^3  \, ,
        \end{equation}
        where $\Gamma^\text{HD}_{ab}$ is the HD correlation, given by
        \begin{equation}\label{eqn:HD_correlation}
            \Gamma^\text{HD}_{ab}={\frac {1}{2}}-{\frac {x_{ab}}{4}}+{\frac {3}{2}}x_{ab}\ln x_{ab} \, .
        \end{equation}
        Here $x_{ab}\equiv (1-\cos \theta _{ab})/2$, where $\theta_{ab}$ is the angle between the Earth-pulsar lines of sight to pulsars $a$ and $b$. Note that the HD-autocorrelation, $\Gamma^\text{HD}_{aa}=\lim_{\theta_{ab}\to0}\Gamma^\text{HD}_{ab} = 1/2.$ %
    \end{enumerate}
\end{itemize}
For the remainder of this work we neglect pulsar-intrinsic red noise, so the total noise PSD is a sum of Eqs.~\eqref{eqn:white_noise_PSD}-\eqref{eqn:GWB_PSD}, 
\begin{equation}\label{eqn:noise_PSD_decomposition}
    S_{ab}(f) = N_a\,\delta_{ab} + S^{\rm (GWB)}_{ab}(f) \, .
\end{equation}
The diagonal element $S_{aa}(f)$ of the PSD in Eq.~\eqref{eqn:noise_PSD_decomposition} is the noise autocorrelation of pulsar $a$, while the off-diagonal elements ($a\neq b$) encode the correlations between pulsars, which arise solely from the HD-correlated part of the GWB in  Eq.~\eqref{eqn:HD_correlation}. For this work, the statistic is constructed from the timing residuals of a single pulsar rather than the full array of residuals. For a signal localized to pulsar $a$, the relevant noise PSD is therefore $S_n(f)\equiv S_{aa}(f)$, the autocorrelation of that pulsar's noise,

\begin{equation} \label{eqn:explicit_Sn}
    S_{n}(f)= 2\Delta t t^2_\text{rms} +  \frac{A^2_\text{GWB}}{12\pi^2} \left(\frac{1\text{yr}^{-1}}{f}\right)^{\gamma_\text{GWB}}\text{yr}^3.
\end{equation}

\subsection{Marginalizing over the Timing Model}
\label{sec:timing_model_marginalization}

To %
accurately estimate the detectability of a new physics signal, one must marginalize over the timing-model parameters. Following Ref.~\cite{vanHaasteren:2012hj}, this is equivalent to projecting the signal onto the subspace orthogonal to the timing model under a noise-weighted inner product; for completeness, we review this equivalence in App.~\ref{app:pulsar_SNRs}. 

Throughout, we work with mean-subtracted timing residuals over the observing span $[0,T]$, so that the constant component of the residuals is removed as part of the timing fit. 
The remaining timing-model subspace relevant for projection is therefore spanned by the linear and quadratic modes. Dropping the pulsar index, the timing residuals from Eq.~\eqref{eqn:time_shift_a} become
\begin{equation} \label{eqn:timing_residual_full}
   \delta t(t) \approx \vec{m}(t) \cdot \vec{\xi} +  \delta t_{\text{sig}}(t) +  \delta n(t) \, ,
\end{equation}
where the timing-model contribution $\vec{m}(t)\cdot \vec{\xi}$ rewrites the quadratic polynomial of Eq.~\eqref{eqn:timing_model} in an orthonormal basis. Although the quadratic model has three coefficients, the constant mode is removed by the mean subtraction above and carries no weight in the noise-weighted inner product below, leaving the two surviving modes --- the linear and quadratic Legendre polynomials $\phi_1$ and $\phi_2$ of Eq.~\eqref{eqn:legendre_def} --- as the design vector $\vec{m}$ with coefficients $\vec{\xi}$: %
\begin{equation}\label{eqn:design vector}
    \vec{m}(t) = \begin{pmatrix}
        \phi_1(t) \\
        \phi_2(t)
    \end{pmatrix}, \quad
    \vec{\xi} = \begin{pmatrix}
        \xi^{(1)} \\
        \xi^{(2)}
    \end{pmatrix}.
\end{equation}
The basis functions are chosen to be Legendre polynomials orthonormal on $[0,T]$,
\begin{equation}\label{eqn:legendre_def}
    \phi_n(t)=\sqrt{2n+1}\,P_{\! n}\!\left(\frac{2t}{T}-1\right), \quad \langle\phi_i, \phi_j  \rangle = 
    \frac{1}{T}\int_0^T \phi_i(t)\,\phi_j(t)\,dt = \delta_{ij}.
\end{equation}
We emphasize that though these Legendre polynomials are orthonormal under the inner product defined above, they are generically not orthonormal under a red-noise weighted inner product. This subtlety becomes relevant later in Sec.~\ref{sec:analytics_static_limit}. As both the basis functions and the projection operation are defined on a finite observation window, the natural frequency representation is discrete, with frequencies $f_k = k/T$. For the analytic estimates in this work, the observation time is sufficiently long that the spacing between frequency bins is small, and the discrete sum can be well approximated by a continuous integral. %
The integrand of this integral, however, must be the analytic continuation of the basis-function transforms \emph{evaluated} on the discrete modes $f_k$, rather than their full continuous Fourier transform on the window: the latter carries oscillatory zeros between the modes that are windowing artifacts, not independent frequency bins.  This continuation is used throughout the computation of the projected noise-weighted norms of the DM signals in App.~\ref{app:phi3perp_norm}.

With the timing-model basis now specified, we define the projection onto the orthogonal subspace using the noise-weighted inner product. %
For arbitrary functions $g$ and $h$ with support on $[0, T]$, we define the noise-weighted inner product $(g|h)$ as
\begin{equation} \label{eqn:inner_prod_def}
\begin{split}
(g|h)  &=
\frac{1}{T^2}\int_0^T \int_0^T dt \, dt' \,g(t) N^{-1}(t-t') h(t') \\
&=\frac{4}{T}\,\mathrm{Re}\left[\sum_{k=1}^\infty \frac{\widetilde{g}^*(f_k)\widetilde{h}(f_k)}{S_n(f_k)}\right]\approx4 \,\mathrm{Re}\left[\int_{1/T}^{\infty} df \frac{\widetilde{g}^*(f)\widetilde{h}(f)}{S_n(f)}\right],
\end{split}
\end{equation}
where the second-equality writes the inner product as a sum over the discrete modes $f_k = k/T$ (factor of $4$ from the one-sided PSD) and the final equality is the continuum approximation introduced above.\footnote{We use the Fourier transform convention $\tilde{x}(f) = \int dt\, x(t)\, e^{-2\pi i f t}$ and $x(t) = \int df\, \tilde{x}(f)\, e^{2\pi i f t}$. The integrands of interest are dominated by low frequencies and fall steeply with $f$, so they are already negligible at the Nyquist frequency $f_{\rm Nyq} = 1/2\Delta t$; we therefore extend the upper limit to infinity.} Here $N(t,t')$ is the covariance function of the noise that is related to the PSD $S_n(f)$ by
\begin{equation}
N(t,t')
= \int_{0}^{\infty} \! df\;S_n(f)\,\cos\!\big[2\pi f (t-t')\big].
\end{equation}
A signal $\delta t^\perp_\text{sig}(t)$ is orthogonal to the timing model if
\begin{equation} \label{eqn:orthogonality}
(\phi_1 | \delta t^\perp_\text{sig}) =(\phi_2 | \delta t^\perp_\text{sig}) = 0.
\end{equation}
Stationarity of the noise implies even/odd decoupling under the inner product, 
\begin{equation}\label{eqn:orthonormal_opp_parity}
(\phi_i|\phi_{j})= 0 \qquad \text{for } i+j \text{ odd},
\end{equation}
which follows from the symmetry of $\phi_i$ under $t\to T-t$; we give the short derivation in App.~\ref{app:pulsar_SNRs}.
In particular, $(\phi_1|\phi_2)=0$ and the component of the signal orthogonal to the timing-model subspace is therefore
\begin{equation}\label{eqn:deterministic_pulsar_proj}
    \delta t^\perp_\text{sig}(t) = \delta t_\text{sig}(t) - \frac{(\phi_1 |\delta t_\text{sig})}{(\phi_1 |\phi_1)} \phi_1(t)- \frac{(\phi_2 |\delta t_\text{sig})}{(\phi_2 |\phi_2)} \phi_2(t).
\end{equation}
Only this projected component contributes to the detection statistic in the subsequent signal-to-noise analysis.

\subsection{The Signal-to-Noise Ratio}
\label{sec:deterministic_signal_statistics}
\label{sec:snr_stochastic}

With the projection established, we write down the SNR for both deterministic and stochastic signals and their sensitivity is set by the spectral index of the projected signal relative to that of the noise.
Because the projection removes low-frequency signal power, the surviving signal competes against red noise with a steeper spectrum than the original.

\subsubsection{Deterministic signals}
The projected signal $\delta t_{\mathrm{sig}}^\perp(t)$ defined in Eq.~\eqref{eqn:deterministic_pulsar_proj} has had its timing-model-degenerate components removed; the SNR is simply its noise-weighted norm (derived in App.~\ref{app:pulsar_SNRs}),
\begin{equation}
\label{eqn:projected_snr_pulsar}
\begin{split}
\mathrm{SNR}^2
&=(\delta t_{\mathrm{sig}}^{\perp} | \delta t_{\mathrm{sig}}^{\perp}) \\
&=\left(\delta t_{\mathrm{sig}} | \delta t_{\mathrm{sig}}\right)
- \sum_{i=1, 2} \frac{(\phi_i |\delta t_{\mathrm{sig}})^2}{(\phi_i |\phi_i)}.
\end{split}
\end{equation}
The second line makes explicit how the projection removes the components of the signal that overlap with the timing-model modes $\phi_i$.

\subsubsection{Stochastic signals}
When the signal cannot be resolved as individual events and instead manifests as a zero-mean Gaussian contribution to the timing residuals, it is fully characterized by its projected covariance,
\begin{equation} \label{eqn:projected_pulsar_covariance}
    S^\perp_{\mathrm{sig}}(t,t') = \langle\delta t^\perp_{\mathrm{sig}}(t)\,\delta t^\perp_{\mathrm{sig}}(t')\rangle.
\end{equation}
For weak signals, $S_{\mathrm{sig}} \ll N$, the optimal SNR is obtained by expanding the log-likelihood ratio to second order (see App.~\ref{app:pulsar_SNRs}),
\begin{equation}  \label{eqn:projected_stoch_SNR_pulsar}
    \text{SNR}^2 = \frac{1}{2} \mathrm{Tr} \left[N^{-1} S^\perp_{\mathrm{sig}}\, N^{-1} S^\perp_{\mathrm{sig}}\right],
\end{equation}
where the trace runs over time indices. For a stationary signal, defining the one-sided signal PSD consistent with the noise convention of Eq.~\eqref{eqn:noise_PSD_def},
\begin{equation}\label{eqn:sig_PSD_onesided}
    \langle\delta\tilde{t}^\perp_{\mathrm{sig}}(f)\,\delta\tilde{t}^{\perp*}_{\mathrm{sig}}(f')\rangle \equiv \frac{1}{2}S^\perp_{\mathrm{sig}}(f)\,\delta(f-f'),
\end{equation}
the SNR becomes
\begin{equation}\label{eqn:stoch_SNR_fourier}
    \mathrm{SNR}^2 = N_PT\int_{1/T}^\infty df\left(\frac{S^\perp_{\mathrm{sig}}(f)}{S_n(f)}\right)^2,
\end{equation}
where the factor of $N_P$ arises because each pulsar sees an independent realization of the signal.

\section{Dark Matter Substructure Signals in Pulsar Timing}
\label{sec:dark_matter_residual}

In this section, we derive the expected signals on pulsar timing from a transiting DM subhalo. A subhalo with mass $M$ induces a small perturbation to spacetime, described by the metric perturbation $h_{\mu\nu}$. This leads to shifts in the \textit{proper} times of arrival as measured by a telescope on Earth, denoted by $\delta t_{\mathrm{DM}}(t)$. %
This proper time observable has been worked out in Ref.~\cite{Lee_2026} and has been explicitly shown to be invariant under general gauge transformations in $h_{\mu\nu}$ (see also Ref.~\cite{Dror:2025nvg} in the context of ultralight dark matter, where the observable is related to the Riemann tensor--- a gauge-invariant quantity in linearized gravity). Moreover, it has been shown that $\delta t_{\rm{DM}}(t)$ can be written as a sum of three distinct contributions: the Doppler effect, which describes the relative motion of the Earth and the pulsar under the geodesic equation; the Shapiro delay, which represents the delay in photon travel time through the perturbed spacetime; and the Einstein delay, which arises from the gravitational redshift in the clock measuring the times-of-arrival%
. In total, the time shift is written as %
\begin{equation}
\label{eqn:delta_t_t}
	\delta t_{\mathrm{DM}}(t) =\,  \delta t_{\mathcal{D}}^{(E)}(t) - \delta t_{\mathcal{D}}^{(P)}(t) + \delta t_{\mathcal{S}}(t) + \delta t_{\mathcal{E}}^{(E)}(t) - \delta t_{\mathcal{E}}^{(P)}(t) \,  ,
\end{equation}
where the $\mathcal{D}$, $\mathcal{S}$, and $\mathcal{E}$ subscripts correspond to the Doppler, Shapiro, and Einstein terms, respectively, while the $(E)$ and $(P)$ superscripts correspond to the Earth and pulsar terms, respectively, \textit{i.e.,} the effect of DM acting on either the Earth or the pulsar. The Shapiro term is an integrated effect over the line-of-sight between Earth and the pulsar, and thus does not distinguish between Earth and pulsar terms. While some of the individual terms in Eq.~\eqref{eqn:delta_t_t} (namely $\delta t_{\mathcal{D}}^{(P)}(t)$, $\delta t_{\mathcal{D}}^{(E)}(t)$, and $\delta t_{\mathcal{S}}(t)$) have been considered in previous works~\cite{Dror:2019twh, Ramani:2020hdo, Lee:2020wfn}, the precise way in which the total DM observable is constructed is novel for PTAs (analogous decompositions of the observables in the context of using other GW experiments to detect DM have been identified for laser~\cite{Du:2023dhk, Lee:2024oxo} and atom interferometers~\cite{Badurina:2024rpp, Badurina:2025xwl}). 

To proceed in deriving the signal, one must choose a gauge. Throughout this work, we choose the Newtonian gauge, where the metric perturbation sourced by a non-relativistic DM subhalo is written as
\begin{align}\label{eqn:subhalo_metric}
	ds^2 = -(1+2\Phi)dt^2 + (1-2\Phi)dx_idx^i \, ,
\end{align}
where $\Phi$ is the DM-induced Newtonian potential, given by
\begin{align}\label{eqn:subhalo_potential}
	\Phi(t,\vec{r}) = -\frac{GM}{|\vec{r}-\vec{r}_{\mathrm{DM}}(t)|} \, .
\end{align}
We emphasize again that the observable, being a proper time quantity, does not depend on this particular gauge choice, as we demonstrate %
in App~\ref{app:gauge_invariance}. Using the general expression for the proper time shift observable as derived in Ref.~\cite{Lee_2026} and the form of the metric in Eq.~\eqref{eqn:subhalo_metric}, each contribution to the proper time shift can be written in integral form (setting $c=1$)
\begin{align}\label{eqn:subhalo_observable}
	\delta t_{\mathcal{D}}^{(E)}(t) &= \int^tdt'\int^{t'}dt''\,\unit{n}\cdot\nabla\Phi(t'',\vec{r}_E) \nonumber \\
	\delta t_{\mathcal{D}}^{(P)}(t) &= \int^{t-L}dt'\int^{t'}dt''\,\unit{n}\cdot\nabla\Phi(t'',\vec{r}_P)\nonumber \\
	\delta t_{\mathcal{S}}(t) &= -2\int_{-L/2}^{L/2} dz\, \Phi\left[t-\left(z+\frac{L}{2}\right),\vec{r}_{\mathrm{mid}}+z\unit{n}\right] \nonumber \\
	\delta t_{\mathcal{E}}^{(E)}(t) &=  \int^{t}dt'\, \Phi(t',\vec{r}_{E}) \nonumber \\
	\delta t_{\mathcal{E}}^{(P)}(t) &=  \int^{t-L}dt'\, \Phi(t',\vec{r}_{P}) \, ,
\end{align}
where $\vec{r}_E$ and $\vec{r}_P=\vec{r}_E+L\,\unit{n}$ are the Earth and pulsar locations, respectively, $L$ is the Earth–pulsar distance, $\unit{n}$ is the unit vector pointing from Earth to the pulsar, $\vec{r}_{\mathrm{mid}}=(\vec{r}_E+\vec{r}_P)/2$ is the midpoint between Earth and the pulsar, and all unspecified lower limits in the time integrals are arbitrary, as they are degenerate with the timing model. %

To evaluate the potential, we first parameterize the DM trajectory. Assuming that DM moves in a straight line with constant velocity, we write %
\begin{equation}\label{eqn:DM_trajectory}
	\vec{r}_{\mathrm{DM}}(t) = \vec{r}_0 + \vec{v}t \, ,
\end{equation}
where $\vec{r}_0$ is the DM initial position and $\vec{v}$ is its velocity. %
Following Ref.~\cite{Dror:2019twh}, the DM trajectory can be re-parameterized in terms of its impact parameters to given reference locations to simplify the integrals in Eq.~\eqref{eqn:subhalo_observable}, and the choice of re-parameterize depends on the particular term in the equation. %
We choose three different parameterizations (see also Ref.~\cite{Du:2023dhk})
\begin{align}\label{eqn:DM_parameterization}
	\vec{r}_{\mathrm{DM}} &= \vec{r}_E + \vec{b}_E + \vec{v}(t-t_E) \nonumber \\
	&= \vec{r}_P+\vec{b}_P + \vec{v}(t-t_P) \nonumber \\
	&= \vec{r}_{\mathrm{mid}} + \left[b_{\parallel}+v_{\parallel}(t-t_{\perp})\right]\unit{n} + \left[\vec{b}_{\perp} + \vec{v}_{\perp}(t-t_{\perp})\right] \, ,
\end{align}
where $\vec{b}_E$, $\vec{b}_P$ and $\vec{b}_{\perp}$ are the DM impact parameters defined relative to Earth, the pulsar, and the Earth-pulsar line-of-sight, $t_E$, $t_P$ and $t_{\perp}$ is the corresponding time at which the %
DM %
reaches said impact parameter, $b_{\parallel}$ is the DM's longitudinal distance from Earth along the Earth-pulsar line of sight, 
and $v_{\parallel}=\vec{v}\cdot\unit{n}$ and $v_{\perp}=\vec{v} - v_{\parallel}\unit{n}$ are the parallel and orthogonal components of $\vec{v}$ relative to $\unit{n}$. The impact parameters are defined such that $\vec{b}_E\cdot \vec{v}=\vec{b}_P\cdot \vec{v} = \vec{b}_{\perp}\cdot \vec{v} = \vec{b}_{\perp}\cdot \unit{n}= \vec{v}_{\perp}\cdot \unit{n}=\vec{b}_{\perp}\cdot \vec{v}_{\perp}=0$. Here the parameterizations in the three lines in Eq.~\eqref{eqn:DM_parameterization} are useful for (from top to bottom) computing $\delta t_{\mathcal{E}}^{(E)}(t)$ and $\delta t_{\mathcal{D}}^{(E)}(t)$ (Earth terms), $\delta t_{\mathcal{E}}^{(P)}(t)$ and $\delta t_{\mathcal{D}}^{(P)}(t)$ (pulsar terms), and $\delta t_{\mathcal{S}}(t)$ (Shapiro term), respectively. These quantities are related to the original DM trajectory in Eq.~\eqref{eqn:DM_trajectory} as follows. The times of closest approach to Earth, the pulsar, and the line of sight are respectively %
\begin{align}\label{eqn:closest_approach_times}
	t_E &= -\frac{(\vec{r}_0 - \vec{r}_E)\cdot\vec{v}}{v^2} \nonumber \\
	t_P &= -\frac{(\vec{r}_0 - \vec{r}_P)\cdot\vec{v}}{v^2} \nonumber \\
	t_{\perp} &= -\frac{\left[(\vec{r}_0 - \vec{r}_{\mathrm{mid}}) - ((\vec{r}_0 - \vec{r}_{\mathrm{mid}})\cdot\unit{n})\unit{n}\right]\cdot\vec{v}_{\perp}}{v_{\perp}^2} \, .
\end{align}
and the corresponding impact parameters are given by 
\begin{align}\label{eqn:impact_parameters}
	\vec{b}_E &= (\vec{r}_0 - \vec{r}_E) - \vec{v}\,t_E \nonumber \\
	\vec{b}_P &= (\vec{r}_0 - \vec{r}_P) - \vec{v}\,t_P \nonumber \\
	\vec{b}_{\perp} &= (\vec{r}_0 - \vec{r}_{\mathrm{mid}}) - ((\vec{r}_0 - \vec{r}_{\mathrm{mid}})\cdot\unit{n})\unit{n} - \vec{v}_{\perp}\,t_{\perp} \nonumber \\
	b_{\parallel} &= (\vec{r}_0 - \vec{r}_{\mathrm{mid}})\cdot\unit{n} + v_{\parallel}\,t_{\perp} \, .
\end{align}
\begin{figure}[h]
    \centering
    \includegraphics[width=\linewidth]%
    {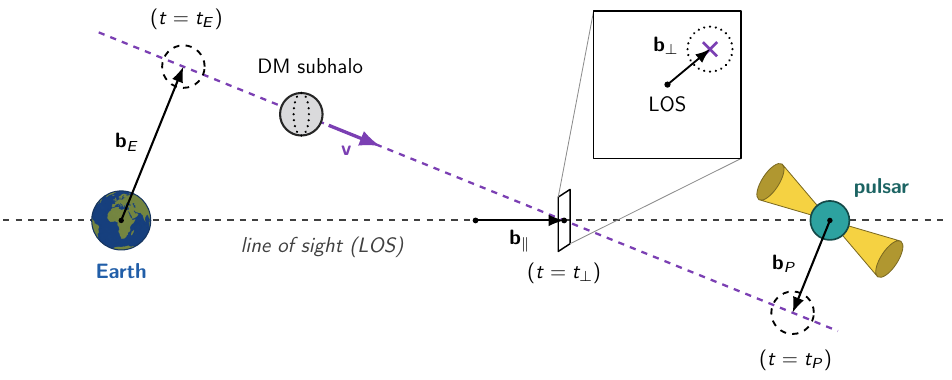}
    \caption{%
    Geometry of a DM subhalo transiting the Earth--pulsar system. The Earth (blue, left) and pulsar (red, right) are connected by the line of sight (dashed line). A DM subhalo (black circle) travels along a straight-line trajectory (purple dashed line) with velocity $\vec{v}$. At different times, the subhalo reaches closest approach to Earth ($t = t_E$, with impact parameter $\vec{b}_E$), to the line of sight ($t = t_\perp$, with impact parameter $\vec{b}_\perp$ at a distance $b_\parallel$ from the midpoint between Earth and pulsar), and to the pulsar ($t = t_P$, with impact parameter $\vec{b}_P$). The Doppler and Einstein contributions to the timing residual arise from the gravitational influence of the subhalo on the Earth and pulsar (Earth and pulsar terms), where the relevant length scales are $b_E$ and $b_P$, respectively, while the Shapiro contribution arises from the photon traversing the perturbed spacetime along the line of sight, where the relevant distances are $b_{\perp}$ and $b_{\parallel}$. %
    }
    \label{fig:PTA_DM_cartoon}
\end{figure}
In Fig.~\ref{fig:PTA_DM_cartoon}, we show the geometry of the system. We consider the trajectory of a typical DM subhalo, and show the relevant distance quantities ($b_E$, $b_P$, and $b_{\perp}$) that contribute to the observable.

Evaluating each individual term in Eq.~\eqref{eqn:subhalo_observable} using Eq.~\eqref{eqn:subhalo_potential} and Eq.~\eqref{eqn:DM_parameterization}, we compute the Doppler and Einstein terms and find
\begin{subequations}\label{eqn:DM_evaluated}
\begin{align}
\delta t_{\mathcal D}^{(E)}(t)
&=
\frac{GM}{v^2}
\left[
\sqrt{1+x_E^2}\,(\unit{n}\cdot \unit{b}_E)
-\sinh^{-1}(x_E)(\unit{n}\cdot \unit{v})
\right],
\label{eqn:DM_evaluated_doppler_earth}
\\
\delta t_{\mathcal D}^{(P)}(t)
&=
\frac{GM}{v^2}
\left[
\sqrt{1+x_P^2}\,(\unit{n}\cdot \unit{b}_P)
-\sinh^{-1}(x_P)(\unit{n}\cdot \unit{v})
\right],
\label{eqn:DM_evaluated_doppler_pulsar}
\\
\delta t_{\mathcal E}^{(E)}(t)
&=
-\frac{GM}{v}\sinh^{-1}(x_E),
\label{eqn:DM_evaluated_einstein_earth}
\\
\delta t_{\mathcal E}^{(P)}(t)
&=
-\frac{GM}{v}\sinh^{-1}(x_P).
\label{eqn:DM_evaluated_einstein_pulsar}
\end{align}
\end{subequations}
where we have defined dimensionless time variables, $x_E\equiv (t-t_E)/\tau_E$ and $x_P\equiv (t-t_{\mathcal{D}, 0}
)/\tau_P$, in which $\tau_E\equiv b_E/v$ and $\tau_P\equiv b_P/v$ characterize the time durations of the interaction and we define $t_{\mathcal{D}, 0} = t_P+L$. The time $t_P$ is the coordinate time at which the subhalo is closest to the pulsar, as defined in Eq.~\eqref{eqn:DM_parameterization}, while $t_{\mathcal D,0}$ is the Earth-arrival time of the pulse most affected by that encounter, where the extra $L$ accounts for the pulsar-to-Earth light travel time. We observe from Eq.~\eqref{eqn:DM_evaluated} that the Einstein term contribution is suppressed by a factor of $v$ compared to the Doppler contribution. The expressions for the Doppler effect have been derived in Ref.~\cite{Dror:2019twh}, but we now put them in the proper context of a gauge-invariant observable, and show explicitly how each term adds to the time shifts being measured. %

To derive the Shapiro contribution, we follow Ref.~\cite{Du:2023dhk} and write the photon-DM distance as  %
\begin{align} \label{eqn:NR_simplification}
	&\vec{r}_{\mathrm{DM}}(t-(z+L/2))-\vec{r}_{\mathrm{mid}}-z\unit{n} \nonumber\\
    &= \left[b_\parallel+v_\parallel\left(t-\frac L2-t_\perp\right)-z \right]\unit n + \left[\vec b_\perp+\vec v_\perp\left(t-\frac L2-t_\perp\right) \right]-z\vec v  \nonumber\\
    &= \left[b_{\parallel}+v_{\parallel}\left(t-\frac{L}{2}-t_{\perp}\right)-z\right]\unit{n} + \left[\vec{b}_{\perp} + \vec{v}_{\perp}\left(t-\frac{L}{2}-t_{\perp}\right)\right] -b_\parallel\vec{v} + (b_{\parallel}-z) \vec{v} \nonumber\\
    &= \left[b_{\parallel}+v_{\parallel}\left(t-\frac{L}{2}-b_\parallel - t_{\perp}\right)-z\right]\unit{n} + \left[\vec{b}_{\perp} + \vec{v}_{\perp}\left(t-\frac{L}{2}-b_\parallel-t_{\perp}\right)\right]
\end{align} 
where we used the parameterization in Eq.~\eqref{eqn:DM_parameterization} and in the last line, used  the non-relativistic approximation, $(b_{\parallel}-z)(\vec{v}+\unit{n})\approx (b_{\parallel}-z)\unit{n}$. The Shapiro term in Eq.~\eqref{eqn:subhalo_observable} can now be rewritten as
\begin{align}\label{eqn:DM_Shapiro}
	\delta t_{\mathcal{S}}(t) &= 2GM\int_{-L/2}^{L/2} dz\, \frac{1}{\sqrt{(r_{\parallel}-z)^2+r_{\perp}^2}} \, ,
\end{align}
where %
\begin{align}\label{eqn:r_par_perp_def}
	r_{\parallel} &\equiv b_{\parallel}+v_{\parallel}\left(t-t_{\mathcal{S}, 0}\right),
    \\
	r_{\perp}^2&\equiv b_{\perp}^2 + v_{\perp}^2\left(t-t_{\mathcal{S}, 0}\right)^2,
\end{align}
where we define $t_{\mathcal{S}, 0}\equiv L/2+b_\parallel+t_\perp$, the Earth-arrival time of the photon whose path passes closest to the subhalo along the line of sight.
The integral in Eq.~\eqref{eqn:DM_Shapiro} can be readily evaluated as
\begin{align}\label{eqn:DM_Shapiro_2}
	\delta t_{\mathcal{S}}(t) &= 2GM\log\left[\frac{r_{\parallel}+(L/2)+\sqrt{r_{\parallel}^2+\left[r_{\parallel}+(L/2)\right]^2}}{r_{\parallel}-(L/2)+\sqrt{r_{\perp}^2+\left[r_{\parallel}-(L/2)\right]^2}}\right] \nonumber\, , \\
	&\approx 2GM \begin{dcases*}
		\log\left(\frac{L^2}{r_{\perp}^2}\right) & if $\sqrt{r_{\perp}^2+r_{\parallel}^2}\lesssim L/2$ \\
		\frac{L}{\sqrt{r_{\perp}^2+r_{\parallel}^2}} & if $\sqrt{r_{\perp}^2+r_{\parallel}^2}\gtrsim L/2$
	\end{dcases*} \, .
\end{align}
Physically, $\sqrt{r_\perp^2+r_\parallel^2}$ is the distance of the subhalo from $\vec{r}_{\rm mid}$ at the retarded time when the photon observed at time $t$ passes the line-of-sight point closest to the subhalo trajectory. The two limits in Eq.~\eqref{eqn:DM_Shapiro_2} then distinguish whether the subhalo lies inside the Earth--pulsar segment or beyond its endpoints: for $\sqrt{r_\perp^2+r_\parallel^2}\lesssim L/2$ the delay is controlled by the perpendicular distance $r_\perp$ and is nearly independent of $r_\parallel$, whereas for $\sqrt{r_\perp^2+r_\parallel^2}\gtrsim L/2$ it falls off as $L/\sqrt{r_\perp^2+r_\parallel^2}$ with the total distance. In each limit we can further simplify Eq.~\eqref{eqn:DM_Shapiro_2} by dropping terms that are degenerate with the timing model
\begin{align}\label{eqn:DM_Shapiro_3}
	\delta t_{\mathcal{S}}(t)\approx 
    -2GM \begin{dcases*}
		\log\left(1+x_{\perp}^2\right) & if $ \sqrt{r_{\perp}^2+r_{\parallel}^2}\lesssim L/2$ \\
		-\frac{L}{\sqrt{r_{\parallel}^2+r_{\perp}^2}} & if $\sqrt{r_{\perp}^2+r_{\parallel}^2}\gtrsim L/2$
	\end{dcases*} \, ,
\end{align}
where we defined 
$x_{\parallel}\equiv v_{\parallel}(t-t_{\mathcal{S}, 0}
)/b_{\parallel}$,
$x_{\perp}\equiv (t-t_{\mathcal{S}, 0}
)/\tau_{\perp}$, with $\tau_\perp\equiv b_\perp/\bar{v}_\perp$ characterizing the duration of the flyby. Note that $r_{\parallel}=b_{\parallel}(1+x_{\parallel})$ and $r_{\perp}^2=b_{\perp}^2(1+x_{\perp}^2)$.
The first line in Eq.~\eqref{eqn:DM_Shapiro_3} is derived and used in Ref.~\cite{Dror:2019twh}, although we now explicitly show how it arises from the limit of a DM subhalo passing close to the Earth-pulsar line-of-sight. %

\section{Analytic Sensitivity to Dark Matter Substructure}
\label{sec:analytical_estimates}

With both the statistical framework for analysing PTA data and the form of the DM signals at hand, we are now ready to compute the sensitivity of PTA experiments to DM substructure. %
Before showing the full result, which requires numerical simulations, we first derive analytic formulae to estimate the projected sensitivity, which allow us to identify the key scaling relations governing detectability and to expand on several subtleties of the problem. %
To do so, we employ three 
simplifying assumptions: First, we only consider the Doppler pulsar term ($\delta t^{(P)}_{\mathcal{D}}$) and Shapiro term ($\delta t_{\mathcal{S}}$) in the gauge-invariant observable in Eq.~\eqref{eqn:delta_t_t} as derived in Sec.~\ref{sec:dark_matter_residual}. %
In particular, we neglect the Einstein delays ($\delta t^{(P)}_{\mathcal{E}}$ and $\delta t^{(E)}_{\mathcal{E}}$), which are subdominant as noted in Sec.~\ref{sec:dark_matter_residual}\footnote{While the Shapiro term also seems to be weaker than the Doppler term by a factor of $v^2$, it is compensated by the long baseline $L$ between the Earth and a pulsar, so its contribution can exceed the Doppler term for certain subhalos~\cite{Dror:2019twh}.}, and we also defer discussion of the Doppler Earth term ($\delta t^{(E)}_{\mathcal{D}}$) to a companion paper~\cite{in_prep}. %
Second, we compute the SNR for each term individually, as we expect that for any given DM geometry, only one term in Eq.~\eqref{eqn:delta_t_t} dominates the signal. %
Third, we separate the analysis into three regimes (static, dynamic, and stochastic) depending on the dynamics of the signals, as we explain below. %
These assumptions are relaxed in Sec.~\ref{sec:results_and_discussion} which uses the full gauge-invariant observable derived in Sec.~\ref{sec:dark_matter_residual} and  handles all encounter timescales continuously. As we show later in our main result in Fig.~\ref{fig:analytic_overlay}%
, the analytic scaling relations derived in this section are in good agreement with the numerical results in the relevant limits.

\begin{figure}
    \includegraphics[width=\linewidth]{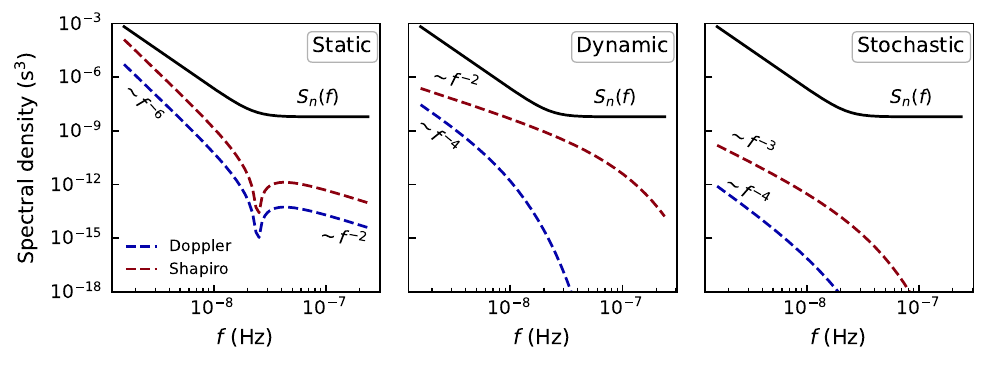}
    \caption{%
   Analytic spectra of the projected DM signal 
   versus the noise PSD $S_n(f)$, assuming a SKA-like PTA benchmark with a SMBHB-baseline GWB (see Sec.~\ref{sec:results_and_discussion}). 
   For the deterministic regime (static, dynamic limits), we plot the single‑event periodogram $\tfrac{4}{T}|\delta t^\perp(f)|^2$ of the loudest event in the array,
   evaluated at the 10th percentile over realizations. 
    In the stochastic regime, we instead plot the pulsar-term auto-PSD; combining the independent pulsar signals gives a $\sqrt{N_P}$ gain in the SNR
    that is not folded into the plotted curves. All curves assume $f_{\rm DM}=1$ and are plotted at representative masses near each regime's peak sensitivity, specified below. On the left panel we show the ~static limit --- both Doppler (mass-independent) and Shapiro ($M=M_\odot$) signal spectra scale as $f^{-6}$ at low frequencies and transition to $f^{-2}$ above the crossover frequency. %
    The sharp notch marks the $f^{-6}\!\to\!f^{-2}$ transition, where the projected signal briefly cancels. %
    On the middle panel we show the dynamic limit --- the Doppler spectrum ($M=10^{-9}M_\odot$) scales as $f^{-4}$ and the Shapiro ($M=10^{-3}M_\odot$) as $f^{-2}$, with exponential suppression at $f\gtrsim 1/\tau$. The
    right panel shows the ~stochastic limit --- the Doppler %
    PSD ($M=10^{-12}M_\odot$) scales as $f^{-4}$ and the Shapiro ($M=10^{-5}M_\odot$) as $f^{-3}$. %
    }
    \label{fig:signal_PSDs}
\end{figure}

We begin by discussing various regimes of the analysis. We also show the spectral density of the signal in each regime in the panels of Fig.~\ref{fig:signal_PSDs}.
\begin{itemize}[leftmargin=*]
    \item \textit{Static limit} ($\tau \gg T$): the flyby is unresolved over the observation window. The signal is well approximated by a low-order Taylor expansion in $t/\tau$, and after marginalization over the quadratic timing model, the leading contribution projects onto the cubic template $\phi_3^\perp$ for both Doppler (Eq.~\eqref{eqn:dop_static_proj_simplified}) and Shapiro (Eq.~\eqref{eqn:shap_static_proj_simplified}). Since $\tilde{\phi}_3^\perp(f) \sim f^{-3}$ at low frequencies (\textit{cf.} App.~\ref{app:phi3perp_norm}), the projected signal power $|\delta\tilde{t}^\perp(f)|^2/T$ scales as $f^{-6}$ in both cases; see the first panel of Fig.~\ref{fig:signal_PSDs}. 
    \item \textit{Dynamic limit} ($\tau \ll T$): the flyby is resolved --- a brief encounter near closest approach --- but the residual it leaves is long-lived, carrying power to low frequencies.
    The projected signal power, $|\delta\tilde{t}^\perp(f)|^2/T$, scales as $f^{-4}$ for Doppler (Eq.~\eqref{eqn:dop_dyn_FT}) and $f^{-2}$ for Shapiro (Eq.~\eqref{eqn:shap_dyn_FT}) at frequencies $f \lesssim 1/\tau$; see the second panel of Fig.~\ref{fig:signal_PSDs}.
    \item \textit{Stochastic limit}: %
    the number of encounters during the observation window that contribute comparably to the variance is large, individual events are unresolved, and their cumulative effect is modeled as a Gaussian process. The projected signal PSD, $S^\perp_{\mathrm{sig}}(f)$, scales as $f^{-4}$ for Doppler (Eq.~\eqref{eqn:dop_stoch_PSD}) and $f^{-3}$ for Shapiro (Eq.~\eqref{eqn:shap_stoch_PSD}); see the third panel of Fig.~\ref{fig:signal_PSDs}. %
\end{itemize}

We now apply the single-pulsar statistic of Sec.~\ref{sec:SNR} to the Doppler pulsar term and the Shapiro line-of-sight signal, where only the single-pulsar autocorrelation $S_{aa}(f)$, denoted $S_n(f)$, enters. In the analytic estimates below, we take the red component of this single-pulsar PSD to be the diagonal GWB contribution. %
To quantify the red-noise suppression, we conveniently write the total single-pulsar noise PSD in Eq.~\eqref{eqn:explicit_Sn} as
\begin{equation} \label{eqn:noise_PSD_fstar}
    S_n(f) = 2\Delta t\, t_{\mathrm{rms}}^2 \left[1 + \left(\frac{f_\star}{f}\right)^\gamma\right],
\end{equation}
where $f_\star$ is the frequency at which the noise transitions from red- to white-dominated and for the rest of the paper, we define the shorthand $\gamma \equiv \gamma_\text{GWB}$. All sensitivity scalings derived below are controlled by the dimensionless combination $f_\star T$, and in particular are valid in the regime $f_\star T \gg 1$. For the SKA-like benchmark parameters and the current NANOGrav 15-year posterior in Sec.~\ref{sec:results_and_discussion}, this assumption is comfortably satisfied,
\begin{equation}
    \label{eqn:fstarT}
    f_\star T = 20 \cdot 0.25^{\frac{1}{\gamma}}\left(\frac{T}{20\,  \mathrm{yr}}\right)\,\left(\frac{A_{\mathrm{GWB}}}{2.4\times 10^{-15}}\right)^{\frac{2}{\gamma}} \left(\frac{2\,\mathrm{wk}}{\Delta t}\right)^{\frac{1}{\gamma}}\left(\frac{50\,\mathrm{ns}}{t_{\mathrm{rms}}}\right)^{\frac{2}{\gamma}}.
\end{equation}
Using the parameterization in Eq.~\eqref{eqn:noise_PSD_fstar} %
, the noise-weighted inner product of Eq.~\eqref{eqn:inner_prod_def} simplifies to
\begin{equation} \label{eqn:inner_prod_simplified}
(g|h)  =\frac{2}{\Delta t t^2_\text{rms}} \,\mathrm{Re}\left[\int_{1/T}^{\infty} df \frac{f^\gamma}{f_\star^\gamma+f^\gamma} \widetilde{g}^*(f)\widetilde{h}(f) \right].
\end{equation}
As we derive below, each DM signal of interest has a low-frequency power-law projected spectrum, $|\delta\tilde{t}^\perp(f)|^2\sim f^{-n}$ %
, for a signal-dependent exponent $n$. Because the projected signal power %
scales as $f^{-n}$ at low frequencies in every case, the SNR reduces to a dimensionless integral whose scaling with $f_\star T$ can be evaluated in closed form. This integral takes one of two forms. For a deterministic signal, the SNR is set by the noise-weighted norm of the signal (Eq.~\eqref{eqn:projected_snr_pulsar}), which involves a single power of the noise suppression factor; for a stochastic signal, the SNR involves the square of the signal-to-noise per frequency bin (Eq.~\eqref{eqn:stoch_SNR_fourier}), doubling the suppression
\begin{equation}\label{eqn:integral_scalings}
    \begin{split}
    \text{det:}\quad I_{n,\gamma} &\equiv \int_{1/T}^{\infty}\frac{df}{f}\,\frac{f^{\gamma}}{f_\star^\gamma+f^\gamma}\,\frac{1}{(fT)^{n-1}} \sim (f_\star T)^{-\min(\gamma,\,n-1)},\\
    \text{stoch:}\quad I_{n,\gamma}^{(2)} &\equiv \int_{1/T}^{\infty}\frac{df}{f}\left(\frac{f^{\gamma}}{f_\star^\gamma+f^\gamma}\right)^{\!2}\frac{1}{(fT)^{2n-1}} \sim (f_\star T)^{-2\min(\gamma,\,n-1/2)},
    \end{split}
\end{equation}
   where the scalings hold for $f_\star T \gg 1$.\footnote{For the analytic estimates below, we use \(\sim\) to denote equality up to order-one numerical factors, and occasionally logarithmic dependence. We use \(\lesssim\) and \(\gtrsim\) in the same parametric sense: the boundary may shift by order-one factors. We use \(\propto\) when only the dependence on a specified parameter is being indicated and other dimensional or parametric factors are being suppressed. By contrast, \(\approx\) denotes an approximation to the actual expression, not merely its parametric scaling.} Both scalings follow from a simple argument (detailed in App.~\ref{sec:appendix_integral}) locating where the weight of the SNR integrand in Eq.~\eqref{eqn:integral_scalings} accumulates ---  at the lowest resolved frequency $f\sim 1/T$ or at the red-to-white crossover $f\sim f_\star$  --- with the $\min$ selecting whichever the signal and noise slopes favor.
   The values of $n$ for each signal type follow from the regime definitions above. In Table~\ref{tab:signal_scalings}, we summarize the scalings of the power and resulting SNR for each signal. %

\begin{table}
\centering
\renewcommand{\arraystretch}{1.6}
\setlength{\tabcolsep}{8pt}
\small
\begin{tabular}{|c|c|c|c|c|c|c|}
\hline
\multirow{2}{*}{\textbf{Signal}} & \multirow{2}{*}{\textbf{Regime}} & \multirow{2}{*}{\textbf{Spectral density}} & \multicolumn{4}{c|}{\textbf{SNR scaling}} \\
\cline{4-7}
 & & & $\boldsymbol{T}$ & $\boldsymbol{f}_{\mathbf{DM}}$ & $\boldsymbol{M}$ & $\boldsymbol{N_P}$ \\
\hline
\multirow{3}{*}{Doppler}
  & Static     & $|\delta \tilde{t}^{\perp}(f)|^2 \sim f^{-6}$ & $T^{\tfrac{7}{2}}\,(f_\star T)^{-\tfrac{\min(\gamma,5)}{2}}$ & $f_{\rm DM}$ & --- & $N_P$ \\
  & Dynamic    & $|\delta \tilde{t}^{\perp}(f)|^2 \sim f^{-4}$ & $T^{2}\,(f_\star T)^{-\tfrac{\min(\gamma,3)}{2}}$ & $f_{\rm DM}^{1/2}$ & $M^{1/2}$ & $N_P^{1/2}$ \\
  & Stochastic & $S^{\perp}_{\rm DM}(f) \sim f^{-4}$ & $T^{4}\,(f_\star T)^{-\min\left(\gamma,\tfrac{7}{2}\right)}$ & $f_{\rm DM}$ & $M$ & $N_P^{1/2}$ \\
\hline
\multirow{3}{*}{Shapiro}
  & Static     & $|\delta \tilde{t}^{\perp}(f)|^2 \sim f^{-6}$ & $T^{\tfrac{7}{2}}\,(f_\star T)^{-\tfrac{\min(\gamma,5)}{2}}$ & $f_{\rm DM}^{3/2}$ & $M^{-1/2}$ & $N_P^{3/2}$ \\
  & Dynamic    & $|\delta \tilde{t}^{\perp}(f)|^2 \sim f^{-2}$ & $T^{\tfrac{1}{2}}\,(f_\star T)^{-\tfrac{\min(\gamma,1)}{2}}$ & --- & $M$ & --- \\
  & Stochastic & $S^{\perp}_{\rm DM}(f) \sim f^{-3}$ & $T^{3}\,(f_\star T)^{-\min\left(\gamma,\tfrac{5}{2}\right)}$ & $f_{\rm DM}$ & $M$ & $N_P^{1/2}$ \\
\hline
\end{tabular}
\caption{
Overview of the low-frequency scaling of the projected spectral density for Doppler and Shapiro signals deep in the static, dynamic, and stochastic regimes, and the corresponding scaling of the SNR with $T$, the substructure fraction $f_{\rm DM}$, the subhalo mass $M$, and the array size $N_P$, in the presence of the GWB, assuming $f_\star T \gg 1$. A dash (---) indicates that the SNR is independent of the corresponding parameter in that regime.
}
\label{tab:signal_scalings}
\end{table}

Throughout the rest of this section, we quote the projected upper limit on the substructure fraction, $f_{\mathrm{DM}}\equiv \rho_{\mathrm{subhalo}}/\rho_{\mathrm{DM}}$ --- the fraction of the total local DM density, $\rho_{\mathrm{DM}}=0.46 $ GeV/cm$^3$, contained in subhalos --- attained for $90\%$ of universes. In other words, this means that they have a number density $\bar{n}=f_\text{DM} \rho_{DM}/M$. We model the subhalos as point masses of mass $M$ moving at a common speed $v = 340\,\mathrm{km/s}$ with isotropically sampled directions. The Doppler signal depends on this common speed, whereas the Shapiro signal depends on the transverse speed $v_\perp$, which varies with the direction of motion relative to the line of sight even at fixed $v$; for the Shapiro estimates we therefore work throughout with its mean value, $\bar{v}_\perp =\langle v_\perp\rangle\approx270\,\mathrm{km/s}$.\footnote{Here $v=340\,\mathrm{km/s}$ is the mean speed of the boosted Maxwell--Boltzmann distribution considered in Ref.~\cite{Ramani:2020hdo}. We expect that a full velocity-distribution treatment would change our reach only at the order-unity level, apart from the directional anisotropy %
arising from Earth's motion through the galaxy.}
The SNR thresholds we adopt follow from requiring a false-alarm probability below $5\%$ under the null hypothesis. For pulsar term deterministic analyses, the detection statistic is the per-pulsar SNR, which is one-sided Gaussian under the null hypothesis. Demanding that none of the $N_P$ pulsars produces a false positive gives $\left[\mathrm{erf}\left(\mathrm{SNR}/\sqrt{2}\right)\right]^{N_P} = 0.95$, yielding $\mathrm{SNR} \approx 4$ for $N_P \sim 200$~\cite{Dror:2019twh}. For the stochastic pulsar-term analyses the pulsars combine into a single array statistic, for which the $5\%$ false-alarm threshold is $\mathrm{SNR}\approx 2$. We nonetheless adopt $\mathrm{SNR}=4$ throughout for simplicity, conservative at the $\mathcal{O}(1)$ level of our estimates.

\subsection{Static Limit}
\label{sec:analytics_static_limit}
In the static limit ($\tau \gg T$) the DM flyby is unresolved over the observation window: the dimensionless time $t/\tau$ remains small throughout $[0,T]$, so we Taylor-expand each signal in powers of $t/\tau$. The constant, linear, and quadratic terms are degenerate with the timing-model basis $\{\phi_0, \phi_1, \phi_2\}$ (the mean, spin frequency, and spin-down fit, respectively) and are absorbed in the marginalization of Sec.~\ref{sec:timing_model_marginalization}; the leading non-degenerate contribution is therefore the cubic term, which projects entirely onto the orthogonalized template $\phi_3^\perp$ for both Doppler and Shapiro. Projecting onto $\phi_3^\perp$ rather than $\phi_3$ is essential because the red-noise weighting spoils the ordinary orthogonality of the Legendre basis, specifically $(\phi_1|\phi_3)\neq 0$: the cubic mode is no longer clean of the timing model, and $\phi_3^\perp$ is the part that remains orthogonal to it under the red-noise-weighted inner product.
The relevant quantity that enters the SNR in this limit is then $(\phi_3^\perp|\phi_3^\perp)$, which we derive in App.~\ref{app:phi3perp_norm} and evaluate to be
\begin{equation} \label{eqn:phi3perp_norm}
    (\phi_3^\perp|\phi_3^\perp)
    \simeq \frac{3150}{\pi^6}\,
    \frac{T}{\Delta t\, t^2_{\mathrm{rms}}}
    \left(f_\star T\right)^{-\min(\gamma,5)} \, .
\end{equation}
This corresponds to $n=6$ as shown in Tab.~\ref{tab:signal_scalings}. The projected cubic template also explains a feature of the static-limit spectra in
Fig.~\ref{fig:signal_PSDs}: the sharp dip where $|\delta\tilde{t}^\perp(f)|^2$ drops as the
spectrum crosses from its low-frequency $f^{-6}$ scaling to the $f^{-2}$ tail. The dip
falls at the node of that template, where the two components cancel. It does not affect
the sensitivity, which is dominated by frequencies well below it.

\subsubsection{Doppler}

We first consider timing signal from the Doppler term. 
The cubic Taylor term of Eq.~\eqref{eqn:DM_evaluated}, $\tfrac{1}{6}\delta t_\mathcal{D}'''(0)\,t^3$ (evaluated explicitly in Ref.~\cite{Dror:2019twh}), decomposes on the Legendre basis of Eq.~\eqref{eqn:legendre_def} as $t^3 = \tfrac{T^3}{20\sqrt{7}}\,\phi_3(t) + (\text{timing-model modes})$, leaving the projected static Doppler signal
\begin{equation}\label{eqn:dop_static_proj_simplified}
\begin{split}
\delta t_\mathcal{D}^{\perp}(t)
&\approx \frac{GMv\,T^3} {120\sqrt{7}\,r_P^3} \, \phi_3^\perp(t)\!\left[\unit{v}+\frac{3vt_{\mathcal{D}, 0}}{r_{ P}}\unit{r}_P\right]\!\cdot\!\unit{n} .
\end{split}
\end{equation}
For randomly oriented closest encounters $\tau_P$ and $t_{\mathcal{D}, 0}$ are parametrically the same scale, so the prefactor $[\unit{v} + 3(vt_{\mathcal{D}, 0}/r_P)\unit{r}_P]\cdot\unit{n} \sim 1$, and we drop it.\footnote{Here and in the analogous static Shapiro estimate in Eq.~\eqref{eqn:shap_static_proj_simplified}, Ref.~\cite{Dror:2019twh} estimates the dropped order-unity prefactor as $\sim 2$ rather than $\sim 1$ (the isotropic-orientation average), accounting for a $\sim 2\times$ difference between our static-signal expressions and theirs.} It is clear that the size of the signal crucially depends on the distance between the DM subhalo and the pulsar. Assuming that the signal is dominated by the subhalo closest to the pulsar with distance $r_P=r_{\min}$,
we use the distribution of $r_{\min}$ across realizations of the subhalo ensemble, derived in Ref.~\cite{Dror:2019twh} assuming a Poisson distribution of subhalos. Let $r_p$ denote its $p$-th percentile, defined by $P(r_{\min} \le r_p) = p$ --- so a fraction $p$ of realizations have a subhalo at least this close. Explicitly,
\begin{equation}\label{eq:static_closest}
    r_p = \left[-\frac{3}{4\pi}\frac{M}{N_P f_{\rm DM}\rho_{\mathrm{DM}}}\ln(1-p)\right]^{1/3}.
\end{equation}
Because $\mathrm{SNR}_{\mathcal{D},\,\mathrm{stat}}$ decreases monotonically with $r_0$, evaluating it at $r_0 = r_{0.9}$ gives a lower bound exceeded in $90\%$ of realizations. Substituting Eq.~\eqref{eq:static_closest} with $p=0.9$ and Eq.~\eqref{eqn:phi3perp_norm} into Eq.~\eqref{eqn:projected_snr_pulsar} yields
\begin{equation}\label{eqn:snr_gdop_stat}
    \text{SNR}_{\mathcal{D},\,\mathrm{stat}} \sim \frac{14}{\left(f_\star T\right)^{{\min(\gamma, 5)}/{2}}}
    \left(\frac{N_Pf_{\rm DM}}{200}\right)
    \left(\frac{T}{20\, \rm{yr}}\right)^{\frac{7}{2}} \left(\frac{2 \, \rm{wk}}{\Delta t}\right)^{\frac{1}{2}} \left(\frac{50 \, \rm{ns}}{t_{\rm rms}}\right) .
\end{equation}
The resulting total SNR scaling in powers of $T$ %
arises from the combination of the explicit $T^3$ growth of the static Doppler signal, an additional $T^{1/2}$ enhancement from the noise-weighted norm of the projected cubic template, and dependence on the red-noise properties of the PTA entering through $(\phi_3^\perp|\phi_3^\perp)$, which yields the $(f_\star T)^{-\min(\gamma,5)/2}$ suppression inherited from the static-limit projection, as summarized in Table~\ref{tab:signal_scalings}. Setting $\text{SNR}_{\mathcal{D},\,\mathrm{stat}}=4$
, we find that the Doppler signal in the static limit is sensitive to substructure with fraction. %
\begin{equation} \label{eqn:dop_stat_sensitivity}
    f_{\mathrm{DM}}^{{\mathcal{D}}, \, \rm{stat}}
    \gtrsim 0.3\, (f_\star T)^\frac{\min(\gamma, 5)}{2}
    \left(\frac{200}{N_P}\right)
    \left(\frac{20\text{yr}}{T}\right)^{\frac{7}{2}} \left(\frac{\Delta t}{2 \, \rm{wk}}\right)^{\frac{1}{2}} \left(\frac{t_{\rm rms}}{50 \, \rm{ns}}\right) .
\end{equation}
This reach is independent of $M$, but applies only where the closest flyby evolves slowly over the observing baseline, $r_{\text{min}} \gtrsim vT$. Using Eq.~\eqref{eq:static_closest}, this restricts the static-limit analysis to
\begin{equation} \label{eqn:dop_stat_validity}
    f_{\rm DM}^{{\mathcal{D}}, \, \rm{stat}}
    \lesssim \left(\frac{200}{N_P}\right)
    \left(\frac{M}{10^{-6} M_\odot}\right)
    \left(\frac{20\text{yr}}{T}\right)^{3}.
\end{equation}
\subsubsection{Shapiro}
Like the Doppler signal, the projected static Shapiro signal has low-frequency scaling $|\delta\tilde{t}^\perp|^2 \sim f^{-6}$ (Table~\ref{tab:signal_scalings}), so the GWB suppression is identical. The sensitivity scaling differs, however, because the relevant impact parameter is the perpendicular distance to the line of sight rather than to the pulsar, the SNR therefore scales as $(N_P f_{\rm DM})^{3/2}$ rather than linearly. In the close-encounter regime $\sqrt{r_\parallel^2+r_\perp^2}\lesssim L/2$, the Shapiro pulsar-term signal of Eq.~\eqref{eqn:DM_Shapiro_3} reads
\begin{equation}
    \delta t_\mathcal{S}(t) = -2GM\,\log \bigl(1+x_\perp^2(t)\bigr),\qquad x_\perp(t)\equiv\frac{t-t_{\mathcal{S}, 0}}{\tau_\perp}.
\end{equation}
Repeating the static-limit Taylor expansion of the Doppler case (evaluated explicitly in Ref.~\cite{Dror:2019twh}), the projected static Shapiro signal is 
\begin{equation}\label{eqn:shap_static_proj_simplified}
\begin{split}
\delta t_\mathcal{S}^{\perp}(t)
&\approx \frac{GM\,\bar{v}_\perp^3}{15\sqrt{7}\,r_\perp^3}\,T^3\phi_3^\perp(t)\left(\frac{t_{\mathcal{S}, 0} \bar{v}_\perp}{r_\perp}\right)^3\left(1-\frac{3\tau_\perp^2}{t_{\mathcal{S}, 0}^2}\right),
\end{split}
\end{equation}
where we replace $v_\perp$ of a subhalo with its typical value $\bar{v}_\perp=\langle v_\perp\rangle$. For randomly oriented closest encounters $\tau_\perp$ and $t_{\mathcal{S}, 0}$ are parametrically the same scale, so the prefactor $(t_{\mathcal{S}, 0} v_\perp/r_\perp)^3 (1 - 3\tau_\perp^2/t_{\mathcal{S}, 0}^2) \sim 1$ and we drop it.
As in the Doppler case, assume the signal is dominated by the subhalo with the smallest initial perpendicular distance to the line of sight, $r_\perp = r_{\perp, \min}$.
Its $p$-th percentile, now from a 2D Poisson distribution projected along the line of sight, is \cite{Dror:2019twh}
\begin{equation}\label{eq:static_closest_shap}
    r_{\perp, p} = \sqrt{-\frac{1}{\pi}\frac{M}{N_P f_{\rm DM}\rho_{\mathrm{DM}}d}\ln(1-p)}.
\end{equation}
Substituting at $p=0.9$ along with Eq.~\eqref{eqn:phi3perp_norm} into Eq.~\eqref{eqn:projected_snr_pulsar} yields 
\begin{equation}\label{eqn:snr_gshap_stat}
    \mathrm{SNR}_{\mathcal{S},\,\mathrm{stat}} \sim \frac{70}{\left(f_\star T\right)^{\min(\gamma, 5)/2}}
    \left(\frac{N_Pf_{\rm DM}}{200}\right)^\frac{3}{2}
    \left(\frac{M_\odot}{M}\right)^\frac{1}{2}
    \left(\frac{T}{20\text{ yr}}\right)^\frac{7}{2} \left(\frac{2 \, \rm{wk}}{\Delta t}\right)^{\frac{1}{2}} \left(\frac{50 \, \rm{ns}}{t_{\rm rms}}\right) \left(\frac{L}{5\,\rm{kpc}}\right)^{\frac{3}{2}}.
\end{equation}
Setting $\mathrm{SNR}_{\mathcal{S},\,\mathrm{stat}}=4$, we find that the Shapiro signal in the static limit is sensitive to substructure with fraction
\begin{equation} \label{eqn:shap_stat_sensitivity}
    f_{\rm DM}^{\mathcal{S}, \, \rm{stat}}
    \gtrsim 0.15\,
    (f_\star T)^\frac{\min(\gamma, 5)}{3}
    \left(\frac{200}{N_P}\right)
    \left(\frac{M}{M_\odot}\right)^{\frac{1}{3}}
    \left(\frac{20\text{yr}}{T}\right)^{\frac{7}{3}} \left(\frac{\Delta t}{2 \, \rm{wk}}\right)^{\frac{1}{3}} \left(\frac{t_{\rm rms}}{50 \, \rm{ns}}\right)^{\frac{2}{3}}\left(\frac{5\,\rm{kpc}}{L}\right).
\end{equation}
As $M$ decreases the reach improves, but applies only where the closest flyby evolves slowly over the observing baseline, $r_{\perp,\min} \gtrsim \bar{v}_\perp T$. This restricts the static-limit analysis to
\begin{equation}  \label{eqn:shap_stat_validity}
    f^\text{S, stat}_{\rm DM}
    \lesssim \left(\frac{200}{N_P}\right)
    \left(\frac{M}{M_\odot}\right)
    \left(\frac{20\text{yr}}{T}\right)^{2}
    \left(\frac{5\, \mathrm{kpc}}{L}\right).
\end{equation}

\subsection{Dynamic Limit}
\label{sec:analytics_dynamic_limit}

In the dynamic limit the flyby is fast, with an interaction timescale $\tau \ll T$, and the signal is no longer captured by the low-order Taylor expansion of the static limit. Although the encounter is brief, the residual it leaves extends well beyond it and is not compactly supported in time, varying slowly enough to carry power down to low frequencies. As we show below in Eq.~\eqref{eqn:dop_dyn_FT} and Eq.~\eqref{eqn:shap_dyn_FT}, this low-frequency content sets the Doppler and Shapiro envelopes, $1/f^2$ and $1/f$, so their spectra scale as $|\delta\tilde t_\mathcal{D}|^2\sim f^{-4}$ and $|\delta\tilde t_\mathcal{S}|^2\sim f^{-2}$ (Table~\ref{tab:signal_scalings}).

\subsubsection{Doppler}
\label{sec:analytics_dop_dyn_pulsar}
The Doppler signal from Eq.~\eqref{eqn:DM_evaluated} in frequency space is given by %
\begin{equation}\label{eqn:dop_dyn_FT_finite}
    \begin{split}
        \delta \tilde{t}_\mathcal{D}(f) &= \int_0^T dt\, e^{-2\pi i f t} \delta t_\mathcal{D}(t) \, ,\\
        &= \frac{GM\tau}{v^2}\,e^{-2\pi i f t_{\mathcal{D}, 0}}
        \int_{-t_{\mathcal{D}, 0}/\tau_P}^{(T-t_{\mathcal{D}, 0})/\tau_P} dx_P\,
        e^{-2\pi i f\tau_P x_P}
        \left[\sqrt{1+x_P^2}\,\unit{b}_P-\operatorname{arcsinh}(x_P)\,\unit{v}\right]\!\cdot\unit{n}.
    \end{split}
\end{equation}
For bulk events with $t_{\mathcal{D}, 0}\gg\tau_P$ and $T-t_{\mathcal{D}, 0}\gg\tau_P$, the finite-window endpoints can be sent to $\pm\infty$ at leading order, and the resulting kernel Fourier transform reduces, %
up to delta functions at $f=0$ that are degenerate with the timing model%
, to\footnote{These Fourier transforms are not absolutely convergent and should be understood distributionally. The antiderivatives that turn the standard \(K_0\) and \(K_1\) transform pairs into transforms of \(\operatorname{arcsinh}x\) and \(\sqrt{1+x^2}\) also generate terms localized at \(f=0\). In the time domain, these are polynomial pieces that are degenerate with the timing model and are removed by the projection.} 
\begin{equation}\label{eqn:dop_dyn_FT}
    \delta \tilde{t}_\mathcal{D}(f)
    \approx
    -\frac{GM}{v^2}\frac{e^{-2\pi i f t_{\mathcal{D}, 0}}}{\pi f}
    \left[K_1(2\pi f\tau_P)\unit{b}_P
    -
    iK_0(2\pi f\tau_P)\unit{v}
    \right]\cdot\unit{n},
\end{equation}
where $K_0$ and $K_1$ are modified Bessel functions of the second kind. The noise-weighted inner product has support at $f \lesssim f_\star$, so in the deep-dynamic regime $f_\star \tau \ll 1$, the Bessel functions are probed at small argument: $K_1(x) \sim 1/x$ dominates the logarithmically softer $K_0(x) \sim \ln(1/x)$, so the $K_0$ term %
is dropped, and $\delta\tilde t_\mathcal{D} \sim 1/f^2$ throughout the $f \lesssim f_\star$ support. The unprojected norm then follows from Eq.~\eqref{eqn:integral_scalings} as $(\delta t_\mathcal{D}|\delta t_\mathcal{D}) \sim (f_\star T)^{-\min(\gamma, 3)}$.

We now turn to the projected norm. As established in App.~\ref{app:proj_dyn_dop}, for $\gamma > 3$, the oscillatory factor $e^{-2\pi i f t_{\mathcal{D}, 0}}$ in Eq.~\eqref{eqn:dop_dyn_FT} decoheres the signal parametrically against the timing-model modes $\phi_1$ and $\phi_2$, leaving it effectively orthogonal to the timing model. On the other hand, for $\gamma < 3$, the signal retains an overlap with the timing model even after the distributional degeneracies are removed, but this overlap is bounded by an order-unity prefactor and does not change the scaling, so $(\delta t^\perp_\mathcal{D}|\delta t^\perp_\mathcal{D}) \sim (\delta t_\mathcal{D}|\delta t_\mathcal{D})$. Replacing the angular factor $(\unit{b}\cdot\unit{n})^2$ by its isotropic average $1/3$, we then have
\begin{equation}\label{eqn:snr_dop_dyn}
\begin{split}
    \text{SNR}_{\mathcal{D}, \text{dyn}} \sim
    \frac{GM}{\pi^2v^2\tau_P}
    \sqrt{\frac{T^3}{6\Delta t t^2_\text{rms}}
    \left(f_\star T\right)^{-\min(\gamma, 3)}}.
\end{split}
\end{equation}
In the dynamic limit, the signal is dominated by the subhalo with the smallest impact parameter $b_{\min}$ across all worldlines passing near the pulsar during the observation window. As in the static case, the distribution of $b_{\min}$ follows from Poisson statistics on the encounter ensemble --- the relevant volume is now the swept tube $\sim \pi b^2_P v T$ around the pulsar, rather than the snapshot volume $\sim r_P^3$. Its $p$-th percentile is~\cite{Ramani:2020hdo}
\begin{equation}\label{eqn:dop_dyn_nearest}
    b_{p} = \sqrt{-\frac{1}{\pi}\frac{M}{N_P f_{\rm DM}\rho_{\mathrm{DM}}vT}\ln(1-p)}.
\end{equation}
Substituting this into Eq.~\eqref{eqn:snr_dop_dyn}, using $\tau_{P}=b_{P}/v$ , we find that the $90^{\text{th}}$ percentile sensitivity of the Doppler signal in the dynamic limit is
\begin{align} \label{eqn:dop_dyn_sensitivity_pulsar}
f_{\rm DM}^{{\mathcal{D}}, \, \rm{dyn}} &\gtrsim
3\,(f_\star T)^{\min(\gamma, 3)}
\left(\frac{200}{N_P}\right)
\left(\frac{10^{-9} M_\odot}{M}\right)
\left(\frac{20\text{yr}}{T}\right)^4\left(\frac{\Delta t}{2 \, \rm{wk}}\right) \left(\frac{t_{\rm rms}}{50 \, \rm{ns}}\right)^2.
\end{align}
Finally, the dynamic parametrization assumes the full blip duration $\sim 2\tau$ fits within the observing baseline, $\tau_{\text{min}} \lesssim T/2$, which restricts the dynamic-limit analysis to
\begin{equation}\label{eqn:dop_dyn_validity_pulsar}
    f_{\rm DM}^{{\mathcal{D}}, \, \rm{dyn}}
    \gtrsim \left(\frac{200}{N_P}\right)
    \left(\frac{M}{10^{-7} M_\odot}\right)
    \left(\frac{20\text{yr}}{T}\right)^{3}.
\end{equation}
We emphasize that this estimate is valid only deep in the dynamic regime. As one approaches the boundary $\tau_P \sim T$, the signal becomes less localized, finite-window effects become important, and the overlap with timing-model components is no longer parametrically suppressed. The Bessel-function suppression at $f\tau_P \gtrsim 1$ that we dropped also becomes relevant, leading to an order of magnitude degradation in sensitivity.

\subsubsection{Shapiro}
\label{sec:analytics_shap_dyn_pulsar}
As in the Doppler case, in the limit $\tau_\perp \ll T$, %
we can Fourier transform the Shapiro signal in the close-encounter regime $\sqrt{r_\parallel^2+r_\perp^2}\lesssim L/2$ from Eq.~\eqref{eqn:DM_Shapiro_3}, by sending the finite-window endpoints to $\pm \infty$ and dropping timing-model-degenerate pieces, to obtain
\begin{equation}\label{eqn:shap_dyn_FT}
    \begin{split}
        \delta \tilde{t}_{\mathcal{S}}(f) &= \int_0^T dt \, e^{-2\pi i f t} \delta t_\mathcal{S}(t)  \\
     &\approx -2 GM\tau_\perp e^{-2\pi i f t_{\mathcal{S}, 0}} \int_{-\infty}^{\infty} dx_\perp \,   e^{-2\pi i f \tau_\perp x_\perp} \ln (1+x_\perp^2) \\
     &= \frac{2GM}{f} e^{-2\pi f\tau_\perp} e^{-2\pi i ft_{\mathcal{S}, 0}}.
    \end{split}
\end{equation}
The exponential factor $e^{-2\pi f\tau_\perp}$ encodes the finite duration of the encounter and provides a smooth cutoff for frequencies $f\gtrsim 1/\tau_\perp$. The noise-weighted inner product is again supported at $f \lesssim f_\star$, where $f_\star \tau_\perp \ll 1$ makes $e^{-4\pi f\tau_\perp} \approx 1$, leaving $\delta\tilde t_\mathcal{S} \sim 1/f$ throughout the support. The unprojected norm then follows from Eq.~\eqref{eqn:integral_scalings} as $(\delta t_\mathcal{S}|\delta t_\mathcal{S}) \sim (f_\star T)^{-\min(\gamma, 1)}$. In App.~\ref{app:proj_dyn_shap}, analogous to the Doppler case, we establish that for $\gamma > 2$, the oscillatory factor $e^{-2\pi i f t_{\mathcal{S},0}}$ in Eq.~\eqref{eqn:shap_dyn_FT} decoheres the signal against the timing model, and for $\gamma < 2$, the overlap with the timing model does not change the scaling and only contributes an order-unity degradation, allowing us to write $(\delta t^\perp_\mathcal{S}|\delta t^\perp_\mathcal{S}) \sim (\delta t_\mathcal{S}|\delta t_\mathcal{S})$. The SNR then works out to be
\begin{equation}\label{eqn:snr_shap_dyn}
    \text{SNR}_{\mathcal{S}, \text{dyn}} \sim
    GM\sqrt{\frac{8T}{\Delta t t^2_\text{rms}}
    \left(f_\star T\right)^{-\min(\gamma, 1)}}.
\end{equation}
Inverting this expression, we find that the Shapiro signal in the dynamic limit is sensitive to subhalos with mass
\begin{equation}
    M_{\mathcal{S}, \text{dyn}} \gtrsim 6\times 10^{-4}\,M_\odot\, (f_\star T)^\frac{\min(\gamma, 1)}{2}
    \left(\frac{20\text{yr}}{T}\right)^\frac{1}{2} \left(\frac{\Delta t}{2 \, \rm{wk}}\right)^{\frac{1}{2}} \left(\frac{t_{\rm rms}}{50 \, \rm{ns}}\right) .
\end{equation}
The dynamic parametrization is valid provided the encounter duration is short compared to the observing baseline. By the same Poisson logic as the Doppler-dynamic case, the smallest Shapiro encounter duration $\tau_{\perp, \min} = b_{\perp,\min}/\bar{v}_\perp$ across the array follows from worldlines sweeping a slab of effective volume $\sim 2b_\perp (\bar{v}_\perp T) L$ along the LOS~\cite{Dror:2019twh},
\begin{equation} \label{eqn:shap_dyn_nearest}
    \tau_{\perp, \text{min}} \simeq
    \frac{2}{v} \frac{M}{N_Pf_{\rm DM} \rho_{\rm DM} v T L}
    \lesssim \frac{T}{2},
\end{equation}
which restricts the dynamic-limit analysis to
\begin{equation}
\label{eqn:validity_shap_dyn_det}
    f_{\rm DM}^{\mathcal{S}, \text{dyn}}
    \gtrsim 10
    \left(\frac{200}{N_P}\right)
    \left(\frac{M}{M_\odot}\right)
    \left(\frac{20\text{yr}}{T}\right)^{2}
    \left(\frac{5\, \mathrm{kpc}}{L}\right).
\end{equation}
We emphasize, as in the Doppler case, that this estimate is valid only deep in the dynamic regime; the same approximations break down near $\tau_\perp \sim T$, but the steeper exponential suppression $e^{-2\pi f\tau_\perp}$ leads to two orders of magnitude degradation in sensitivity rather than one.

\subsection{Stochastic Limit}
\label{sec:analytics_stochastic_regime}

In Sec.~\ref{sec:SNR} we computed the reach for a weak signal with a known Gaussian covariance. The DM substructure signal, however, is a sum over a Poisson population of subhalo flybys, which is  Gaussian only when the number of flybys contributing comparably to its variance is large %
--- a count we quantify by $N_Q$ in App.~\ref{app:stochastic_limit_cutoff}. %
This fails when the variance is dominated by the subhalo with the loudest encounter. For instance, the variance of the flyby population in the static limit for both Doppler and Shapiro signals are closest-event dominated, so the static population is already described by the deterministic analysis above. The stochastic calculation below therefore applies only to the dynamic, many-event limit ($N_Q \gg 1$).  

The dynamic Shapiro covariance is finite and insensitive to the closest flyby, so it requires no special treatment. The dynamic Doppler covariance, however, is logarithmically sensitive to the closest encounter, which we regulate with its $90^{\rm th}$-percentile value. Because this sensitivity is only logarithmic --- unlike the power-law closest-event domination of the static limit --- the regulated population still Gaussianizes; we confirm this in App.~\ref{app:stochastic_limit_cutoff} %
by verifying that $N_Q \gg 1$ throughout the regime where our analytic estimates apply.%
Note that the stochastic estimate is valid for masses when the closest encounter that sets the cutoff is deep in the dynamic regime.%

This stochastic regime was previously considered in Ref.~\cite{Ramani:2020hdo}, where the signal was described by a two-dimensional covariance function in the time
domain. While correct, interfacing it with standard PTA
data analysis software is numerically challenging. Here we instead derive a clean analytic expression for
the PSD of the stochastic DM signal, finding that the Doppler and Shapiro contributions scale as $f^{-4}$ (Eq.~\eqref{eqn:dop_stoch_PSD}) and $f^{-3}$
(Eq.~\eqref{eqn:shap_stoch_PSD}), respectively. We then estimate the sensitivity by evaluating the SNR of Eq.~\eqref{eqn:stoch_SNR_fourier} using the scaling relation from Eq.~\eqref{eqn:integral_scalings}. As in Ref.~\cite{Ramani:2020hdo}, we restrict our attention to the auto-correlation signal at each pulsar, for which the stochasticity arises from the 1-halo term --- the contribution from a single subhalo perturbing the pulsar, summed over the subhalo population. The cross-correlation of the stochastic \textit{pulsar} term between two distant pulsars vanishes at this 1-halo level, since no single subhalo can lie close to both, and is instead carried by the 2-halo term. This extension is beyond the scope of the present work; we leave it to future study as a potential avenue for disentangling the DM signal from the stochastic GWB.

\subsubsection{Doppler}

We start from the expression for $\delta t_\mathcal{D}^\perp$ from Eq.~\eqref{eqn:dop_dyn_FT},
\begin{equation}
	\begin{split}
		\widetilde{\delta t}^{\perp}_{\mathcal{D}}(f)
		&\approx - \frac{GM}{b_{P}v} \frac{e^{-2\pi i ft_{\mathcal{D},0}}}{(2\pi f)^2}\, \tilde{\vec{s}}\!\left(\frac{2\pi fb_P}{v}\right) \cdot \unit{n},
	\end{split}
\end{equation}
where we define $\tilde{\vec{s}}(x) \equiv 2x\bigl[K_1(x)\,\unit{b}_P - i K_0(x)\,\unit{v}\bigr]$. As in Sec.~\ref{sec:analytics_dop_dyn_pulsar}, we approximate $\widetilde{\delta t}^{\perp}_{\mathcal{D}} \approx \widetilde{\delta t}_{\mathcal{D}}$. 
As we assume each subhalo has a common speed, a flyby is parametrized by the direction of its velocity $\Omega_v$ (drawn isotropically) and the impact parameter $\vec b_P = (b_P, \phi_b)$ in the plane perpendicular to $\vec v$, and $t_{\mathcal{D}, 0}$; in this parametrization the spatial volume element is $d^3 \vec{r}_P = v\,dt_{\mathcal{D}, 0}\,d^2 \vec{b}_P$.
Now we compute the covariance of this signal over the DM ensemble, %
\begin{align}
		\langle \delta \tilde{t}^\perp_\mathcal{D} (f) \delta \tilde{t}^{\perp, *}_\mathcal{D} (f') \rangle \nonumber &= \frac{\bar{n}}{(2\pi f)^2(2\pi f')^2} \int v dt_{\mathcal{D}, 0} \, e^{-2\pi i (f-f')t_{\mathcal{D}, 0}} \\ &\times\int \frac{d^2 \Omega_v}{4\pi} \int d^2 \vec{b}_P  \,  \frac{G^2M^2}{b^2_P v^2} \, \left|\tilde{\vec{s}}\!\left(\frac{2\pi fb_P}{v}\right) \cdot \unit{n}\right|\left|\tilde{\vec{s}}\!\left(\frac{2\pi f'b_P}{v}\right) \cdot \unit{n}\right|^*, \nonumber \\[0.2em]
        &=\frac{\bar{n} }{(2\pi f)^4}\int \frac{d^2 \Omega_v}{4\pi}\int \frac{d^2 \vec{b}_P}{b^2_P} \,\frac{G^2M^2}{v} \, \left|\tilde{\vec{s}}\!\left(\frac{2\pi fb_P}{v}\right) \cdot \unit{n}\right|^2 \delta(f-f')
\end{align}
where the integral over $t_0$ enforces stationarity and we can define an one-sided PSD as in Eq.~\eqref{eqn:sig_PSD_onesided},
\begin{equation}
	\begin{split}
		S^\perp_\mathcal{D} (f) = \frac{2\bar{n}}{(2\pi f)^4} \int \frac{d^2 \Omega_v}{4\pi} \int \frac{d^2 \vec{b}_P}{b^2_P}\, \frac{G^2M^2}{v} \, \left|\tilde{\vec{s}}\!\left(\frac{2\pi fb_P}{v}\right) \cdot \unit{n}\right|^2.
	\end{split}
\end{equation}
Assigning all subhalos a common speed $v$ and averaging over the isotropic directions of $\hat v$ and $\hat b$,\footnote{We have the isotropic average, for $\hat{b}_P \equiv \phi_b $ in the plane orthogonal to the velocity $\mathbf{v}$

$$\int \frac{d^2\Omega_v}{4\pi}\int d \phi_b \,  (\unit{v}\cdot \unit{n}_a)(\unit{v}\cdot \unit{n}_b) = \int \frac{d^2\Omega_v}{4\pi}\int d \phi_b  (\unit{b}_P\cdot \unit{n}_a)(\unit{b}_P\cdot \unit{n}_b)= \frac{2\pi}{3} \unit{n}_a\cdot \unit{n}_b,$$ and $\int d^2\Omega_v\int d \phi_b  (\unit{v}\cdot \unit{n}_a)(\unit{b}_P\cdot \unit{n}_b) = 0$. See, Ref.~\cite{Du:2023dhk}.} the cross terms in $|\tilde{\vec s}\cdot\unit n|^2$ vanish and only the diagonal $K_0^2(x)+K_1^2(x)$ combination survives. With $x \equiv 2\pi fb_P/v$, the PSD reduces to
\begin{equation}
\label{eqn:dop_stoch_PSD_with_limits}
    \begin{split}
        S^{\perp}_{\mathcal{D}}(f) &= \frac{G^2M\,f_{\rm DM}\rho_{\rm DM}}{3\pi^3 v\,f^4}\int_{x_{\rm min}}^{x_{\rm max}} dx\; x \left[K_0^2(x)+K_1^2(x)\right], \\
        &= -\frac{G^2M\,f_{\rm DM}\rho_{\rm DM}}{3\pi^3 v\,f^4} \left[\frac{2\pi fb_P}{v} K_0\left(\frac{2\pi fb_P}{v}\right)K_1\left(\frac{2\pi fb_P}{v}\right)\right]\Bigg|_{b_\text{min}}^{b_\text{max}}
    \end{split}
\end{equation}
where we regularize the upper limit of the integral by the dynamic limit validity bound $b_\text{max}= vT $ and the lower limit by the $90^\text{th}$ percentile closest expected encounter from Eq.~\eqref{eqn:dop_dyn_nearest} as discussed above, evaluated at $N_P=1$, since each pulsar's PSD is regulated by the closest approach to that pulsar alone; the $N_P$ independent per-pulsar signals are then combined in the SNR. The SNR integral in Eq.~\eqref{eqn:stoch_SNR_fourier} has support for $f \lesssim f_\star$. The upper limit in Eq.~\eqref{eqn:dop_stoch_PSD_with_limits} vanishes once $2\pi f b_\text{max}/v = 2\pi fT \gg 1$ as $x K_0(x)K_1(x)$ falls off exponentially as $\frac{\pi}{2} e^{-2x}$ for $x\gg1$. Instead, the covariance is set by the lower-limit. Using the complementary small-argument asymptotic $x K_0(x)K_1(x) \approx \ln \left(2/(x e^{\gamma_{\rm EM}})\right)$ for $x \ll 1$, we write
\begin{equation} \label{eqn:dop_stoch_PSD}
\begin{split}
    S^{\perp}_{\mathcal{D}}(f) &\approx \frac{G^2M\,f_{\rm DM}\rho_{\rm DM}}{3\pi^3 v\,f^4} \left(\frac{2\pi f b_\text{min}}{v}\right) K_0\left(\frac{2\pi fb_\text{min}}{v}\right)K_1\left(\frac{2\pi fb_\text{min}}{v}\right)\\[0.3em]
    &\sim \frac{G^2M\,f_{\rm DM}\rho_{\rm DM}}{3\pi^3 v\,f^4} 
    \begin{dcases*}
        \ln\left(\frac{v}{\pi e^{\gamma_{\rm EM}} f b_{\text{min}}}\right),
            & if $f \ll \frac{v}{2\pi b_\text{min}}$\\
        \frac{\pi}{2} \exp\left(-\frac{4\pi fb_\text{min}}{v}\right), & if $f \gg \frac{v}{2\pi b_\text{min}}$
    \end{dcases*},
\end{split}
\end{equation}
where $\gamma_{\rm EM} \approx 0.577$ is the Euler-Mascheroni constant. Because the integrand of the SNR in Eq.~\eqref{eqn:stoch_SNR_fourier} is supported at frequencies $f\lesssim f_\star$, assuming the closest event is deep in the dynamic limit, $2\pi f_\star b_{\rm min}/v \ll 1$, we are in the logarithmic (first) case of Eq.~\eqref{eqn:dop_stoch_PSD}.
Unlike the deterministic case, the PSD does not take a simple power-law form due to the logarithm. As discussed in Sec.~\ref{sec:analytics_dynamic_limit}, for $\gamma > 7/2$ the SNR integrand in Eq.~\eqref{eqn:stoch_SNR_fourier} is broadly peaked near $f_\star$, whereas for $\gamma < 7/2$ it is peaked near $1/T$. Since the logarithm is slowly varying we evaluate it at the peak of the integrand, $f_\text{eff}$ defined as 
\begin{equation}
    f_\text{eff}\sim \begin{dcases*}
        & $f_\star$, for $\gamma > \frac{7}{2}$ \\
        & $\frac{1}{T}$, for $\gamma < \frac{7}{2}$
    \end{dcases*}.
\end{equation}
Substituting into Eq.~\eqref{eqn:stoch_SNR_fourier} and applying Eq.~\eqref{eqn:integral_scalings}, we obtain
\begin{equation}\label{eqn:snr_dop_stoch}
    \text{SNR}_{\mathcal{D},\,\rm stoch} \sim \frac{\sqrt{N_P}G^2Mf_{\rm DM}\rho_{\rm DM} T^4}{6\pi^3 v\Delta t t^2_\text{rms}} \ln \left(\!\frac{v}{\pi e^{\gamma_{\rm EM}} f_\text{eff} \,b_{\text{min}}}\!\right) \left(f_\star T\right)^{-\min(\gamma, \frac{7}{2})}
\end{equation}
Due to the non-linear dependence of $f_{\rm DM}$ inside the logarithm, we cannot readily invert this SNR and instead numerically solve the condition $\text{SNR}=4$. Here we provide an order-of-magnitude estimate on the upper bound on $f_{\mathrm{DM}}$ by dropping the weak log dependence in Eq.~\eqref{eqn:snr_dop_stoch}
\begin{equation}\label{eqn:dop_stoc_sensitivity}
    f_{\mathrm{DM}}^{D,\,\mathrm{stoch}} \gtrsim 4 \left(f_*T\right)^{\min(\gamma,\frac{7}{2})}\left(\frac{10^{-9}\,M_{\odot}}{M}\right)\left(\frac{200}{N_P}\right)^{1/2}\left(\frac{20\,\mathrm{yr}}{T}\right)^4\left(\frac{\Delta t}{2\,{\rm wk}}\right)
    \left(\frac{t_{\rm rms}}{50\,{\rm ns}}\right)^{2} \, .
\end{equation}

\subsubsection{Shapiro}
\label{sec:analytics_shap_stoch}
An ensemble of subhalos distributed along the Earth--pulsar line of sight produces a stochastic Shapiro signal. Starting from $\delta t_\mathcal{S}^\perp$ in Eq.~\eqref{eqn:shap_dyn_FT} and treating the flybys as a Poisson process, we average over epochs $t_0$, impact parameters $b_\perp$, and the line-of-sight coordinate $z$; in this parametrization the spatial volume element is $d^3 \vec{r}_0 = 2v_\perp db_\parallel \,d b_\perp\,dt_0$,%
\footnote{The factor of 2 counts both sides of approach to the line of sight.}
to obtain the covariance %
\begin{equation}
    \begin{split}
     \langle \delta \tilde{t}^\perp_S(f)\delta \tilde{t}^{\perp *} _S(f') \rangle &= \bar n \int \frac{d^2 \Omega_v}{4\pi} \int db_\perp \int db_\parallel \, \int 2 v_\perp dt_{\mathcal{S}, 0} \frac{4G^2M^2}{ff'} e^{-2\pi (f+f')b_\perp/v_\perp}e^{2\pi i (f-f')t_{\mathcal{S}, 0}}\\
     &\approx \frac{8 G^2M f_{\rm DM}\rho_{\rm DM}  \bar{v}_\perp L}{f^2}\delta(f-f') \int_{b_{\perp, \text{min}}}^{\bar{v}_\perp T} db_\perp \, e^{-4\pi fb_\perp/v_\perp},
    \end{split}
\end{equation}
where we do the approximation $ \langle f(v_\perp)\rangle_{\Omega_v} \approx f(\bar{v}_\perp)$ and only consider subhalos with $b_\parallel < L/2$. %
As in the doppler case $b_{\perp, \text{min}}$ is the 90th percentile closest impact parameter from Eq.~\eqref{eqn:shap_dyn_nearest}, with $N_P=1$. The one-sided PSD is therefore
\begin{equation}
    \begin{split}
      S^{\perp}_{\mathcal{S}}(f) = \frac{4G^2Mf_{\rm DM}\rho_{\rm DM} \bar{v}_\perp^2 L }{\pi f^3} \left[e^{-4\pi f b_{\perp, \text{min}}/v_\perp}-e^{-4\pi f T}\right]
    \end{split}
\end{equation}
Assuming the closest subhalo is deep in the dynamic limit, $4\pi f_\star b_{\perp,\rm min}/\bar{v}_\perp \ll 1$, we can approximate $e^{-4\pi f b_{\perp,\rm min}/\bar{v}_\perp} \approx 1$. Dropping the upper limit $e^{-4\pi f T} \ll 1$, the PSD can be approximated as 
\begin{equation} \label{eqn:shap_stoch_PSD}
    \begin{split}
      S^{\perp}_{\mathcal{S}}(f) \approx \frac{4G^2M f_{\rm DM}\rho_{\rm DM} \bar{v}_\perp^2 L}{\pi f^3}.
    \end{split}
\end{equation}
Substituting into Eq.~\eqref{eqn:stoch_SNR_fourier} and applying Eq.~\eqref{eqn:integral_scalings}, we obtain
\begin{equation}\label{eqn:snr_shap_stoch}
    \text{SNR}_{\mathcal{S},\,\rm stoch} \sim \frac{2\sqrt{N_P}G^2M f_{\rm DM}\rho_{\rm DM} \bar{v}_\perp^2 L \, T^3}{ \pi \Delta t t^2_\text{rms}}  \left(f_\star T\right)^{-\min(\gamma, \frac{5}{2})}
\end{equation}
Inverting for $f_{\rm DM}$ gives the sensitivity reach
\begin{equation}
\label{eqn:shap_stoch_sensitivity}
    f_{\rm DM}^{\mathcal{S},\,\rm stoch} 
    \gtrsim 0.5 \, (f_\star T)^{\min(\gamma,\,5/2)}
    \left(\frac{10^{-4} \, M_\odot}{M}\right)
    \left(\frac{200}{N_P}\right)^{1/2}
    \left(\frac{20\,{\rm yr}}{T}\right)^{3}
    \left(\frac{\Delta t}{2\,{\rm wk}}\right)
    \left(\frac{t_{\rm rms}}{50\,{\rm ns}}\right)^{2}
    \left(\frac{5\,{\rm kpc}}{L}\right).
\end{equation}
However, this sensitivity estimate is only valid if the closest subhalo is truly in the dynamic limit, i.e. Eq.~\eqref{eqn:validity_shap_dyn_det} with $N_P=1$ instead of $N_P=200$. %

\section{Results and Discussion}
\label{sec:results_and_discussion}

\begin{figure}
    \centering
    \includegraphics[width=\linewidth]{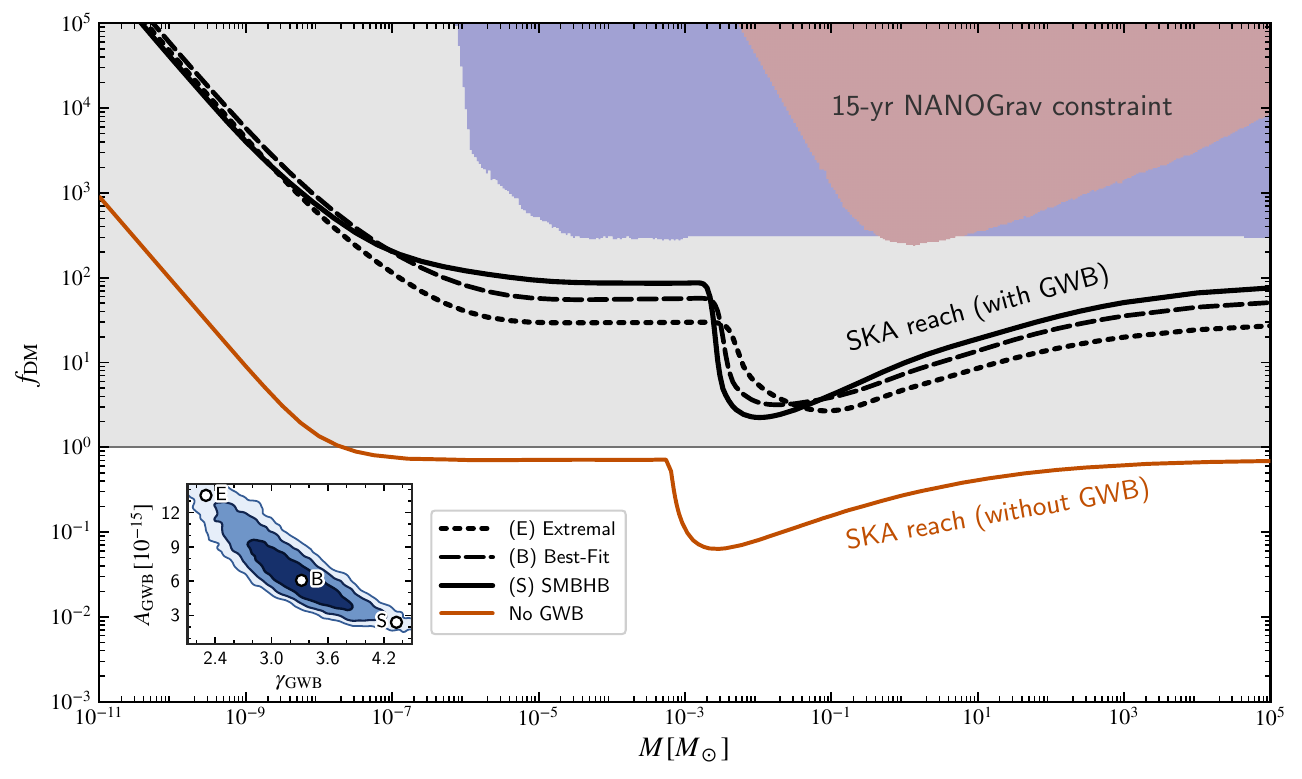}
    \caption{
    Projected upper limits on the substructure fraction $f_\text{DM}$ as a function of subhalo mass $M$, assuming an SKA-like PTA benchmark (see main text for details). Solid, dashed, and dotted curves show the reach at three representative points of the NANOGrav 15-year GWB posterior~\cite{NANOGrav:2023gor} (inset): (S) the SMBHB baseline, (B) the posterior best-fit, and (E) an extremal $3\sigma$ corner. The orange solid curve shows the projected reach in the absence of a GWB, as in earlier forecasts~\cite{Ramani:2020hdo, Lee:2020wfn}. Blue and red shaded regions show $(M, f_\text{DM})$ parameter space already excluded by NANOGrav 15-year Doppler and Shapiro static-signal searches~\cite{NANOGrav:2023hvm}. Each curve traces the better of the deterministic and stochastic reach at every mass. The GWB suppresses the reach by one to three orders of magnitude relative to the no-GWB case. This suppression is robust to GWB parameter uncertainty, varying by only order-unity factors across the posterior.}
    \label{fig:money_plot}
\end{figure}

To forecast the reach, we adopt the Square Kilometre Array (SKA) Phase-II benchmark used in prior DM-substructure forecasts~\cite{Dror:2019twh, Ramani:2020hdo, Lee:2020wfn}, originally tabulated in Ref.~\cite{Rosado:2015epa}: $N_P = 200$ millisecond pulsars at typical distance $L = 5$ kpc, observed for a baseline of $T = 20$ yr with cadence $\Delta t = 2$ wk, and per-TOA white-noise residual $t_\text{rms} = 50$ ns. We retain this choice both for direct comparability with the earlier literature and because next-generation radio facilities --- SKA-Mid~\cite{Janssen:2014dka}, DSA-2000~\cite{Berghaus:2025kvn}, ngVLA~\cite{NANOGrav:2018vfs}, or a combined IPTA-style dataset~\cite{Hobbs:2009yy} --- could plausibly deliver such an array within the next two decades. As for the GWB parameters, %
the SMBHB prediction for the spectral index, $\gamma = 13/3$~\cite{Phinney:2001di}, lies more than $2\sigma$ from the NANOGrav 15-year posterior median~\cite{NANOGrav:2023gor}. With current data not sharply constraining $(\gamma, A_\text{GWB})$, no single point on the posterior is strongly favored. We therefore compute the reach at three characteristic points on the posterior (labeled in the inset of Fig.~\ref{fig:money_plot}): (S) the SMBHB baseline at $(\gamma, A_\text{GWB}) = (13/3, 2.4\times10^{-15})$, (B) the posterior best-fit at $(3.3, 6.1\times10^{-15})$, and (E) a representative $3 \sigma$ extremal corner at $(2.3, 13.5\times10^{-15})$. 

Across most of the mass range the GWB suppresses the reach by two to three orders of magnitude relative to the white-noise-only case, with the size of the penalty varying with mass and signal regime as dissected below.%
\footnote{Consistent with the Bayesian forecast of Ref.~\cite{Lee:2021zqw}, which reports a similar worsening of reach with red noise that agrees with the $(f_\star T)^{-\min(\gamma, n-1)/2}$ scaling derived in Sec.~\ref{sec:analytical_estimates}.} Despite the parameter uncertainty, the suppression varies by at most an order-unity factor across the three points, so our qualitative conclusions are robust to the current ambiguity in the inferred GWB. %

\begin{figure}
    \centering
    \includegraphics[width=\linewidth]{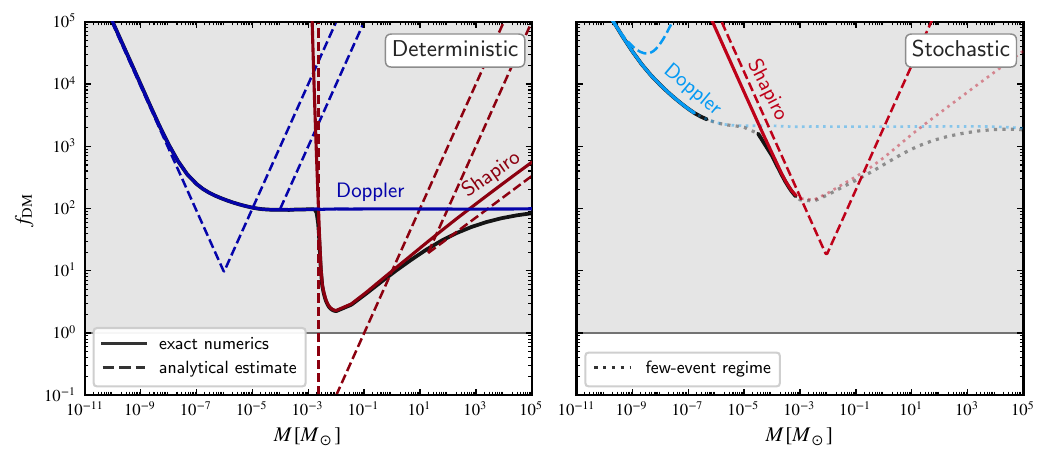}
    \caption{
    Decomposition of the projected reach $f_\text{DM}(M)$ at the SMBHB-baseline GWB into the two independent geometric regions identified in %
    the main text: the pulsar-Doppler sphere (Doppler-P) %
    and the Shapiro line-of-sight cylinder (Shapiro) for deterministic signals on the left panel and stochastic signals on the right panel. Solid lines are the full gauge-invariant reach computed numerically; dashed lines are the closed-form analytic estimates of Sec.~\ref{sec:analytical_estimates}. Each signal type on the left has two dashed wedges corresponding to the dynamic regime for lower masses and the static regime for larger masses. Dotted lines in the right panel indicate where a few events dominate, the assumption of Gaussianity of the stochastic signal breaks down (see App.~\ref{app:stochastic_limit_cutoff}) and our sensitivity estimate is no longer valid. %
    The exact numerics and our analytical estimates agree up to order-unity factors deep in each regime. %
    }
    \label{fig:analytic_overlay}
\end{figure}

To produce the curves in Fig.~\ref{fig:money_plot}, we now describe the numerical evaluation of the reach, in which we work with the full gauge-invariant delay of  Sec.~\ref{sec:dark_matter_residual} %
and drop the static and dynamic limit splits of Sec.~\ref{sec:analytical_estimates}. 
To evaluate the deterministic reach, rather than simulating full subhalo realizations, we recast the loudest-event reach as the substructure fraction at which a Poisson population yields at least one detectable event. A subhalo worldline, labeled by $\theta=(\vec r_0,\vec v)$ where $\vec r_0$ and $\vec v$ are its initial position and velocity as in Eq.~\eqref{eqn:DM_trajectory} respectively, produces the single-event $\text{SNR}^2_{\rm ev}(\theta)=\left(\delta\vec{t}^\perp(\theta)|\,\delta\vec{t}^\perp(\theta)\right)$
where $\delta\vec{t}^\perp(\theta)$ is the proper time-shift of Eq.~\eqref{eqn:delta_t_t} projected orthogonal to the timing model. In the numerics we fold the projection into one reusable inverse-noise operator rather than projecting each signal (App.~\ref{app:time_domain}).
The number of events with $\text{SNR}_{\rm ev}$ above a threshold $\text{SNR}_{\rm th}=4$ is Poisson\footnote{Follows from the thinning property of Poisson processes.} with mean
\begin{equation}\label{eqn:N_ev}
\bar{N}_{\mathrm{th}}
= \frac{\rho_{\mathrm{DM}}f_{\mathrm{DM}}}{M}\int d^3\vec{r}_0\int d^3\vec{v}\,f_{\vec{v}}(\vec{v})\, 
\, \Theta\left[\text{SNR}_{\rm ev}(\theta)-\text{SNR}_{\rm th}\right] \, ,
\end{equation}
where we assume that every DM subhalo has the same speed $\bar{v} = 340\, \text{km/s}$ with isotropically sampled direction, i.e. $f_\vec{v}(\vec{v})\propto \delta(v-\bar{v})$, as in Sec.~\ref{sec:analytical_estimates}. The reach is the smallest $f_{\rm DM}$ at which $90\%$ of universes contain at least one event above threshold, i.e. $1-e^{-\bar{N}_{\mathrm{th}}}=0.9$, or $\bar{N}_{\mathrm{th}}=\ln 10$.

We evaluate the integral in Eq.~\eqref{eqn:N_ev} for a single pulsar over two regions: a Doppler sphere around it and a Shapiro cylinder along its line of sight. Because both signals act on a single pulsar, the $N_P$ pulsars contribute independently to the integral (the single-pulsar statistic of Sec.~\ref{sec:SNR}), and the total is $N_P$ times this contribution.\footnote{Over the full mass range $10^{-11}$--$10^{5}\,M_\odot$, less than $1\%$ of these regions' volume overlaps for $N_P=200$ isotropically distributed pulsars at $L=5\,\mathrm{kpc}$, so we treat them as independent.} This also allows us to compare the full numerics against the closed-form analytic estimates of Sec.~\ref{sec:analytical_estimates} separately for the Doppler and Shapiro signals (Fig.~\ref{fig:analytic_overlay}). The region construction and Monte Carlo sampling are detailed in App.~\ref{app:sampling_geometries}. Note that we ignore contributions from subhalos close to the Earth and defer their discussion to a companion paper \cite{in_prep}.

The stochastic reach follows from the many-event Gaussian limit. As noted in Sec.~\ref{sec:analytics_stochastic_regime}, the ensemble-averaged covariance of the Doppler stochastic signal diverges logarithmically, and we regulate it with a percentile cut that excludes the loudest events, yielding a well-defined population covariance $\mathbf{S}_{(10)}(M, f_\text{DM})$ (App.~\ref{app:stochastic_limit_cutoff}). In this limit the optimal quadratic statistic depends only on the covariance, with the expected SNR
\begin{equation}\label{eq:DM_stoch_SNR}
  \mathrm{SNR}^2 = \frac{1}{2}\,\mathrm{Tr}\,[\mathbf{N}^{-1} \mathbf{S}^\perp_{(10)}\mathbf{N}^{-1}_\perp \mathbf{S}^\perp_{(10)}].
\end{equation}
This mirrors the stochastic statistic of Sec.~\ref{sec:analytical_estimates} and is valid only
in the many-event regime. Where only few events contribute comparably (dotted, Fig.~\ref{fig:analytic_overlay}) the Gaussian estimate breaks down; we quantify this threshold in App.~\ref{app:numerics}.

With the deterministic and stochastic estimates in hand, we now dissect the reach. The shape of each envelope in Fig.~\ref{fig:money_plot} is set by the dominating contribution at any given mass; Fig.~\ref{fig:analytic_overlay} shows this decomposition at the SMBHB-baseline GWB, where the full gauge-invariant reach (solid) tracks the analytic estimates of Sec.~\ref{sec:analytical_estimates} (dashed) to order unity deep in each regime. %
As $M$ grows, the deterministic Doppler-pulsar contribution first appears in its dynamic regime, then settles into the static plateau at $f_\text{DM} \sim 100$ across $M \sim 10^{-7}\text{--}10^{-3}\,M_\odot$. In a narrow window near $M \sim 10^{-3}\,M_\odot$ the stochastic Shapiro signal briefly takes over, followed by a sharp cliff in the deterministic Shapiro contribution down to $f_\text{DM} \sim 2$ at $M \sim 10^{-2}\,M_\odot$. At higher masses the Shapiro sensitivity degrades into its static limit, and by $M \sim 10^{5}\,M_\odot$ the envelope returns to the static Doppler reach. 

The hierarchy of contributions in Fig.~\ref{fig:analytic_overlay} and the modest order-unity spread in Fig.~\ref{fig:money_plot} are both organized by Table~\ref{tab:signal_scalings}. Among the deterministic signals, the dynamic Shapiro contribution has the shallowest $(f_\star T)$ suppression in the table, with the exponent fixed at $\min(\gamma, 1) = 1$ across the posterior; it loses only an order of magnitude in $f_\text{DM}$ relative to white noise. %
The static signals fare worse: they sit at the maximal exponent $\min(\gamma, 5)/2$, suffering roughly two orders of magnitude of suppression and inheriting a stronger (still order-unity) sensitivity to the choice of $(\gamma, A_\text{GWB})$. The dynamic Doppler signal has a shallower $(f_\star T)$ exponent than the static cases, but takes the worst $f_\text{DM}$ hit of all four deterministic regimes --- close to three orders of magnitude --- because its SNR scales only as $f_\text{DM}^{1/2}$. The stochastic Doppler and Shapiro signals take a comparable hit despite their larger $(f_\star T)$-suppression exponents, because their SNR scales linearly with $f_\text{DM}$.

To further improve the sensitivity, %
one could consider %
an alternative dataset with different observation baseline $T$, timing quality $t_\text{rms}$ and $\Delta t$, or number of pulsars $N_P$. The dependence on the first two is substantially weakened by the GWB. For the static signals that set the envelope across most of our mass range, the SNR grows as $T^{7/2}$ in the white-noise limit but only as $T^{(7-\gamma)/2}$ in the GWB regime, leaving $T^{4/3}$ at the SMBHB-baseline, $\gamma = 13/3$. For these static signals, timing quality is neutralized entirely: in the GWB regime the white-noise factor $t_\text{rms}^2\,\Delta t$ cancels exactly against the $(f_\star T)^{-\gamma/2}$ suppression, so the SNR is independent of $t_\text{rms}$ and $\Delta t$ as long as $f_\star T \gg 1$.
At the SMBHB-baseline, this means timing precision can be degraded from $t_\text{rms} \sim 50 \, \mathrm{ns}$ to $10\,\mu\mathrm{s}$ before $f_\star T$ falls to unity and white noise re-enters. Only the array size retains full leverage: the $N_P$ scaling is unaffected by the GWB, with $f_\text{DM} \sim 1/N_P$ in the deterministic regime (Table~\ref{tab:signal_scalings}).

Despite this GWB suppression, it is important to stress that $f_{\rm DM}$ is normalized to the local dark matter density $\rho_{\rm DM} = 0.46~\mathrm{GeV/cm^3}$ as inferred from Galactic kinematics, whereas pulsar timing is sensitive to the subhalo abundance in the immediate vicinity of the Earth and pulsars, where the dark matter density is not directly measured. As emphasized in Ref.~\cite{Kim:2023kyy}, typical determinations of the local DM abundance rely on large-scale kinematic properties of the Milky Way and probe volumes of order $\mathcal{O}(10^6~\mathrm{pc^3})$ or larger (for reviews, see Refs.~\cite{Read:2014qva, deSalas:2020hbh}), so the inferred value $\rho_{\rm DM} \sim 0.4~\mathrm{GeV/cm^3}$ does not exclude overdensities on the sub-pc scales relevant here. As a consequence, $f_{\rm DM} > 1$ is physically admissible whenever the density near the Earth or pulsars exceeds this large-volume average --- for instance, if enhanced small-scale power leaves the halo with large sub-pc density fluctuations. Direct constraints (\textit{e.g.} from solar-system ephemerides) still permit local overdensities as large as $\sim 2\times10^4$~\cite{2013AstL...39..141P}, and no analogous constraint exists at the pulsar locations; the reach at $f_{\rm DM} \sim 10$--$10^3$ therefore remains a competitive probe of substructure in this mass range.

A possible path towards probing the $\Lambda$CDM benchmark $f_\text{DM} < 1$ lies in a large array of pulsars not timed at the millisecond-pulsar level. The Shapiro deterministic knee at $M \sim 10^{-2}\,M_\odot$ is a specialized target: it already sits at $f_\text{DM} \sim 2$ in Fig.~\ref{fig:money_plot}, so a modest factor of $\sim 2$ in $N_P$ pushes it below $f_\text{DM}=1$. Across the bulk of the mass range, however, the static-signal envelope is suppressed by roughly two orders of magnitude, and the $f_\text{DM} \sim 1/N_P$ pulsar scaling would require a substantial increase of two orders of magnitude 
in array size to compensate.

\section{Conclusion and Outlook}\label{sec:conclusion}

In this work we have evaluated the sensitivity to DM substructure signals in future PTA surveys in the presence of a red noise source such as a stochastic GWB. Determining the impact of the degeneracy between the timing model and the DM signal types, we present simple analytic scaling relations for the achievable sensitivity and their respective regimes of validity (see Tab.~\ref{tab:signal_scalings} for a summary). 

In addition to analytic scaling relations in the limiting regimes, we derive the full proper-time gauge-invariant observable and numerically compute the expected reach for SKA benchmarks for three different GWB realizations. We find that the sensitivity to DM substructure is diminished by one to three orders of magnitude compared to the white noise only expectation, with the largest penalty for lighter DM substructure (see Fig.~\ref{fig:money_plot}). The dynamic Shapiro signal suffers the least suppression of only one order of magnitude, giving peak sensitivity to DM substructure with $M\sim 10^{-2} M_{\odot}$.
 Our results indicate dependence on the properties of the stochastic GWB when evaluating future survey's sensitivity to DM substructure within an order-unity %
 factor, given current three sigma inference of the GWB's measured amplitude and spectral index. %

Even in the presence of the GWB, the SKA-like benchmark improves substantially on the existing NANOGrav 15-year static-signal exclusions shown as the blue and red regions in Fig.~\ref{fig:money_plot}. The projected reach improves on the NANOGrav 15-year exclusions by up to an order of magnitude for the deterministic Doppler signal, and by roughly two orders of magnitude for the deterministic Shapiro signal near $M \sim 10^{-2}\,M_\odot$. Note that the dynamic Shapiro signal was not implemented in the NANOGrav 15-year data analysis~\cite{NANOGrav:2023hvm}.

Due to the large signal suppression, it remains a challenge to probe the benchmark $f_{\text{DM}} < 1$ for the signals considered in this work, even with future surveys that allow for a larger number of pulsars and longer observation times.
We find that larger pulsar array size
should be prioritized over timing precision but the scaling with $N_P$ for upcoming and next-generation experiments alone will not reach the  $f_{\text{DM}} <1 $ for the majority of parameter space. It is therefore important to investigate alternative approaches to searching for DM substructure with PTAs that have less susceptibility to the stochastic GWB contribution.

A promising avenue, and a caveat to the above, is that we have exclusively focused on single-pulsar signals. In a companion paper \cite{in_prep} we generalize our framework to signals that exploit correlations across the array. Notably, we consider DM subhalo flybys near pulsar pairs close enough that the GWB imprint on them is correlated, which can be leveraged to improve sensitivity. Beyond the strategies discussed in \cite{in_prep}, correlations may offer another handle: the stochastic DM analyses so far use only the 1-halo term~\cite{Ramani:2020hdo}, which vanishes in the cross-correlation between distant pulsars; the 2-halo term, sourced by correlations between distinct subhalos, survives and could help separate the DM signal from %
the GWB. Separately, if a loud individual SMBHB could be resolved and subtracted as a continuous wave~\cite{NANOGrav:2023pdq}, the effective GWB amplitude would reduce. We leave a systematic study of these directions to future work.

\textit{Note added}--- While this work was in its final stages of completion Ref.~\cite{
Foster:2026kfg} appeared and reaches compatible conclusions using machine-learned surrogate likelihoods rather than analytic scaling relations.
Our approach delivers closed-form expressions that expose how the sensitivity depends on the number of pulsars, observation time, and properties of the GWB, making it applicable to a wide variety of benchmark scenarios and future surveys. We additionally derive the full gauge-invariant proper-time observable for DM substructure,
which is specific to our analysis.

\begingroup
\renewcommand\addcontentsline[3]{}
\begin{acknowledgments}

AC was supported by the Margaret Leong and Michael P. Checca Summer Undergraduate Research Fellowship. The computations presented here were conducted on the Caltech High Performance Cluster, partially supported by a grant from the Gordon and Betty Moore Foundation. KB is funded by the Deutsche Forschungsgemeinschaft (DFG, German Research Foundation) through the Emmy Noether Programme Project No.~548044346.
KB thanks the U.S.~Department of Energy, Office of Science, Office of High Energy Physics, under Award Number DE-SC0011632, and the Walter Burke Institute for Theoretical Physics. VL is supported by the Network for Neutrinos, Nuclear Astrophysics and Symmetries (N3AS) through the National Science Foundation Physics Frontier Center, Grant No. PHY-2020275. KZ is supported by the U.S. Department of Energy, Office of Science, Office of High Energy Physics, under Award No. DE-SC0011632, the Walter Burke Institute for Theoretical
Physics, the Heising-Simons Foundation, and a Simons Investigator award.

\end{acknowledgments}
\endgroup

\let\oldaddcontentsline\addcontentsline
\renewcommand{\addcontentsline}[3]{}%
\bibliography{07_biblio}

\begin{thebibliography}{59}%
\makeatletter
\providecommand \@ifxundefined [1]{%
 \@ifx{#1\undefined}
}%
\providecommand \@ifnum [1]{%
 \ifnum #1\expandafter \@firstoftwo
 \else \expandafter \@secondoftwo
 \fi
}%
\providecommand \@ifx [1]{%
 \ifx #1\expandafter \@firstoftwo
 \else \expandafter \@secondoftwo
 \fi
}%
\providecommand \natexlab [1]{#1}%
\providecommand \enquote  [1]{``#1''}%
\providecommand \bibnamefont  [1]{#1}%
\providecommand \bibfnamefont [1]{#1}%
\providecommand \citenamefont [1]{#1}%
\providecommand \href@noop [0]{\@secondoftwo}%
\providecommand \href [0]{\begingroup \@sanitize@url \@href}%
\providecommand \@href[1]{\@@startlink{#1}\@@href}%
\providecommand \@@href[1]{\endgroup#1\@@endlink}%
\providecommand \@sanitize@url [0]{\catcode `\\12\catcode `\$12\catcode
  `\&12\catcode `\#12\catcode `\^12\catcode `\_12\catcode `\%12\relax}%
\providecommand \@@startlink[1]{}%
\providecommand \@@endlink[0]{}%
\providecommand \url  [0]{\begingroup\@sanitize@url \@url }%
\providecommand \@url [1]{\endgroup\@href {#1}{\urlprefix }}%
\providecommand \urlprefix  [0]{URL }%
\providecommand \Eprint [0]{\href }%
\providecommand \doibase [0]{https://doi.org/}%
\providecommand \selectlanguage [0]{\@gobble}%
\providecommand \bibinfo  [0]{\@secondoftwo}%
\providecommand \bibfield  [0]{\@secondoftwo}%
\providecommand \translation [1]{[#1]}%
\providecommand \BibitemOpen [0]{}%
\providecommand \bibitemStop [0]{}%
\providecommand \bibitemNoStop [0]{.\EOS\space}%
\providecommand \EOS [0]{\spacefactor3000\relax}%
\providecommand \BibitemShut  [1]{\csname bibitem#1\endcsname}%
\let\auto@bib@innerbib\@empty
\bibitem [{\citenamefont {Hogan}\ and\ \citenamefont
  {Rees}(1988)}]{Hogan:1988mp}%
  \BibitemOpen
  \bibfield  {author} {\bibinfo {author} {\bibfnamefont {C.~J.}\ \bibnamefont
  {Hogan}}\ and\ \bibinfo {author} {\bibfnamefont {M.~J.}\ \bibnamefont
  {Rees}},\ }\bibfield  {title} {\bibinfo {title} {{Axion miniclusters}},\
  }\href {https://doi.org/10.1016/0370-2693(88)91655-3} {\bibfield  {journal}
  {\bibinfo  {journal} {Phys. Lett. B}\ }\textbf {\bibinfo {volume} {205}},\
  \bibinfo {pages} {228} (\bibinfo {year} {1988})}\BibitemShut {NoStop}%
\bibitem [{\citenamefont {Kolb}\ and\ \citenamefont
  {Tkachev}(1993)}]{Kolb:1993zz}%
  \BibitemOpen
  \bibfield  {author} {\bibinfo {author} {\bibfnamefont {E.~W.}\ \bibnamefont
  {Kolb}}\ and\ \bibinfo {author} {\bibfnamefont {I.~I.}\ \bibnamefont
  {Tkachev}},\ }\bibfield  {title} {\bibinfo {title} {{Axion miniclusters and
  Bose stars}},\ }\href {https://doi.org/10.1103/PhysRevLett.71.3051}
  {\bibfield  {journal} {\bibinfo  {journal} {Phys. Rev. Lett.}\ }\textbf
  {\bibinfo {volume} {71}},\ \bibinfo {pages} {3051} (\bibinfo {year}
  {1993})},\ \Eprint {https://arxiv.org/abs/hep-ph/9303313}
  {arXiv:hep-ph/9303313} \BibitemShut {NoStop}%
\bibitem [{\citenamefont {Zurek}\ \emph {et~al.}(2007)\citenamefont {Zurek},
  \citenamefont {Hogan},\ and\ \citenamefont {Quinn}}]{Zurek:2006sy}%
  \BibitemOpen
  \bibfield  {author} {\bibinfo {author} {\bibfnamefont {K.~M.}\ \bibnamefont
  {Zurek}}, \bibinfo {author} {\bibfnamefont {C.~J.}\ \bibnamefont {Hogan}},\
  and\ \bibinfo {author} {\bibfnamefont {T.~R.}\ \bibnamefont {Quinn}},\
  }\bibfield  {title} {\bibinfo {title} {{Astrophysical Effects of Scalar Dark
  Matter Miniclusters}},\ }\href {https://doi.org/10.1103/PhysRevD.75.043511}
  {\bibfield  {journal} {\bibinfo  {journal} {Phys. Rev. D}\ }\textbf {\bibinfo
  {volume} {75}},\ \bibinfo {pages} {043511} (\bibinfo {year} {2007})},\
  \Eprint {https://arxiv.org/abs/astro-ph/0607341} {arXiv:astro-ph/0607341}
  \BibitemShut {NoStop}%
\bibitem [{\citenamefont {Buschmann}\ \emph {et~al.}(2020)\citenamefont
  {Buschmann}, \citenamefont {Foster},\ and\ \citenamefont
  {Safdi}}]{Buschmann:2019icd}%
  \BibitemOpen
  \bibfield  {author} {\bibinfo {author} {\bibfnamefont {M.}~\bibnamefont
  {Buschmann}}, \bibinfo {author} {\bibfnamefont {J.~W.}\ \bibnamefont
  {Foster}},\ and\ \bibinfo {author} {\bibfnamefont {B.~R.}\ \bibnamefont
  {Safdi}},\ }\bibfield  {title} {\bibinfo {title} {{Early-Universe Simulations
  of the Cosmological Axion}},\ }\href
  {https://doi.org/10.1103/PhysRevLett.124.161103} {\bibfield  {journal}
  {\bibinfo  {journal} {Phys. Rev. Lett.}\ }\textbf {\bibinfo {volume} {124}},\
  \bibinfo {pages} {161103} (\bibinfo {year} {2020})},\ \Eprint
  {https://arxiv.org/abs/1906.00967} {arXiv:1906.00967 [astro-ph.CO]}
  \BibitemShut {NoStop}%
\bibitem [{\citenamefont {Eggemeier}\ \emph {et~al.}(2020)\citenamefont
  {Eggemeier}, \citenamefont {Redondo}, \citenamefont {Dolag}, \citenamefont
  {Niemeyer},\ and\ \citenamefont {Vaquero}}]{Eggemeier:2019khm}%
  \BibitemOpen
  \bibfield  {author} {\bibinfo {author} {\bibfnamefont {B.}~\bibnamefont
  {Eggemeier}}, \bibinfo {author} {\bibfnamefont {J.}~\bibnamefont {Redondo}},
  \bibinfo {author} {\bibfnamefont {K.}~\bibnamefont {Dolag}}, \bibinfo
  {author} {\bibfnamefont {J.~C.}\ \bibnamefont {Niemeyer}},\ and\ \bibinfo
  {author} {\bibfnamefont {A.}~\bibnamefont {Vaquero}},\ }\bibfield  {title}
  {\bibinfo {title} {{First Simulations of Axion Minicluster Halos}},\ }\href
  {https://doi.org/10.1103/PhysRevLett.125.041301} {\bibfield  {journal}
  {\bibinfo  {journal} {Phys. Rev. Lett.}\ }\textbf {\bibinfo {volume} {125}},\
  \bibinfo {pages} {041301} (\bibinfo {year} {2020})},\ \Eprint
  {https://arxiv.org/abs/1911.09417} {arXiv:1911.09417 [astro-ph.CO]}
  \BibitemShut {NoStop}%
\bibitem [{\citenamefont {Xiao}\ \emph {et~al.}(2021)\citenamefont {Xiao},
  \citenamefont {Williams},\ and\ \citenamefont {McQuinn}}]{Xiao:2021nkb}%
  \BibitemOpen
  \bibfield  {author} {\bibinfo {author} {\bibfnamefont {H.}~\bibnamefont
  {Xiao}}, \bibinfo {author} {\bibfnamefont {I.}~\bibnamefont {Williams}},\
  and\ \bibinfo {author} {\bibfnamefont {M.}~\bibnamefont {McQuinn}},\
  }\bibfield  {title} {\bibinfo {title} {{Simulations of axion minihalos}},\
  }\href {https://doi.org/10.1103/PhysRevD.104.023515} {\bibfield  {journal}
  {\bibinfo  {journal} {Phys. Rev. D}\ }\textbf {\bibinfo {volume} {104}},\
  \bibinfo {pages} {023515} (\bibinfo {year} {2021})},\ \Eprint
  {https://arxiv.org/abs/2101.04177} {arXiv:2101.04177 [astro-ph.CO]}
  \BibitemShut {NoStop}%
\bibitem [{\citenamefont {Peccei}\ and\ \citenamefont
  {Quinn}(1977)}]{Peccei:1977hh}%
  \BibitemOpen
  \bibfield  {author} {\bibinfo {author} {\bibfnamefont {R.~D.}\ \bibnamefont
  {Peccei}}\ and\ \bibinfo {author} {\bibfnamefont {H.~R.}\ \bibnamefont
  {Quinn}},\ }\bibfield  {title} {\bibinfo {title} {{CP Conservation in the
  Presence of Instantons}},\ }\href
  {https://doi.org/10.1103/PhysRevLett.38.1440} {\bibfield  {journal} {\bibinfo
   {journal} {Phys. Rev. Lett.}\ }\textbf {\bibinfo {volume} {38}},\ \bibinfo
  {pages} {1440} (\bibinfo {year} {1977})}\BibitemShut {NoStop}%
\bibitem [{\citenamefont {Green}\ \emph {et~al.}(2005)\citenamefont {Green},
  \citenamefont {Hofmann},\ and\ \citenamefont {Schwarz}}]{Green:2005fa}%
  \BibitemOpen
  \bibfield  {author} {\bibinfo {author} {\bibfnamefont {A.~M.}\ \bibnamefont
  {Green}}, \bibinfo {author} {\bibfnamefont {S.}~\bibnamefont {Hofmann}},\
  and\ \bibinfo {author} {\bibfnamefont {D.~J.}\ \bibnamefont {Schwarz}},\
  }\bibfield  {title} {\bibinfo {title} {{The First wimpy halos}},\ }\href
  {https://doi.org/10.1088/1475-7516/2005/08/003} {\bibfield  {journal}
  {\bibinfo  {journal} {JCAP}\ }\textbf {\bibinfo {volume} {08}},\ \bibinfo
  {pages} {003}},\ \Eprint {https://arxiv.org/abs/astro-ph/0503387}
  {arXiv:astro-ph/0503387} \BibitemShut {NoStop}%
\bibitem [{\citenamefont {Siegel}\ \emph {et~al.}(2007)\citenamefont {Siegel},
  \citenamefont {Hertzberg},\ and\ \citenamefont {Fry}}]{Siegel:2007fz}%
  \BibitemOpen
  \bibfield  {author} {\bibinfo {author} {\bibfnamefont {E.~R.}\ \bibnamefont
  {Siegel}}, \bibinfo {author} {\bibfnamefont {M.~P.}\ \bibnamefont
  {Hertzberg}},\ and\ \bibinfo {author} {\bibfnamefont {J.~N.}\ \bibnamefont
  {Fry}},\ }\bibfield  {title} {\bibinfo {title} {{Probing Dark Matter
  Substructure with Pulsar Timing}},\ }\href
  {https://doi.org/10.1111/j.1365-2966.2007.12435.x} {\bibfield  {journal}
  {\bibinfo  {journal} {Mon. Not. Roy. Astron. Soc.}\ }\textbf {\bibinfo
  {volume} {382}},\ \bibinfo {pages} {879} (\bibinfo {year} {2007})},\ \Eprint
  {https://arxiv.org/abs/astro-ph/0702546} {arXiv:astro-ph/0702546}
  \BibitemShut {NoStop}%
\bibitem [{\citenamefont {Baghram}\ \emph {et~al.}(2011)\citenamefont
  {Baghram}, \citenamefont {Afshordi},\ and\ \citenamefont
  {Zurek}}]{Baghram:2011is}%
  \BibitemOpen
  \bibfield  {author} {\bibinfo {author} {\bibfnamefont {S.}~\bibnamefont
  {Baghram}}, \bibinfo {author} {\bibfnamefont {N.}~\bibnamefont {Afshordi}},\
  and\ \bibinfo {author} {\bibfnamefont {K.~M.}\ \bibnamefont {Zurek}},\
  }\bibfield  {title} {\bibinfo {title} {{Prospects for Detecting Dark Matter
  Halo Substructure with Pulsar Timing}},\ }\href
  {https://doi.org/10.1103/PhysRevD.84.043511} {\bibfield  {journal} {\bibinfo
  {journal} {Phys. Rev. D}\ }\textbf {\bibinfo {volume} {84}},\ \bibinfo
  {pages} {043511} (\bibinfo {year} {2011})},\ \Eprint
  {https://arxiv.org/abs/1101.5487} {arXiv:1101.5487 [astro-ph.CO]}
  \BibitemShut {NoStop}%
\bibitem [{\citenamefont {Clark}\ \emph {et~al.}(2016)\citenamefont {Clark},
  \citenamefont {Lewis},\ and\ \citenamefont {Scott}}]{Clark:2015sha}%
  \BibitemOpen
  \bibfield  {author} {\bibinfo {author} {\bibfnamefont {H.~A.}\ \bibnamefont
  {Clark}}, \bibinfo {author} {\bibfnamefont {G.~F.}\ \bibnamefont {Lewis}},\
  and\ \bibinfo {author} {\bibfnamefont {P.}~\bibnamefont {Scott}},\ }\bibfield
   {title} {\bibinfo {title} {{Investigating dark matter substructure with
  pulsar timing {\textendash} I. Constraints on ultracompact minihaloes}},\
  }\href {https://doi.org/10.1093/mnras/stv2743} {\bibfield  {journal}
  {\bibinfo  {journal} {Mon. Not. Roy. Astron. Soc.}\ }\textbf {\bibinfo
  {volume} {456}},\ \bibinfo {pages} {1394} (\bibinfo {year} {2016})},\
  \bibinfo {note} {[Erratum: Mon.Not.Roy.Astron.Soc. 464, 2468 (2017)]},\
  \Eprint {https://arxiv.org/abs/1509.02938} {arXiv:1509.02938 [astro-ph.CO]}
  \BibitemShut {NoStop}%
\bibitem [{\citenamefont {Schutz}\ and\ \citenamefont
  {Liu}(2017)}]{Schutz:2016khr}%
  \BibitemOpen
  \bibfield  {author} {\bibinfo {author} {\bibfnamefont {K.}~\bibnamefont
  {Schutz}}\ and\ \bibinfo {author} {\bibfnamefont {A.}~\bibnamefont {Liu}},\
  }\bibfield  {title} {\bibinfo {title} {{Pulsar timing can constrain
  primordial black holes in the LIGO mass window}},\ }\href
  {https://doi.org/10.1103/PhysRevD.95.023002} {\bibfield  {journal} {\bibinfo
  {journal} {Phys. Rev. D}\ }\textbf {\bibinfo {volume} {95}},\ \bibinfo
  {pages} {023002} (\bibinfo {year} {2017})},\ \Eprint
  {https://arxiv.org/abs/1610.04234} {arXiv:1610.04234 [astro-ph.CO]}
  \BibitemShut {NoStop}%
\bibitem [{\citenamefont {Kashiyama}\ and\ \citenamefont
  {Oguri}(2018)}]{Kashiyama:2018gsh}%
  \BibitemOpen
  \bibfield  {author} {\bibinfo {author} {\bibfnamefont {K.}~\bibnamefont
  {Kashiyama}}\ and\ \bibinfo {author} {\bibfnamefont {M.}~\bibnamefont
  {Oguri}},\ }\bibfield  {title} {\bibinfo {title} {{Detectability of
  Small-Scale Dark Matter Clumps with Pulsar Timing Arrays}},\ }\href@noop {}
  {\  (\bibinfo {year} {2018})},\ \Eprint {https://arxiv.org/abs/1801.07847}
  {arXiv:1801.07847 [astro-ph.CO]} \BibitemShut {NoStop}%
\bibitem [{\citenamefont {Dror}\ \emph {et~al.}(2019)\citenamefont {Dror},
  \citenamefont {Ramani}, \citenamefont {Trickle},\ and\ \citenamefont
  {Zurek}}]{Dror:2019twh}%
  \BibitemOpen
  \bibfield  {author} {\bibinfo {author} {\bibfnamefont {J.~A.}\ \bibnamefont
  {Dror}}, \bibinfo {author} {\bibfnamefont {H.}~\bibnamefont {Ramani}},
  \bibinfo {author} {\bibfnamefont {T.}~\bibnamefont {Trickle}},\ and\ \bibinfo
  {author} {\bibfnamefont {K.~M.}\ \bibnamefont {Zurek}},\ }\bibfield  {title}
  {\bibinfo {title} {{Pulsar Timing Probes of Primordial Black Holes and
  Subhalos}},\ }\href {https://doi.org/10.1103/PhysRevD.100.023003} {\bibfield
  {journal} {\bibinfo  {journal} {Phys. Rev. D}\ }\textbf {\bibinfo {volume}
  {100}},\ \bibinfo {pages} {023003} (\bibinfo {year} {2019})},\ \Eprint
  {https://arxiv.org/abs/1901.04490} {arXiv:1901.04490 [astro-ph.CO]}
  \BibitemShut {NoStop}%
\bibitem [{\citenamefont {Ramani}\ \emph {et~al.}(2020)\citenamefont {Ramani},
  \citenamefont {Trickle},\ and\ \citenamefont {Zurek}}]{Ramani:2020hdo}%
  \BibitemOpen
  \bibfield  {author} {\bibinfo {author} {\bibfnamefont {H.}~\bibnamefont
  {Ramani}}, \bibinfo {author} {\bibfnamefont {T.}~\bibnamefont {Trickle}},\
  and\ \bibinfo {author} {\bibfnamefont {K.~M.}\ \bibnamefont {Zurek}},\
  }\bibfield  {title} {\bibinfo {title} {{Observability of Dark Matter
  Substructure with Pulsar Timing Correlations}},\ }\href
  {https://doi.org/10.1088/1475-7516/2020/12/033} {\bibfield  {journal}
  {\bibinfo  {journal} {JCAP}\ }\textbf {\bibinfo {volume} {12}},\ \bibinfo
  {pages} {033}},\ \Eprint {https://arxiv.org/abs/2005.03030} {arXiv:2005.03030
  [astro-ph.CO]} \BibitemShut {NoStop}%
\bibitem [{\citenamefont {Lee}\ \emph {et~al.}(2021{\natexlab{a}})\citenamefont
  {Lee}, \citenamefont {Mitridate}, \citenamefont {Trickle},\ and\
  \citenamefont {Zurek}}]{Lee:2020wfn}%
  \BibitemOpen
  \bibfield  {author} {\bibinfo {author} {\bibfnamefont {V.~S.~H.}\
  \bibnamefont {Lee}}, \bibinfo {author} {\bibfnamefont {A.}~\bibnamefont
  {Mitridate}}, \bibinfo {author} {\bibfnamefont {T.}~\bibnamefont {Trickle}},\
  and\ \bibinfo {author} {\bibfnamefont {K.~M.}\ \bibnamefont {Zurek}},\
  }\bibfield  {title} {\bibinfo {title} {{Probing Small-Scale Power Spectra
  with Pulsar Timing Arrays}},\ }\href
  {https://doi.org/10.1007/JHEP06(2021)028} {\bibfield  {journal} {\bibinfo
  {journal} {JHEP}\ }\textbf {\bibinfo {volume} {06}},\ \bibinfo {pages}
  {028}},\ \Eprint {https://arxiv.org/abs/2012.09857} {arXiv:2012.09857
  [astro-ph.CO]} \BibitemShut {NoStop}%
\bibitem [{\citenamefont {Lee}\ \emph {et~al.}(2021{\natexlab{b}})\citenamefont
  {Lee}, \citenamefont {Taylor}, \citenamefont {Trickle},\ and\ \citenamefont
  {Zurek}}]{Lee:2021zqw}%
  \BibitemOpen
  \bibfield  {author} {\bibinfo {author} {\bibfnamefont {V.~S.~H.}\
  \bibnamefont {Lee}}, \bibinfo {author} {\bibfnamefont {S.~R.}\ \bibnamefont
  {Taylor}}, \bibinfo {author} {\bibfnamefont {T.}~\bibnamefont {Trickle}},\
  and\ \bibinfo {author} {\bibfnamefont {K.~M.}\ \bibnamefont {Zurek}},\
  }\bibfield  {title} {\bibinfo {title} {{Bayesian Forecasts for Dark Matter
  Substructure Searches with Mock Pulsar Timing Data}},\ }\href
  {https://doi.org/10.1088/1475-7516/2021/08/025} {\bibfield  {journal}
  {\bibinfo  {journal} {JCAP}\ }\textbf {\bibinfo {volume} {08}},\ \bibinfo
  {pages} {025}},\ \Eprint {https://arxiv.org/abs/2104.05717} {arXiv:2104.05717
  [astro-ph.CO]} \BibitemShut {NoStop}%
\bibitem [{\citenamefont {Gresham}\ \emph {et~al.}(2023)\citenamefont
  {Gresham}, \citenamefont {Lee},\ and\ \citenamefont
  {Zurek}}]{Gresham:2022biw}%
  \BibitemOpen
  \bibfield  {author} {\bibinfo {author} {\bibfnamefont {M.~I.}\ \bibnamefont
  {Gresham}}, \bibinfo {author} {\bibfnamefont {V.~S.~H.}\ \bibnamefont
  {Lee}},\ and\ \bibinfo {author} {\bibfnamefont {K.~M.}\ \bibnamefont
  {Zurek}},\ }\bibfield  {title} {\bibinfo {title} {{Astrophysical observations
  of a dark matter-Baryon fifth force}},\ }\href
  {https://doi.org/10.1088/1475-7516/2023/02/048} {\bibfield  {journal}
  {\bibinfo  {journal} {JCAP}\ }\textbf {\bibinfo {volume} {02}},\ \bibinfo
  {pages} {048}},\ \Eprint {https://arxiv.org/abs/2209.03963} {arXiv:2209.03963
  [astro-ph.HE]} \BibitemShut {NoStop}%
\bibitem [{\citenamefont {Afzal}\ \emph {et~al.}(2023)\citenamefont {Afzal}
  \emph {et~al.}}]{NANOGrav:2023hvm}%
  \BibitemOpen
  \bibfield  {author} {\bibinfo {author} {\bibfnamefont {A.}~\bibnamefont
  {Afzal}} \emph {et~al.} (\bibinfo {collaboration} {NANOGrav}),\ }\bibfield
  {title} {\bibinfo {title} {{The NANOGrav 15 yr Data Set: Search for Signals
  from New Physics}},\ }\href {https://doi.org/10.3847/2041-8213/acdc91}
  {\bibfield  {journal} {\bibinfo  {journal} {Astrophys. J. Lett.}\ }\textbf
  {\bibinfo {volume} {951}},\ \bibinfo {pages} {L11} (\bibinfo {year}
  {2023})},\ \bibinfo {note} {[Erratum: Astrophys.J.Lett. 971, L27 (2024),
  Erratum: Astrophys.J. 971, L27 (2024)]},\ \Eprint
  {https://arxiv.org/abs/2306.16219} {arXiv:2306.16219 [astro-ph.HE]}
  \BibitemShut {NoStop}%
\bibitem [{\citenamefont {Berghaus}\ \emph {et~al.}(2025)\citenamefont
  {Berghaus}, \citenamefont {Du}, \citenamefont {Lee}, \citenamefont {Prabhu},
  \citenamefont {Reischke}, \citenamefont {Connor},\ and\ \citenamefont
  {Zurek}}]{Berghaus:2025kvn}%
  \BibitemOpen
  \bibfield  {author} {\bibinfo {author} {\bibfnamefont {K.~V.}\ \bibnamefont
  {Berghaus}}, \bibinfo {author} {\bibfnamefont {Y.}~\bibnamefont {Du}},
  \bibinfo {author} {\bibfnamefont {V.~S.~H.}\ \bibnamefont {Lee}}, \bibinfo
  {author} {\bibfnamefont {A.}~\bibnamefont {Prabhu}}, \bibinfo {author}
  {\bibfnamefont {R.}~\bibnamefont {Reischke}}, \bibinfo {author}
  {\bibfnamefont {L.}~\bibnamefont {Connor}},\ and\ \bibinfo {author}
  {\bibfnamefont {K.~M.}\ \bibnamefont {Zurek}},\ }\bibfield  {title} {\bibinfo
  {title} {{Physics beyond the Standard Model with the DSA-2000}},\ }\href
  {https://doi.org/10.1088/1475-7516/2025/12/035} {\bibfield  {journal}
  {\bibinfo  {journal} {JCAP}\ }\textbf {\bibinfo {volume} {12}},\ \bibinfo
  {pages} {035}},\ \Eprint {https://arxiv.org/abs/2505.23892} {arXiv:2505.23892
  [hep-ph]} \BibitemShut {NoStop}%
\bibitem [{\citenamefont {Goldreich}\ and\ \citenamefont
  {Julian}(1969)}]{Goldreich:1969sb}%
  \BibitemOpen
  \bibfield  {author} {\bibinfo {author} {\bibfnamefont {P.}~\bibnamefont
  {Goldreich}}\ and\ \bibinfo {author} {\bibfnamefont {W.~H.}\ \bibnamefont
  {Julian}},\ }\bibfield  {title} {\bibinfo {title} {{Pulsar
  electrodynamics}},\ }\href {https://doi.org/10.1086/150119} {\bibfield
  {journal} {\bibinfo  {journal} {Astrophys. J.}\ }\textbf {\bibinfo {volume}
  {157}},\ \bibinfo {pages} {869} (\bibinfo {year} {1969})}\BibitemShut
  {NoStop}%
\bibitem [{\citenamefont {Sturrock}(1971)}]{Sturrock:1971zc}%
  \BibitemOpen
  \bibfield  {author} {\bibinfo {author} {\bibfnamefont {P.~A.}\ \bibnamefont
  {Sturrock}},\ }\bibfield  {title} {\bibinfo {title} {{A Model of pulsars}},\
  }\href {https://doi.org/10.1086/150865} {\bibfield  {journal} {\bibinfo
  {journal} {Astrophys. J.}\ }\textbf {\bibinfo {volume} {164}},\ \bibinfo
  {pages} {529} (\bibinfo {year} {1971})}\BibitemShut {NoStop}%
\bibitem [{\citenamefont {Taylor}(2021)}]{Taylor:2021yjx}%
  \BibitemOpen
  \bibfield  {author} {\bibinfo {author} {\bibfnamefont {S.~R.}\ \bibnamefont
  {Taylor}},\ }\bibfield  {title} {\bibinfo {title} {{The Nanohertz
  Gravitational Wave Astronomer}},\ }\href@noop {} {\  (\bibinfo {year}
  {2021})},\ \Eprint {https://arxiv.org/abs/2105.13270} {arXiv:2105.13270
  [astro-ph.HE]} \BibitemShut {NoStop}%
\bibitem [{\citenamefont {Phinney}(2001)}]{Phinney:2001di}%
  \BibitemOpen
  \bibfield  {author} {\bibinfo {author} {\bibfnamefont {E.~S.}\ \bibnamefont
  {Phinney}},\ }\bibfield  {title} {\bibinfo {title} {{A Practical theorem on
  gravitational wave backgrounds}},\ }\href@noop {} {\  (\bibinfo {year}
  {2001})},\ \Eprint {https://arxiv.org/abs/astro-ph/0108028}
  {arXiv:astro-ph/0108028} \BibitemShut {NoStop}%
\bibitem [{\citenamefont {Agazie}\ \emph
  {et~al.}(2023{\natexlab{a}})\citenamefont {Agazie} \emph
  {et~al.}}]{NANOGrav:2023gor}%
  \BibitemOpen
  \bibfield  {author} {\bibinfo {author} {\bibfnamefont {G.}~\bibnamefont
  {Agazie}} \emph {et~al.} (\bibinfo {collaboration} {NANOGrav}),\ }\bibfield
  {title} {\bibinfo {title} {{The NANOGrav 15 yr Data Set: Evidence for a
  Gravitational-wave Background}},\ }\href
  {https://doi.org/10.3847/2041-8213/acdac6} {\bibfield  {journal} {\bibinfo
  {journal} {Astrophys. J. Lett.}\ }\textbf {\bibinfo {volume} {951}},\
  \bibinfo {pages} {L8} (\bibinfo {year} {2023}{\natexlab{a}})},\ \Eprint
  {https://arxiv.org/abs/2306.16213} {arXiv:2306.16213 [astro-ph.HE]}
  \BibitemShut {NoStop}%
\bibitem [{\citenamefont {Lee}(2026)}]{Lee_2026}%
  \BibitemOpen
  \bibfield  {author} {\bibinfo {author} {\bibfnamefont {V.~S.~H.}\
  \bibnamefont {Lee}},\ }\bibfield  {title} {\bibinfo {title} {Proper time
  shifts in pulsar timing arrays}} (\bibinfo {year} {2026}),\ \bibinfo {note}
  {in preparation}\BibitemShut {NoStop}%
\bibitem [{\citenamefont {Mitridate}\ \emph {et~al.}(2023)\citenamefont
  {Mitridate}, \citenamefont {Wright}, \citenamefont {von Eckardstein},
  \citenamefont {Schr{\"o}der}, \citenamefont {Nay}, \citenamefont {Olum},
  \citenamefont {Schmitz},\ and\ \citenamefont {Trickle}}]{Mitridate:2023oar}%
  \BibitemOpen
  \bibfield  {author} {\bibinfo {author} {\bibfnamefont {A.}~\bibnamefont
  {Mitridate}}, \bibinfo {author} {\bibfnamefont {D.}~\bibnamefont {Wright}},
  \bibinfo {author} {\bibfnamefont {R.}~\bibnamefont {von Eckardstein}},
  \bibinfo {author} {\bibfnamefont {T.}~\bibnamefont {Schr{\"o}der}}, \bibinfo
  {author} {\bibfnamefont {J.}~\bibnamefont {Nay}}, \bibinfo {author}
  {\bibfnamefont {K.}~\bibnamefont {Olum}}, \bibinfo {author} {\bibfnamefont
  {K.}~\bibnamefont {Schmitz}},\ and\ \bibinfo {author} {\bibfnamefont
  {T.}~\bibnamefont {Trickle}},\ }\bibfield  {title} {\bibinfo {title}
  {{PTArcade}},\ }\href@noop {} {\  (\bibinfo {year} {2023})},\ \Eprint
  {https://arxiv.org/abs/2306.16377} {arXiv:2306.16377 [hep-ph]} \BibitemShut
  {NoStop}%
\bibitem [{\citenamefont {Ellis}\ and\ \citenamefont
  {Lewicki}(2021)}]{Ellis:2020ena}%
  \BibitemOpen
  \bibfield  {author} {\bibinfo {author} {\bibfnamefont {J.}~\bibnamefont
  {Ellis}}\ and\ \bibinfo {author} {\bibfnamefont {M.}~\bibnamefont
  {Lewicki}},\ }\bibfield  {title} {\bibinfo {title} {{Cosmic String
  Interpretation of NANOGrav Pulsar Timing Data}},\ }\href
  {https://doi.org/10.1103/PhysRevLett.126.041304} {\bibfield  {journal}
  {\bibinfo  {journal} {Phys. Rev. Lett.}\ }\textbf {\bibinfo {volume} {126}},\
  \bibinfo {pages} {041304} (\bibinfo {year} {2021})},\ \Eprint
  {https://arxiv.org/abs/2009.06555} {arXiv:2009.06555 [astro-ph.CO]}
  \BibitemShut {NoStop}%
\bibitem [{\citenamefont {Blasi}\ \emph {et~al.}(2021)\citenamefont {Blasi},
  \citenamefont {Brdar},\ and\ \citenamefont {Schmitz}}]{Blasi:2020mfx}%
  \BibitemOpen
  \bibfield  {author} {\bibinfo {author} {\bibfnamefont {S.}~\bibnamefont
  {Blasi}}, \bibinfo {author} {\bibfnamefont {V.}~\bibnamefont {Brdar}},\ and\
  \bibinfo {author} {\bibfnamefont {K.}~\bibnamefont {Schmitz}},\ }\bibfield
  {title} {\bibinfo {title} {{Has NANOGrav found first evidence for cosmic
  strings?}},\ }\href {https://doi.org/10.1103/PhysRevLett.126.041305}
  {\bibfield  {journal} {\bibinfo  {journal} {Phys. Rev. Lett.}\ }\textbf
  {\bibinfo {volume} {126}},\ \bibinfo {pages} {041305} (\bibinfo {year}
  {2021})},\ \Eprint {https://arxiv.org/abs/2009.06607} {arXiv:2009.06607
  [astro-ph.CO]} \BibitemShut {NoStop}%
\bibitem [{\citenamefont {Ellis}\ \emph {et~al.}(2023)\citenamefont {Ellis},
  \citenamefont {Lewicki}, \citenamefont {Lin},\ and\ \citenamefont
  {Vaskonen}}]{Ellis:2023tsl}%
  \BibitemOpen
  \bibfield  {author} {\bibinfo {author} {\bibfnamefont {J.}~\bibnamefont
  {Ellis}}, \bibinfo {author} {\bibfnamefont {M.}~\bibnamefont {Lewicki}},
  \bibinfo {author} {\bibfnamefont {C.}~\bibnamefont {Lin}},\ and\ \bibinfo
  {author} {\bibfnamefont {V.}~\bibnamefont {Vaskonen}},\ }\bibfield  {title}
  {\bibinfo {title} {{Cosmic superstrings revisited in light of NANOGrav
  15-year data}},\ }\href {https://doi.org/10.1103/PhysRevD.108.103511}
  {\bibfield  {journal} {\bibinfo  {journal} {Phys. Rev. D}\ }\textbf {\bibinfo
  {volume} {108}},\ \bibinfo {pages} {103511} (\bibinfo {year} {2023})},\
  \Eprint {https://arxiv.org/abs/2306.17147} {arXiv:2306.17147 [astro-ph.CO]}
  \BibitemShut {NoStop}%
\bibitem [{\citenamefont {Graham}\ \emph {et~al.}(2016)\citenamefont {Graham},
  \citenamefont {Kaplan}, \citenamefont {Mardon}, \citenamefont {Rajendran},\
  and\ \citenamefont {Terrano}}]{Graham:2015ifn}%
  \BibitemOpen
  \bibfield  {author} {\bibinfo {author} {\bibfnamefont {P.~W.}\ \bibnamefont
  {Graham}}, \bibinfo {author} {\bibfnamefont {D.~E.}\ \bibnamefont {Kaplan}},
  \bibinfo {author} {\bibfnamefont {J.}~\bibnamefont {Mardon}}, \bibinfo
  {author} {\bibfnamefont {S.}~\bibnamefont {Rajendran}},\ and\ \bibinfo
  {author} {\bibfnamefont {W.~A.}\ \bibnamefont {Terrano}},\ }\bibfield
  {title} {\bibinfo {title} {{Dark Matter Direct Detection with
  Accelerometers}},\ }\href {https://doi.org/10.1103/PhysRevD.93.075029}
  {\bibfield  {journal} {\bibinfo  {journal} {Phys. Rev. D}\ }\textbf {\bibinfo
  {volume} {93}},\ \bibinfo {pages} {075029} (\bibinfo {year} {2016})},\
  \Eprint {https://arxiv.org/abs/1512.06165} {arXiv:1512.06165 [hep-ph]}
  \BibitemShut {NoStop}%
\bibitem [{\citenamefont {Porayko}\ \emph {et~al.}(2018)\citenamefont {Porayko}
  \emph {et~al.}}]{Porayko:2018sfa}%
  \BibitemOpen
  \bibfield  {author} {\bibinfo {author} {\bibfnamefont {N.~K.}\ \bibnamefont
  {Porayko}} \emph {et~al.},\ }\bibfield  {title} {\bibinfo {title} {{Parkes
  Pulsar Timing Array constraints on ultralight scalar-field dark matter}},\
  }\href {https://doi.org/10.1103/PhysRevD.98.102002} {\bibfield  {journal}
  {\bibinfo  {journal} {Phys. Rev. D}\ }\textbf {\bibinfo {volume} {98}},\
  \bibinfo {pages} {102002} (\bibinfo {year} {2018})},\ \Eprint
  {https://arxiv.org/abs/1810.03227} {arXiv:1810.03227 [astro-ph.CO]}
  \BibitemShut {NoStop}%
\bibitem [{\citenamefont {Kaplan}\ \emph {et~al.}(2022)\citenamefont {Kaplan},
  \citenamefont {Mitridate},\ and\ \citenamefont {Trickle}}]{Kaplan:2022lmz}%
  \BibitemOpen
  \bibfield  {author} {\bibinfo {author} {\bibfnamefont {D.~E.}\ \bibnamefont
  {Kaplan}}, \bibinfo {author} {\bibfnamefont {A.}~\bibnamefont {Mitridate}},\
  and\ \bibinfo {author} {\bibfnamefont {T.}~\bibnamefont {Trickle}},\
  }\bibfield  {title} {\bibinfo {title} {{Constraining fundamental constant
  variations from ultralight dark matter with pulsar timing arrays}},\ }\href
  {https://doi.org/10.1103/PhysRevD.106.035032} {\bibfield  {journal} {\bibinfo
   {journal} {Phys. Rev. D}\ }\textbf {\bibinfo {volume} {106}},\ \bibinfo
  {pages} {035032} (\bibinfo {year} {2022})},\ \Eprint
  {https://arxiv.org/abs/2205.06817} {arXiv:2205.06817 [hep-ph]} \BibitemShut
  {NoStop}%
\bibitem [{\citenamefont {Kim}\ and\ \citenamefont
  {Mitridate}(2024)}]{Kim:2023kyy}%
  \BibitemOpen
  \bibfield  {author} {\bibinfo {author} {\bibfnamefont {H.}~\bibnamefont
  {Kim}}\ and\ \bibinfo {author} {\bibfnamefont {A.}~\bibnamefont
  {Mitridate}},\ }\bibfield  {title} {\bibinfo {title} {{Stochastic ultralight
  dark matter fluctuations in pulsar timing arrays}},\ }\href
  {https://doi.org/10.1103/PhysRevD.109.055017} {\bibfield  {journal} {\bibinfo
   {journal} {Phys. Rev. D}\ }\textbf {\bibinfo {volume} {109}},\ \bibinfo
  {pages} {055017} (\bibinfo {year} {2024})},\ \Eprint
  {https://arxiv.org/abs/2312.12225} {arXiv:2312.12225 [hep-ph]} \BibitemShut
  {NoStop}%
\bibitem [{\citenamefont {Antoniadis}\ \emph {et~al.}(2024)\citenamefont
  {Antoniadis} \emph {et~al.}}]{EPTA:2023xxk}%
  \BibitemOpen
  \bibfield  {author} {\bibinfo {author} {\bibfnamefont {J.}~\bibnamefont
  {Antoniadis}} \emph {et~al.} (\bibinfo {collaboration} {EPTA, InPTA}),\
  }\bibfield  {title} {\bibinfo {title} {{The second data release from the
  European Pulsar Timing Array - IV. Implications for massive black holes, dark
  matter, and the early Universe}},\ }\href
  {https://doi.org/10.1051/0004-6361/202347433} {\bibfield  {journal} {\bibinfo
   {journal} {Astron. Astrophys.}\ }\textbf {\bibinfo {volume} {685}},\
  \bibinfo {pages} {A94} (\bibinfo {year} {2024})},\ \Eprint
  {https://arxiv.org/abs/2306.16227} {arXiv:2306.16227 [astro-ph.CO]}
  \BibitemShut {NoStop}%
\bibitem [{\citenamefont {Smarra}\ \emph {et~al.}(2023)\citenamefont {Smarra}
  \emph {et~al.}}]{EuropeanPulsarTimingArray:2023egv}%
  \BibitemOpen
  \bibfield  {author} {\bibinfo {author} {\bibfnamefont {C.}~\bibnamefont
  {Smarra}} \emph {et~al.} (\bibinfo {collaboration} {European Pulsar Timing
  Array}),\ }\bibfield  {title} {\bibinfo {title} {{Second Data Release from
  the European Pulsar Timing Array: Challenging the Ultralight Dark Matter
  Paradigm}},\ }\href {https://doi.org/10.1103/PhysRevLett.131.171001}
  {\bibfield  {journal} {\bibinfo  {journal} {Phys. Rev. Lett.}\ }\textbf
  {\bibinfo {volume} {131}},\ \bibinfo {pages} {171001} (\bibinfo {year}
  {2023})},\ \Eprint {https://arxiv.org/abs/2306.16228} {arXiv:2306.16228
  [astro-ph.HE]} \BibitemShut {NoStop}%
\bibitem [{\citenamefont {Smarra}\ \emph {et~al.}(2024)\citenamefont {Smarra}
  \emph {et~al.}}]{Smarra:2024kvv}%
  \BibitemOpen
  \bibfield  {author} {\bibinfo {author} {\bibfnamefont {C.}~\bibnamefont
  {Smarra}} \emph {et~al.},\ }\bibfield  {title} {\bibinfo {title}
  {{Constraints on conformal ultralight dark matter couplings from the European
  Pulsar Timing Array}},\ }\href {https://doi.org/10.1103/PhysRevD.110.043033}
  {\bibfield  {journal} {\bibinfo  {journal} {Phys. Rev. D}\ }\textbf {\bibinfo
  {volume} {110}},\ \bibinfo {pages} {043033} (\bibinfo {year} {2024})},\
  \Eprint {https://arxiv.org/abs/2405.01633} {arXiv:2405.01633 [astro-ph.HE]}
  \BibitemShut {NoStop}%
\bibitem [{\citenamefont {Gan}\ \emph {et~al.}(2026)\citenamefont {Gan},
  \citenamefont {Kim},\ and\ \citenamefont {Mitridate}}]{Gan:2025icr}%
  \BibitemOpen
  \bibfield  {author} {\bibinfo {author} {\bibfnamefont {X.}~\bibnamefont
  {Gan}}, \bibinfo {author} {\bibfnamefont {H.}~\bibnamefont {Kim}},\ and\
  \bibinfo {author} {\bibfnamefont {A.}~\bibnamefont {Mitridate}},\ }\bibfield
  {title} {\bibinfo {title} {{Probing quadratically coupled ultralight dark
  matter with pulsar timing arrays}},\ }\href
  {https://doi.org/10.1103/15g1-gklx} {\bibfield  {journal} {\bibinfo
  {journal} {Phys. Rev. D}\ }\textbf {\bibinfo {volume} {113}},\ \bibinfo
  {pages} {063034} (\bibinfo {year} {2026})},\ \Eprint
  {https://arxiv.org/abs/2510.13945} {arXiv:2510.13945 [hep-ph]} \BibitemShut
  {NoStop}%
\bibitem [{\citenamefont {Boddy}\ \emph {et~al.}(2025)\citenamefont {Boddy},
  \citenamefont {Dror},\ and\ \citenamefont {Lam}}]{Boddy:2025oxn}%
  \BibitemOpen
  \bibfield  {author} {\bibinfo {author} {\bibfnamefont {K.~K.}\ \bibnamefont
  {Boddy}}, \bibinfo {author} {\bibfnamefont {J.~A.}\ \bibnamefont {Dror}},\
  and\ \bibinfo {author} {\bibfnamefont {A.}~\bibnamefont {Lam}},\ }\bibfield
  {title} {\bibinfo {title} {{Ultralight Dark Matter Statistics for Pulsar
  Timing Detection}},\ }\href {https://doi.org/10.1103/hgnx-w1dn} {\bibfield
  {journal} {\bibinfo  {journal} {Phys. Rev. Lett.}\ }\textbf {\bibinfo
  {volume} {135}},\ \bibinfo {pages} {101001} (\bibinfo {year} {2025})},\
  \Eprint {https://arxiv.org/abs/2502.15874} {arXiv:2502.15874 [hep-ph]}
  \BibitemShut {NoStop}%
\bibitem [{\citenamefont {Dror}\ and\ \citenamefont
  {Wei}(2025)}]{Dror:2025nvg}%
  \BibitemOpen
  \bibfield  {author} {\bibinfo {author} {\bibfnamefont {J.~A.}\ \bibnamefont
  {Dror}}\ and\ \bibinfo {author} {\bibfnamefont {Q.}~\bibnamefont {Wei}},\
  }\bibfield  {title} {\bibinfo {title} {{Pulsar timing detection of ultralight
  vector dark matter}},\ }\href {https://doi.org/10.1103/hh8p-gmxl} {\bibfield
  {journal} {\bibinfo  {journal} {Phys. Rev. D}\ }\textbf {\bibinfo {volume}
  {112}},\ \bibinfo {pages} {075024} (\bibinfo {year} {2025})},\ \Eprint
  {https://arxiv.org/abs/2505.22719} {arXiv:2505.22719 [hep-ph]} \BibitemShut
  {NoStop}%
\bibitem [{\citenamefont {Hu}\ \emph {et~al.}(2026)\citenamefont {Hu} \emph
  {et~al.}}]{Hu:2026yop}%
  \BibitemOpen
  \bibfield  {author} {\bibinfo {author} {\bibfnamefont {X.-S.}\ \bibnamefont
  {Hu}} \emph {et~al.},\ }\bibfield  {title} {\bibinfo {title} {{Constraints on
  ultralight scalar and dark photon dark matter from PPTA DR3 and EPTA DR2}},\
  }\href {https://doi.org/10.1103/wff3-b1fs} {\bibfield  {journal} {\bibinfo
  {journal} {Phys. Rev. D}\ }\textbf {\bibinfo {volume} {113}},\ \bibinfo
  {pages} {123019} (\bibinfo {year} {2026})},\ \Eprint
  {https://arxiv.org/abs/2605.02172} {arXiv:2605.02172 [astro-ph.CO]}
  \BibitemShut {NoStop}%
\bibitem [{\citenamefont {Cherukupalli}\ \emph {et~al.}(2026)\citenamefont
  {Cherukupalli}, \citenamefont {Lee}, \citenamefont {Berghaus},\ and\
  \citenamefont {Zurek}}]{in_prep}%
  \BibitemOpen
  \bibfield  {author} {\bibinfo {author} {\bibfnamefont {A.}~\bibnamefont
  {Cherukupalli}}, \bibinfo {author} {\bibfnamefont {V.~S.~H.}\ \bibnamefont
  {Lee}}, \bibinfo {author} {\bibfnamefont {K.~V.}\ \bibnamefont {Berghaus}},\
  and\ \bibinfo {author} {\bibfnamefont {K.}~\bibnamefont {Zurek}}} (\bibinfo
  {year} {2026}),\ \bibinfo {note} {in preparation}\BibitemShut {NoStop}%
\bibitem [{\citenamefont {Lentati}\ \emph {et~al.}(2016)\citenamefont {Lentati}
  \emph {et~al.}}]{Lentati:2016ygu}%
  \BibitemOpen
  \bibfield  {author} {\bibinfo {author} {\bibfnamefont {L.}~\bibnamefont
  {Lentati}} \emph {et~al.},\ }\bibfield  {title} {\bibinfo {title} {{From Spin
  Noise to Systematics: Stochastic Processes in the First International Pulsar
  Timing Array Data Release}},\ }\href {https://doi.org/10.1093/mnras/stw395}
  {\bibfield  {journal} {\bibinfo  {journal} {Mon. Not. Roy. Astron. Soc.}\
  }\textbf {\bibinfo {volume} {458}},\ \bibinfo {pages} {2161} (\bibinfo {year}
  {2016})},\ \Eprint {https://arxiv.org/abs/1602.05570} {arXiv:1602.05570
  [astro-ph.IM]} \BibitemShut {NoStop}%
\bibitem [{\citenamefont {Hellings}\ and\ \citenamefont
  {Downs}(1983)}]{Hellings:1983fr}%
  \BibitemOpen
  \bibfield  {author} {\bibinfo {author} {\bibfnamefont {R.~W.}\ \bibnamefont
  {Hellings}}\ and\ \bibinfo {author} {\bibfnamefont {G.~S.}\ \bibnamefont
  {Downs}},\ }\bibfield  {title} {\bibinfo {title} {{Upper limits on the
  isotropic gravitational radiation background from pulsar timing analysis}},\
  }\href {https://doi.org/10.1086/183954} {\bibfield  {journal} {\bibinfo
  {journal} {Astrophys. J. Lett.}\ }\textbf {\bibinfo {volume} {265}},\
  \bibinfo {pages} {L39} (\bibinfo {year} {1983})}\BibitemShut {NoStop}%
\bibitem [{\citenamefont {van Haasteren}\ and\ \citenamefont
  {Levin}(2013)}]{vanHaasteren:2012hj}%
  \BibitemOpen
  \bibfield  {author} {\bibinfo {author} {\bibfnamefont {R.}~\bibnamefont {van
  Haasteren}}\ and\ \bibinfo {author} {\bibfnamefont {Y.}~\bibnamefont
  {Levin}},\ }\bibfield  {title} {\bibinfo {title} {{Understanding and
  analysing time-correlated stochastic signals in pulsar timing}},\ }\href
  {https://doi.org/10.1093/mnras/sts097} {\bibfield  {journal} {\bibinfo
  {journal} {Mon. Not. Roy. Astron. Soc.}\ }\textbf {\bibinfo {volume} {428}},\
  \bibinfo {pages} {1147} (\bibinfo {year} {2013})},\ \Eprint
  {https://arxiv.org/abs/1202.5932} {arXiv:1202.5932 [astro-ph.IM]}
  \BibitemShut {NoStop}%
\bibitem [{\citenamefont {Du}\ \emph {et~al.}(2023)\citenamefont {Du},
  \citenamefont {Lee}, \citenamefont {Wang},\ and\ \citenamefont
  {Zurek}}]{Du:2023dhk}%
  \BibitemOpen
  \bibfield  {author} {\bibinfo {author} {\bibfnamefont {Y.}~\bibnamefont
  {Du}}, \bibinfo {author} {\bibfnamefont {V.~S.~H.}\ \bibnamefont {Lee}},
  \bibinfo {author} {\bibfnamefont {Y.}~\bibnamefont {Wang}},\ and\ \bibinfo
  {author} {\bibfnamefont {K.~M.}\ \bibnamefont {Zurek}},\ }\bibfield  {title}
  {\bibinfo {title} {{Macroscopic dark matter detection with gravitational wave
  experiments}},\ }\href {https://doi.org/10.1103/PhysRevD.108.122003}
  {\bibfield  {journal} {\bibinfo  {journal} {Phys. Rev. D}\ }\textbf {\bibinfo
  {volume} {108}},\ \bibinfo {pages} {122003} (\bibinfo {year} {2023})},\
  \Eprint {https://arxiv.org/abs/2306.13122} {arXiv:2306.13122 [astro-ph.CO]}
  \BibitemShut {NoStop}%
\bibitem [{\citenamefont {Lee}\ and\ \citenamefont
  {Zurek}(2025)}]{Lee:2024oxo}%
  \BibitemOpen
  \bibfield  {author} {\bibinfo {author} {\bibfnamefont {V.~S.~H.}\
  \bibnamefont {Lee}}\ and\ \bibinfo {author} {\bibfnamefont {K.~M.}\
  \bibnamefont {Zurek}},\ }\bibfield  {title} {\bibinfo {title} {{Proper time
  observables of general gravitational perturbations in laser
  interferometry-based gravitational wave detectors}},\ }\href
  {https://doi.org/10.1103/6q7d-jz26} {\bibfield  {journal} {\bibinfo
  {journal} {Phys. Rev. D}\ }\textbf {\bibinfo {volume} {111}},\ \bibinfo
  {pages} {124037} (\bibinfo {year} {2025})},\ \Eprint
  {https://arxiv.org/abs/2408.03363} {arXiv:2408.03363 [hep-ph]} \BibitemShut
  {NoStop}%
\bibitem [{\citenamefont {Badurina}\ \emph
  {et~al.}(2025{\natexlab{a}})\citenamefont {Badurina}, \citenamefont {Du},
  \citenamefont {Lee}, \citenamefont {Wang},\ and\ \citenamefont
  {Zurek}}]{Badurina:2024rpp}%
  \BibitemOpen
  \bibfield  {author} {\bibinfo {author} {\bibfnamefont {L.}~\bibnamefont
  {Badurina}}, \bibinfo {author} {\bibfnamefont {Y.}~\bibnamefont {Du}},
  \bibinfo {author} {\bibfnamefont {V.~S.~H.}\ \bibnamefont {Lee}}, \bibinfo
  {author} {\bibfnamefont {Y.}~\bibnamefont {Wang}},\ and\ \bibinfo {author}
  {\bibfnamefont {K.~M.}\ \bibnamefont {Zurek}},\ }\bibfield  {title} {\bibinfo
  {title} {{Signatures of linearized gravity in atom interferometers: A
  simplified computational framework}},\ }\href
  {https://doi.org/10.1103/PhysRevD.111.042002} {\bibfield  {journal} {\bibinfo
   {journal} {Phys. Rev. D}\ }\textbf {\bibinfo {volume} {111}},\ \bibinfo
  {pages} {042002} (\bibinfo {year} {2025}{\natexlab{a}})},\ \Eprint
  {https://arxiv.org/abs/2409.03828} {arXiv:2409.03828 [gr-qc]} \BibitemShut
  {NoStop}%
\bibitem [{\citenamefont {Badurina}\ \emph
  {et~al.}(2025{\natexlab{b}})\citenamefont {Badurina}, \citenamefont {Du},
  \citenamefont {Lee}, \citenamefont {Wang},\ and\ \citenamefont
  {Zurek}}]{Badurina:2025xwl}%
  \BibitemOpen
  \bibfield  {author} {\bibinfo {author} {\bibfnamefont {L.}~\bibnamefont
  {Badurina}}, \bibinfo {author} {\bibfnamefont {Y.}~\bibnamefont {Du}},
  \bibinfo {author} {\bibfnamefont {V.~S.~H.}\ \bibnamefont {Lee}}, \bibinfo
  {author} {\bibfnamefont {Y.}~\bibnamefont {Wang}},\ and\ \bibinfo {author}
  {\bibfnamefont {K.~M.}\ \bibnamefont {Zurek}},\ }\bibfield  {title} {\bibinfo
  {title} {{Detecting gravitational signatures of dark matter with atom
  gradiometers}},\ }\href {https://doi.org/10.1103/xs7b-zgtj} {\bibfield
  {journal} {\bibinfo  {journal} {Phys. Rev. D}\ }\textbf {\bibinfo {volume}
  {112}},\ \bibinfo {pages} {063014} (\bibinfo {year} {2025}{\natexlab{b}})},\
  \Eprint {https://arxiv.org/abs/2505.00781} {arXiv:2505.00781 [hep-ph]}
  \BibitemShut {NoStop}%
\bibitem [{\citenamefont {Rosado}\ \emph {et~al.}(2015)\citenamefont {Rosado},
  \citenamefont {Sesana},\ and\ \citenamefont {Gair}}]{Rosado:2015epa}%
  \BibitemOpen
  \bibfield  {author} {\bibinfo {author} {\bibfnamefont {P.~A.}\ \bibnamefont
  {Rosado}}, \bibinfo {author} {\bibfnamefont {A.}~\bibnamefont {Sesana}},\
  and\ \bibinfo {author} {\bibfnamefont {J.}~\bibnamefont {Gair}},\ }\bibfield
  {title} {\bibinfo {title} {{Expected properties of the first gravitational
  wave signal detected with pulsar timing arrays}},\ }\href
  {https://doi.org/10.1093/mnras/stv1098} {\bibfield  {journal} {\bibinfo
  {journal} {Mon. Not. Roy. Astron. Soc.}\ }\textbf {\bibinfo {volume} {451}},\
  \bibinfo {pages} {2417} (\bibinfo {year} {2015})},\ \Eprint
  {https://arxiv.org/abs/1503.04803} {arXiv:1503.04803 [astro-ph.HE]}
  \BibitemShut {NoStop}%
\bibitem [{\citenamefont {Janssen}\ \emph {et~al.}(2015)\citenamefont {Janssen}
  \emph {et~al.}}]{Janssen:2014dka}%
  \BibitemOpen
  \bibfield  {author} {\bibinfo {author} {\bibfnamefont {G.}~\bibnamefont
  {Janssen}} \emph {et~al.},\ }\bibfield  {title} {\bibinfo {title}
  {{Gravitational wave astronomy with the SKA}},\ }\href
  {https://doi.org/10.22323/1.215.0037} {\bibfield  {journal} {\bibinfo
  {journal} {PoS}\ }\textbf {\bibinfo {volume} {AASKA14}},\ \bibinfo {pages}
  {037} (\bibinfo {year} {2015})},\ \Eprint {https://arxiv.org/abs/1501.00127}
  {arXiv:1501.00127 [astro-ph.IM]} \BibitemShut {NoStop}%
\bibitem [{NAN(2018)}]{NANOGrav:2018vfs}%
  \BibitemOpen
  \bibfield  {title} {\bibinfo {title} {{Science with the Next-Generation VLA
  and Pulsar Timing Arrays}},\ }\href@noop {} {\  (\bibinfo {year} {2018})},\
  \Eprint {https://arxiv.org/abs/1810.06594} {arXiv:1810.06594 [astro-ph.IM]}
  \BibitemShut {NoStop}%
\bibitem [{\citenamefont {Hobbs}\ \emph {et~al.}(2010)\citenamefont {Hobbs}
  \emph {et~al.}}]{Hobbs:2009yy}%
  \BibitemOpen
  \bibfield  {author} {\bibinfo {author} {\bibfnamefont {G.}~\bibnamefont
  {Hobbs}} \emph {et~al.},\ }\bibfield  {title} {\bibinfo {title} {{The
  international pulsar timing array project: using pulsars as a gravitational
  wave detector}},\ }\href {https://doi.org/10.1088/0264-9381/27/8/084013}
  {\bibfield  {journal} {\bibinfo  {journal} {Class. Quant. Grav.}\ }\textbf
  {\bibinfo {volume} {27}},\ \bibinfo {pages} {084013} (\bibinfo {year}
  {2010})},\ \Eprint {https://arxiv.org/abs/0911.5206} {arXiv:0911.5206
  [astro-ph.SR]} \BibitemShut {NoStop}%
\bibitem [{\citenamefont {Read}(2014)}]{Read:2014qva}%
  \BibitemOpen
  \bibfield  {author} {\bibinfo {author} {\bibfnamefont {J.~I.}\ \bibnamefont
  {Read}},\ }\bibfield  {title} {\bibinfo {title} {{The Local Dark Matter
  Density}},\ }\href {https://doi.org/10.1088/0954-3899/41/6/063101} {\bibfield
   {journal} {\bibinfo  {journal} {J. Phys. G}\ }\textbf {\bibinfo {volume}
  {41}},\ \bibinfo {pages} {063101} (\bibinfo {year} {2014})},\ \Eprint
  {https://arxiv.org/abs/1404.1938} {arXiv:1404.1938 [astro-ph.GA]}
  \BibitemShut {NoStop}%
\bibitem [{\citenamefont {de~Salas}\ and\ \citenamefont
  {Widmark}(2021)}]{deSalas:2020hbh}%
  \BibitemOpen
  \bibfield  {author} {\bibinfo {author} {\bibfnamefont {P.~F.}\ \bibnamefont
  {de~Salas}}\ and\ \bibinfo {author} {\bibfnamefont {A.}~\bibnamefont
  {Widmark}},\ }\bibfield  {title} {\bibinfo {title} {{Dark matter local
  density determination: recent observations and future prospects}},\ }\href
  {https://doi.org/10.1088/1361-6633/ac24e7} {\bibfield  {journal} {\bibinfo
  {journal} {Rept. Prog. Phys.}\ }\textbf {\bibinfo {volume} {84}},\ \bibinfo
  {pages} {104901} (\bibinfo {year} {2021})},\ \Eprint
  {https://arxiv.org/abs/2012.11477} {arXiv:2012.11477 [astro-ph.GA]}
  \BibitemShut {NoStop}%
\bibitem [{\citenamefont {{Pitjev}}\ and\ \citenamefont
  {{Pitjeva}}(2013)}]{2013AstL...39..141P}%
  \BibitemOpen
  \bibfield  {author} {\bibinfo {author} {\bibfnamefont {N.~P.}\ \bibnamefont
  {{Pitjev}}}\ and\ \bibinfo {author} {\bibfnamefont {E.~V.}\ \bibnamefont
  {{Pitjeva}}},\ }\bibfield  {title} {\bibinfo {title} {{Constraints on dark
  matter in the solar system}},\ }\href
  {https://doi.org/10.1134/S1063773713020060} {\bibfield  {journal} {\bibinfo
  {journal} {Astronomy Letters}\ }\textbf {\bibinfo {volume} {39}},\ \bibinfo
  {pages} {141} (\bibinfo {year} {2013})},\ \Eprint
  {https://arxiv.org/abs/1306.5534} {arXiv:1306.5534 [astro-ph.EP]}
  \BibitemShut {NoStop}%
\bibitem [{\citenamefont {Agazie}\ \emph
  {et~al.}(2023{\natexlab{b}})\citenamefont {Agazie} \emph
  {et~al.}}]{NANOGrav:2023pdq}%
  \BibitemOpen
  \bibfield  {author} {\bibinfo {author} {\bibfnamefont {G.}~\bibnamefont
  {Agazie}} \emph {et~al.} (\bibinfo {collaboration} {NANOGrav}),\ }\bibfield
  {title} {\bibinfo {title} {{The NANOGrav 15 yr Data Set: Bayesian Limits on
  Gravitational Waves from Individual Supermassive Black Hole Binaries}},\
  }\href {https://doi.org/10.3847/2041-8213/ace18a} {\bibfield  {journal}
  {\bibinfo  {journal} {Astrophys. J. Lett.}\ }\textbf {\bibinfo {volume}
  {951}},\ \bibinfo {pages} {L50} (\bibinfo {year} {2023}{\natexlab{b}})},\
  \Eprint {https://arxiv.org/abs/2306.16222} {arXiv:2306.16222 [astro-ph.HE]}
  \BibitemShut {NoStop}%
\bibitem [{\citenamefont {Foster}\ \emph {et~al.}(2026)\citenamefont {Foster},
  \citenamefont {Trickle},\ and\ \citenamefont {Vassallo}}]{Foster:2026kfg}%
  \BibitemOpen
  \bibfield  {author} {\bibinfo {author} {\bibfnamefont {J.~W.}\ \bibnamefont
  {Foster}}, \bibinfo {author} {\bibfnamefont {T.}~\bibnamefont {Trickle}},\
  and\ \bibinfo {author} {\bibfnamefont {F.}~\bibnamefont {Vassallo}},\
  }\bibfield  {title} {\bibinfo {title} {{Projecting the ultimate pulsar timing
  sensitivity to dark matter substructure in a stochastic gravitational wave
  background}},\ }\href@noop {} {\  (\bibinfo {year} {2026})},\ \Eprint
  {https://arxiv.org/abs/2606.18329} {arXiv:2606.18329 [astro-ph.CO]}
  \BibitemShut {NoStop}%
\bibitem [{\citenamefont {Allen}\ and\ \citenamefont
  {Romano}(1999)}]{Allen:1997ad}%
  \BibitemOpen
  \bibfield  {author} {\bibinfo {author} {\bibfnamefont {B.}~\bibnamefont
  {Allen}}\ and\ \bibinfo {author} {\bibfnamefont {J.~D.}\ \bibnamefont
  {Romano}},\ }\bibfield  {title} {\bibinfo {title} {{Detecting a stochastic
  background of gravitational radiation: Signal processing strategies and
  sensitivities}},\ }\href {https://doi.org/10.1103/PhysRevD.59.102001}
  {\bibfield  {journal} {\bibinfo  {journal} {Phys. Rev. D}\ }\textbf {\bibinfo
  {volume} {59}},\ \bibinfo {pages} {102001} (\bibinfo {year} {1999})},\
  \Eprint {https://arxiv.org/abs/gr-qc/9710117} {arXiv:gr-qc/9710117}
  \BibitemShut {NoStop}%
\end{thebibliography}%
\let\addcontentsline\oldaddcontentsline

\onecolumngrid
\appendix

\section{Statistical Framework of PTA Signals}
\label{app:pulsar_SNRs}

This appendix derives the results quoted in Sec.~\ref{sec:SNR}: that marginalizing over the timing model is equivalent to projecting it out, and that the deterministic and stochastic SNRs, Eqs.~\eqref{eqn:projected_snr_pulsar} and \eqref{eqn:projected_stoch_SNR_pulsar}, follow from two special cases of the likelihood ratio. We treat signals localized in a single pulsar here; the generalization to signals correlated across the array follows in \cite{in_prep}. %

Starting from Eq.~\eqref{eqn:timing_residual_full} in Sec.~\ref{sec:SNR}, the timing residuals of a single pulsar are the sum of a timing model, noise, and the signal we wish to detect,
\begin{equation}
    \delta t(t) \approx \vec{m}(t)\cdot\vec{\xi} + \delta t_\text{sig}(t) + \delta n(t) ,
\end{equation}
where $\vec m(t)=[\phi_1(t),\phi_2(t)]$ is the timing-model design vector of
Eq.~\eqref{eqn:design vector}, $\vec\xi$ its unknown coefficients, and $\delta n(t)$ the noise of Sec.~\ref{sec:timing_model_marginalization} with covariance $\langle\delta n(t)\,\delta n(t')\rangle = N(t,t')$. Assuming that the noise $\delta n(t)$ is Gaussian, the likelihood of the data conditioned on a particular value of $\vec{\xi}$ is a Gaussian in the noise-weighted inner product,
\begin{equation}
    p\left[\delta t|\delta t_\text{sig}; \vec{\xi}\right] \propto \exp \!\left[-\frac{1}{2} \left(\delta t - \delta t_\text{sig}- \vec{m}\cdot \vec{\xi} \,\big|\, \delta t - \delta t_\text{sig}-\vec{m}\cdot \vec{\xi}\right)\right],
\end{equation}
where $(g|h)$ denotes the inner product defined in Eq.~\eqref{eqn:inner_prod_def}.\footnote{The functional-determinant normalization cancels in likelihood ratios and is not written explicitly.} Marginalizing over $\vec{\xi}$ with a flat prior~\cite{vanHaasteren:2012hj} reduces to a
Gaussian integral. Completing the square in $\vec\xi$ and integrating,
\begin{equation}
\begin{split}
    p\left[\delta t|\delta t_\text{sig}\right] &\propto \int d\vec{\xi} \,
    p\left[\delta t|\delta t_\text{sig}; \vec{\xi}\right] \\
    &\propto \exp \!\left[-\frac{1}{2} \left(\delta t^\perp -\delta t^\perp_\text{sig}\,\middle|\,
    \delta t^\perp-\delta t^\perp_\text{sig}\right)\right],
\end{split}
\end{equation}
where the projected residual is
\begin{equation}\label{eqn:appendix_projection_formula}
    \delta t^\perp(t)= \delta t(t) - \sum_{ij} \phi_i(t)\, G^{-1}_{ij}\, (\phi_j | \delta t),
\end{equation}
with $G_{ij}=(\phi_i|\phi_j)$ the Gram matrix of the timing model, and $\delta t^\perp_\text{sig}$
defined identically. Equation~\eqref{eqn:appendix_projection_formula} is the projection of the data onto the orthogonal complement of the timing-model subspace $\{\phi_1,\phi_2\}$. Its Gram matrix is
diagonal by parity: under $t\to T-t$ the Legendre modes obey $\phi_n\to(-1)^n\phi_n$ while the stationary kernel depends only on $|t-t'|$ and is therefore even, so $(\phi_i|\phi_j)=(-1)^{i+j}(\phi_i|\phi_j)$ vanishes for $i+j$ odd; in particular $(\phi_1|\phi_2)=0$. The double sum in Eq.~\eqref{eqn:appendix_projection_formula} then collapses to the diagonal, term-by-term projection of Eq.~\eqref{eqn:deterministic_pulsar_proj}, in which each mode $\phi_i$ is subtracted independently.\footnote{The constant mode $\phi_0$ is dropped by working with mean-subtracted residuals,
equivalent to the cutoff $f\ge 1/T$.}

Following the Neyman–Pearson criterion, we use the marginalized likelihood ratio as the detection statistic. A signal hypothesis $\mathcal H_1$ is specified by a prior $\pi[\delta t_\text{sig}]$ on the waveform, with marginal likelihood
\begin{equation}\label{eqn:appendix_evidence}
    p\left[\delta t \,\middle|\, \mathcal H_1\right]
    = \int \mathcal{D}[\delta t_\text{sig}]\;\pi[\delta t_\text{sig}]\;
    p\left[\delta t \,\middle|\, \delta t_\text{sig}\right],
\end{equation}
with the null hypothesis $\mathcal H_0$ corresponding to the absence of a signal,
$p[\delta t|\mathcal H_0]\propto\exp[-\tfrac12(\delta t^\perp|\delta t^\perp)]$. The
detection significance of the log-likelihood ratio
$\ln\Lambda \equiv \ln p[\delta t|\mathcal H_1]-\ln p[\delta t|\mathcal H_0]$ is quantified by the SNR,
\begin{equation}\label{eqn:appendix_snr_def}
    \mathrm{SNR}\equiv
    \frac{\langle\ln\Lambda\rangle_{\mathcal H_1}-\langle\ln\Lambda\rangle_{\mathcal H_0}}
    {\sqrt{\mathrm{Var}_{\mathcal H_0}(\ln\Lambda)}}.
\end{equation}
Two choices of prior exhaust the cases of interest. The first is a deterministic signal with a known waveform, a sharp prior fixing
$\delta t_\text{sig}=\delta t_\text{sig}^\star$, for which Eq.~\eqref{eqn:appendix_snr_def}
gives the exact result
\begin{equation}\label{eqn:appendix_snr_det}
    \mathrm{SNR}^2 = (\delta t_\text{sig}^{\star,\perp}
    \,|\, \delta t_\text{sig}^{\star,\perp}),
\end{equation}
the noise-weighted norm of the projected template, recovering
Eq.~\eqref{eqn:projected_snr_pulsar}. The second is a stochastic, unresolved signal, described by a zero-mean Gaussian prior
$\delta t_\text{sig}\sim\mathcal N(0,S_\text{sig})$. Evaluating Eq.~\eqref{eqn:appendix_snr_def} with this prior, to leading order in the weak-signal limit $S_\text{sig}^\perp\ll N$, i.e., expanding $(N+S^\perp_{\rm sig})^{-1}=N^{-1}-N^{-1}S^\perp_{\rm sig}N^{-1}+\dots$, gives the standard optimal-statistic result~\cite{Allen:1997ad}
\begin{equation}\label{eqn:appendix_snr_stoch}
    \mathrm{SNR}^2 = \frac{1}{2}\,\mathrm{Tr}\!\left[
    N^{-1} S_\text{sig}^\perp\, N^{-1} S_\text{sig}^\perp\right],
\end{equation}
with $N^{-1}$ understood on the timing-model-orthogonal subspace. For stationary signal and noise the trace diagonalizes in the Fourier basis on $[0,T]$, reducing Eq.~\eqref{eqn:appendix_snr_stoch} to 
\begin{equation}
    \label{eqn:stoch_SNR_fourier_one_pulsar}
    \mathrm{SNR}^2 = T\int_{1/T}^{\infty}df\, \left(\frac{S^\perp_{\rm sig}(f)}{S_n(f)}\right)^2,
\end{equation}
with the one-sided PSD of Eq.~\eqref{eqn:sig_PSD_onesided}. The $N_P$ pulsars carry independent noise and signal realizations, so their $\mathrm{SNR}^2$ add incoherently, giving Eq.~\eqref{eqn:stoch_SNR_fourier}.

\section{Demonstration of Gauge Invariance of the DM-induced Proper Time Shifts}\label{app:gauge_invariance}

The pulsar timing array signal due to a transiting DM object as shown in Sec.~\ref{sec:dark_matter_residual} is derived in Newtonian gauge. As argued in Ref.~\cite{Lee_2026}, the proper time shift in a pulsar timing measurement is independent of the gauge choice. In this appendix, we explicitly demonstrate this by performing a gauge transformation on the metric perturbation in Eq.~\eqref{eqn:subhalo_metric}, and show that the expression for the proper time shift as measured by PTAs that is used in our analysis remains unchanged. An analogous proof also for DM signals, but in the context of laser and atom interferometers is given in Refs.~\cite{Du:2023dhk, Badurina:2024rpp}.

In Newtonian gauge, the metric perturbation due to the DM subhalo is given by Eq.~\eqref{eqn:subhalo_metric}
\begin{equation}\label{eqn:Newtonian_gauge_app}
    h_{00} = -2\Phi\,, \quad h_{0i} = 0\,, \quad h_{ij} = -2\Phi\,\delta_{ij}\,,
\end{equation}
with $\Phi$ being the Newtonian potential induced by the DM. In particular, in this gauge choice, the cross components between spatial and temporal coordinates, $h_{0i}$, are taken to be zero. This no longer holds if we now consider the metric in a new gauge obtained by performing a gauge transformation parameterized by, \textit{e.g.},
\begin{equation}\label{eqn:gauge_transformation_app}
    \xi_0 = 0\,, \qquad \xi_i = \partial_i\chi(t,\vec{x})\,,
\end{equation}
where $\chi(t,\vec{x})$ is an arbitrary scalar function. Under $h_{\mu\nu} \to h'_{\mu\nu} = h_{\mu\nu} - \partial_\mu\xi_\nu - \partial_\nu\xi_\mu$, the transformed metric components are
\begin{equation}\label{eqn:transformed_metric_app}
    h'_{00} = -2\Phi\,, \quad h'_{0i} = -\partial_i\dot\chi\,, \quad h'_{ij} = -2\Phi\,\delta_{ij} - 2\partial_i\partial_j\chi\,,
\end{equation}
where $\dot\chi \equiv \partial_t\chi$. 
We now show that the total proper time shift $\delta t(t)$ is unchanged under this transformation. Let $\vec{x}_E$ and $\vec{x}_P$ denote the positions of Earth and the pulsar, respectively, $\unit{n}$ the unit vector pointing from Earth to the pulsar, and $L = |\vec{x}_P - \vec{x}_E|$ the Earth-pulsar distance. The proper time shift observable for a general metric perturbation $h_{\mu\nu}$ decomposes into Doppler, Shapiro, and Einstein contributions as Eq.~\eqref{eqn:delta_t_t}, with~\cite{Lee_2026} %
\begin{align}\label{eqn:contributions_app}
    \delta t^{(\mathcal{D})}_E(t) &= -\frac{1}{2}\int^t dt'\int^{t'}dt''\,n^i\partial_i h_{00}(t'',\vec{x}_E) + \int^t dt'\,n^i h_{0i}(t',\vec{x}_E) \nonumber \\
    \delta t^{(\mathcal{D})}_P(t) &= -\frac{1}{2}\int^{t-L} dt'\int^{t'}dt''\,\hat{n}^i\partial_i h_{00}(t'',\vec{x}_P) + \int^{t-L} dt'\,n^i h_{0i}(t',\vec{x}_P) \nonumber \\
    \delta t^{(\mathcal{S})}(t) &= +\frac{1}{2}\int_0^L dz\,\left[h_{00} - 2n^i h_{0i} + n^in^j h_{ij}\right](t-z,\vec{x}_E+z\hat{n}) \nonumber \\
    \delta t^{(\mathcal{E})}_E(t) &= -\frac{1}{2}\int^t dt'\,h_{00}(t',\vec{x}_E) \nonumber \\
    \delta t^{(\mathcal{E})}_P(t) &= -\frac{1}{2}\int^{t-L} dt'\,h_{00}(t',\vec{x}_P) \, .
\end{align}
We now compute the change in each contribution under the gauge transformation in Eq.~\eqref{eqn:transformed_metric_app}. We denote the difference in proper time shift measured between the two gauges as $\Delta \delta t\equiv \delta t'-\delta t$. Since $h'_{00} = h_{00}$, the Einstein delay terms, which depend only on $h_{00}$, are immediately unchanged (as evident from Eq.~\eqref{eqn:contributions_app})
\begin{equation}\label{eqn:Einstein_unchanged_app}
    \Delta\delta t^{(\mathcal{E})}_E = \Delta\delta t^{(\mathcal{E})}_P = 0\,.
\end{equation}

For the Doppler terms, the $h_{00}$ contribution is likewise unchanged. The nonzero $h'_{0i} = -\partial_i\dot\chi$ generates an additional contribution:
\begin{align}\label{eqn:Doppler_change_app}
    \Delta\delta t^{(\mathcal{D})}_E(t) &= -n^i\partial_i\chi(t,\vec{x}_E)\,, \nonumber \\
    \Delta\delta t^{(\mathcal{D})}_P(t) &= -n^i\partial_i\chi(t-L,\vec{x}_P)\,.
\end{align}
For the Shapiro term, the change in the integrand due to $h'_{0i} = -\partial_i\dot\chi$ and $h'_{ij} = -2\Phi\,\delta_{ij} - 2\partial_i\partial_j\chi$ is
\begin{equation}\label{eqn:Shapiro_integrand_change}
    \Delta\left(h_{00} - 2n^i h_{0i} + n^in^j h_{ij}\right) = 2n^i\partial_i\dot\chi - 2n^in^j\partial_i\partial_j\chi\,.
\end{equation}
Along the unperturbed photon path, the total derivative of any function $f$ with respect to $z$ is
\begin{equation}\label{eqn:total_deriv_def}
    \frac{d}{dz}f(t-z,\, \vec{x}_E + z\hat{n}) = -\partial_t f + n^i\partial_i f\,.
\end{equation}
Applying this to $\hat{n}^j\partial_j\chi$:
\begin{equation}\label{eqn:total_deriv_applied}
    \frac{d}{dz}(n^j\partial_j\chi) = -n^j\partial_j\dot\chi + n^in^j\partial_i\partial_j\chi\,,
\end{equation}
from which we identify Eq.~\eqref{eqn:Shapiro_integrand_change} as $-2(d/dz)(\hat{n}^j\partial_j\chi)$. The change in the Shapiro term therefore reduces to a total derivative: 
\begin{align}\label{eqn:Shapiro_change_app}
    \Delta\delta t^{(\mathcal{S})}(t) &= \frac{1}{2}\int_0^L dz\,\left[-2\frac{d}{dz}(n^i\partial_i\chi)\right] = \left[n^i\partial_i\chi\right]_0^L \nonumber \\
    &=  n^i\partial_i\chi(t,\vec{x}_E) - n^i\partial_i\chi(t-L,\vec{x}_P)\,.
\end{align}
Combining all contributions using Eq.~\eqref{eqn:delta_t_t}, Eq.~\eqref{eqn:Doppler_change_app} and Eq.~\eqref{eqn:Shapiro_change_app}, it is clear that gauge-dependent terms cancel, confirming that the total observable $\delta t(t)$ is invariant under the gauge transformation in Eq.~\eqref{eqn:gauge_transformation_app}.

\section{Red-noise Suppression of the Sensitivity Integrals}
\label{sec:appendix_integral}
Every sensitivity estimate in Sec.~\ref{sec:analytical_estimates} %
is controlled by one of two dimensionless integrals, 
\begin{equation}\label{eqn:app:Ingamma_det_stoch_def}
\begin{split}
    I_{n,\gamma} &= \int_{1/T}^\infty \frac{df}{f}\,\frac{f^\gamma}{f_\star^\gamma + f^\gamma}\,\frac{1}{(fT)^{n-1}},\\
    I^{(2)}_{n,\gamma} &= \int_{1/T}^\infty \frac{df}{f}\,\left(\frac{f^\gamma}{f_\star^\gamma + f^\gamma}\right)^2\,\frac{1}{(fT)^{2n-1}},
\end{split}
\end{equation}
the first for signals in the deterministic regime and the second for signals in the stochastic regime. Following the argument of Ref.~\cite{Berghaus:2025kvn}, we show how these integrals scale with $f_\star T$ in the limit $f_\star T \gg 1$. Because both integrals carry the measure $df/f$, it is natural to work with the integrand per logarithmic frequency bin. For the deterministic case we have 
\begin{equation}
\begin{split}
    \frac{d I_{n, \gamma}}{d \ln f} &= \left(\frac{f^\gamma}{f_\star^\gamma + f^\gamma}\right) \times (fT)^{-(n-1)} \\
    &\propto \begin{dcases*}
        f^{\gamma-(n-1)} & for $f \ll f_\star$\\
        f^{-(n-1)} & for $f \gg f_\star$
    \end{dcases*}.
\end{split}
\end{equation}
When $\gamma > n-1$, the sub-$f_\star$ integrand rises toward $f_\star$ and the weight piles up near $f_\star$. Therefore, up to an order-unity factor, the integral is set by the integrand at the crossover, i.e., $I_{n, \gamma} \sim (f_\star T)^{-(n-1)}$. Conversely, when $\gamma < n-1$ the weight piles up near $1/T$ and $I_{n, \gamma} \sim (f_\star T)^{-\gamma}$. In the marginal case where $\gamma = n-1$, for frequencies $1/T \leq f \ll f_\star$, all bins contribute equally, giving us $(f_\star T)^{-\gamma} \ln(f_\star T)$, the logarithm coming from the width of the red-noise band. Putting it together, along with the order-unity prefactors computed below, we find the following expression for $I_{n, \gamma}$ in the limit $f_\star T\gg 1$
\begin{equation}
    I_{n, \gamma} \approx c_{n, \gamma} (f_\star T)^{-\min(\gamma, \, n-1)}, \quad  c_{n, \gamma}=\begin{cases}
        \displaystyle\frac{\pi}{\gamma\sin(\pi (n-1)/\gamma)}, & \gamma > n-1, \\[8pt]
        \displaystyle \ln(f_\star T), & \gamma = n-1, \\[8pt]
        \displaystyle\frac{1}{n-1-\gamma}, & \gamma < n-1.
    \end{cases}
\end{equation}
We can compute $c_{n, \gamma}$ by substituting $t = f/f_\star$ to factor out the overall scale,
\begin{equation}
    I_{n,\gamma} = (f_\star T)^{-(n-1)} \int_{\epsilon}^\infty \frac{t^{\gamma-n}}{1+t^\gamma}\,dt, \qquad \epsilon \equiv \frac{1}{f_\star T} \to 0,
\end{equation}
when $\gamma > n-1$, the integral is convergent at the lower limit, which can then be moved to zero, leaving a standard integral to leading order.
\begin{align}
    I_{n,\gamma} &=(f_\star T)^{-(n-1)} \int_{0}^\infty \frac{t^{\gamma-n}}{1+t^\gamma}\,dt +\mathcal{O}\left({f_\star T}\right)^{-\gamma} \nonumber\\
    &= \frac{\pi}{\gamma\sin[\pi (n-1)/\gamma]} (f_\star T)^{-(n-1)}  +\mathcal{O}\left({f_\star T}\right)^{-\gamma}.
\end{align}
Conversely, when $\gamma < n-1$, the integral is controlled by the lower limit $\epsilon \to 0$
\begin{align}
    I_{n,\gamma} &=(f_\star T)^{-(n-1)} \int_{\epsilon}^1 \frac{t^{\gamma-n}}{1+t^\gamma}\,dt +\mathcal{O}\left({f_\star T}\right)^{-\min(n-1, 2\gamma)} \nonumber \\[0.2em]
    &= \frac{1}{n-1-\gamma} \left({f_\star T}\right)^{-\gamma}+\mathcal{O}\left({f_\star T}\right)^{-\min(n-1, 2\gamma)},
\end{align}
and in the marginal case $\gamma = n-1$, the integral can be performed exactly
\begin{align}
    I_{n,\gamma}&=  \left({f_\star T}\right)^{-\gamma}\int_{\epsilon}^\infty\frac{dt}{t(1+t^\gamma)}\nonumber
    \\
    &= \left({f_\star T}\right)^{-\gamma}
    \left[\ln\left(f_\star T\right) + \frac{1}{\gamma}\ln\left(1+(f_\star T)^{-\gamma}\right)\right]\nonumber\\[0.2em]
    &=\left({f_\star T}\right)^{-\gamma}\ln\left(f_\star T\right) +\mathcal{O}\left({f_\star T}\right)^{-2\gamma}.
\end{align}
Whether $c_{n,\gamma}$ is order-unity is not automatic; it depends on how far $\gamma$ sits from the boundary $\gamma=n-1$. For a fixed signal, $n$ is set by the template ($n=2,4,6$ for the dynamic Shapiro, dynamic Doppler, and static cases), while $\gamma$ ranges over the GWB posterior in Sec.~\ref{sec:results_and_discussion}, $\gamma\in\{2.3,\,3.3,\,13/3\}$. Across these values $0.4 \lesssim c_{n,\gamma} \lesssim 3$. %
Using an analogous scaling argument, one finds the following expression for $I^{(2)}_{n, \gamma}$
\begin{equation}
    I^{(2)}_{n, \gamma} \approx c^{(2)}_{n, \gamma} (f_\star T)^{-2\min(\gamma, \, n-1/2)}, \quad  c^{(2)}_{n, \gamma}=\begin{cases}
        \displaystyle\frac{1}{\gamma}\, \Gamma\left(\frac{2n-1}{\gamma}\right)\Gamma\left(2-\frac{2n-1}{\gamma}\right), & \gamma > n-\frac{1}{2}, \\[10pt]
        \displaystyle \ln(f_\star T) -\frac{1}{\gamma}, & \gamma = n-\frac{1}{2}, \\[10pt]
        \displaystyle \frac{1}{2n-1-2\gamma}, & \gamma < n-\frac{1}{2}.
    \end{cases}
\end{equation}
The prefactor $c^{(2)}_{n, \gamma}$ is computed in a similar way, starting from the same substitution $t = f/f_\star$,
\begin{equation}
    I^{(2)}_{n,\gamma} = (f_\star T)^{-(2n-1)} \int_{\epsilon}^\infty \frac{t^{2(\gamma-n)}}{(1+t^\gamma)^2}\,dt, \qquad \epsilon \equiv \frac{1}{f_\star T} \to 0,
\end{equation}
when $\gamma > n-1/2$, the lower limit can again be moved to zero, and to leading order
\begin{align}
    I^{(2)}_{n,\gamma} &= (f_\star T)^{-(2n-1)} \int_{0}^\infty \frac{t^{2(\gamma-n)}}{(1+t^\gamma)^2}\,dt + \mathcal{O}\left({f_\star T}\right)^{-2\gamma},\nonumber \\[0.8em]
    &= \frac{1}{\gamma}\, \Gamma\left(\frac{2n-1}{\gamma}\right)\Gamma\left(2-\frac{2n-1}{\gamma}\right)(f_\star T)^{-(2n-1)}  + \mathcal{O}\left({f_\star T}\right)^{-2\gamma},
\end{align}
where we substituted $s=t^\gamma$ to identify this integral with a Beta function. When $\gamma < n-1/2$, as in the deterministic case, the integral is controlled by the lower limit $\epsilon \to 0$
\begin{align}
    I^{(2)}_{n,\gamma} &=(f_\star T)^{-(2n-1)} \int_{\epsilon}^1 \frac{t^{2(\gamma-n)}}{(1+t^\gamma)^2}\,dt +\mathcal{O}\left({f_\star T}\right)^{-\min(2n-1, 3\gamma)} \nonumber \\[0.2em]
    &= \frac{1}{2n-1-2\gamma} \left({f_\star T}\right)^{-2\gamma}+\mathcal{O}\left({f_\star T}\right)^{-\min(2n-1, 3\gamma)},
\end{align}
and in the marginal case $\gamma = n-1/2$, the integral can once again be performed exactly
\begin{align}
    I^{(2)}_{n,\gamma}&=  \left({f_\star T}\right)^{-2\gamma}\int_{\epsilon}^\infty\frac{dt}{t(1+t^\gamma)^2}\nonumber
    \\
    &= \left({f_\star T}\right)^{-2\gamma}
    \left[\ln\left(f_\star T\right) + \frac{1}{\gamma}\left[\ln\left(1+(f_\star T)^{-\gamma}\right) -\frac{1}{1+(f_\star T)^{-\gamma}}\right]\right]\nonumber\\[0.2em]
    &=\left({f_\star T}\right)^{-2\gamma}\left[\ln\left(f_\star T\right) -\frac{1}{\gamma}\right]+\mathcal{O}\left({f_\star T}\right)^{-3\gamma}.
\end{align}
The same reasoning applies to $c^{(2)}_{n,\gamma}$, with the boundary now at $\gamma=n-1/2$. Here $n=3,4$ for the stochastic Shapiro and Doppler signals, and across the same posterior $\gamma\in\{2.3,\,3.3,\,13/3\}$ one finds $0.2\lesssim c^{(2)}_{n,\gamma}\lesssim 3$. %

These prefactors are not universal constants but carry a genuine $n$- and $\gamma$-dependence; since they stay order-unity across our range, we ignore them and set $c_{n,\gamma},\,c^{(2)}_{n,\gamma}\to 1$, recovering the parametric scalings of Eq.~\eqref{eqn:integral_scalings} used throughout the analytic estimates of Sec.~\ref{sec:analytical_estimates}.

\section{Noise-weighted Norm of the projected Signals}
\label{app:noise_weighted_norm}

\subsection{Static Limit}
\label{app:phi3perp_norm}
Because the timing model fits away the constant, linear, and quadratic components of the residuals, a slowly varying DM flyby contributes only through the part of the signal orthogonal to that subspace. In the static limit the leading non-degenerate piece is cubic in time, so the detection statistic depends on the projected cubic template $\phi_3^\perp$ and its noise-weighted norm $(\phi_3^\perp|\phi_3^\perp)$, where $\phi_n$ are the Legendre polynomial basis from Eq.~\eqref{eqn:legendre_def}.
Because the red-noise weighting mixes $\phi_1$ and $\phi_3$ ($(\phi_1|\phi_3)\neq 0$; Sec.~\ref{sec:analytics_static_limit}), the projection genuinely modifies the cubic template, $\phi_3^\perp \neq \phi_3$. To estimate $(\phi_3^\perp| \phi_3^\perp)$, we compute the Fourier transform
\begin{align}\label{eq:legendre_FT}
     \widetilde{\phi}_n(f) &= \int_0^Tdt e^{-2\pi i ft}  \phi_n(t)= T \sqrt{2n+1} (-i)^n (-1)^{fT} j_n(\pi fT).
\end{align}
The likelihood samples this transform on the discrete Fourier-series modes $fT\in \mathbb{Z}$, as in Eq.~\eqref{eqn:inner_prod_def}. For these discrete frequencies, we simplify Eq.~\eqref{eq:legendre_FT} and compute the first three basis functions
\begin{align} \label{eqn:legendre_FT_examples}
     \widetilde{\phi}_1(f) &= \frac{\sqrt{3}i}{\pi f}, \nonumber \\
     \widetilde{\phi}_2(f) &= \frac{3\sqrt{5}}{\pi f}\frac{1}{\pi fT}, \nonumber \\
     \widetilde{\phi}_3(f) &= \frac{\sqrt{7}i}{\pi f}\left[1-\frac{15}{(\pi fT)^2}\right] \, .
\end{align}
For the scaling estimates that follow, we approximate this discrete sum by a continuous integral; however, we still need to use the values at discrete frequencies in Eq.~\eqref{eqn:legendre_FT_examples} and not the continuous one in Eq.~\eqref{eq:legendre_FT}. The oscillatory zeros of the continuous transform between these modes are windowing artifacts and not additional independent frequency bins in the statistic. 

It is convenient to separate the piece of $\tilde{\phi}_3$ that is exactly proportional to $\tilde{\phi}_1$ (and hence degenerate) from the remainder that controls the projected norm. Working with the Fourier transforms from Eq.~\eqref{eqn:legendre_FT_examples}, we define the remainder
\begin{equation}
    \tilde{\chi}(f) = -\frac{15\sqrt{7}\, i}{\pi f}\,\frac{1}{(\pi f T)^2}
\end{equation}
and write
\begin{equation}
    \tilde{\phi}_3(f) = \sqrt{\frac{7}{3}}\tilde{\phi}_1(f) + \tilde{\chi}(f).
\end{equation}
Using $(\phi_2|\phi_3)=0$ from Eq.~\eqref{eqn:orthonormal_opp_parity}, we have
\begin{equation}
\begin{split}
    \tilde{\phi}_3^\perp(f)
    &= \tilde{\phi}_3(f) - \frac{(\phi_1|\phi_3)}{(\phi_1|\phi_1)}\,\tilde{\phi}_1(f) \\
    &= \tilde{\chi}(f) - \frac{(\phi_1|\chi)}{(\phi_1|\phi_1)}\,\tilde{\phi}_1(f).
\end{split}
\end{equation}
The norm of the projected cubic mode is therefore
\begin{equation}
    (\phi_3^\perp|\phi_3^\perp)
    = (\chi|\chi)
    - \frac{(\phi_1|\chi)^2}{(\phi_1|\phi_1)} .
\end{equation}
Because $\tilde{\chi}\sim f^{-3}$ and $\tilde{\phi}_1\sim f^{-1}$, $(\chi|\chi)$ maps onto the $n=6$ case of Eq.~\eqref{eqn:integral_scalings} (see Tab.~\ref{tab:signal_scalings}), while $(\phi_1|\phi_1)$ and $(\chi|\phi_1)$ map onto the $n=2$ and $n=4$ cases, respectively. This yields %
\begin{equation}
\begin{split}
    (\chi | \chi)\simeq \frac{3150}{\pi^6}\,
    \frac{T}{\Delta t\, t^2_{\mathrm{rms}}}
    \left(f_\star T\right)^{-\min(\gamma,5)},
\end{split}
\end{equation}
and
\begin{equation}
    \frac{(\phi_1|\chi)^2}{(\phi_1|\phi_1)}
    \simeq \frac{3150}{\pi^6}
    \frac{T}{\Delta t\, t^2_{\mathrm{rms}}}
    \left(f_\star T\right)^{-2\min(\gamma,3)+\min(\gamma,1)}.
\end{equation}
Because $2\min(\gamma, 3) \geq \min(\gamma,1)+\min(\gamma,5)$, the second term is parametrically subdominant to $(\chi|\chi)$ in the limit
$f_\star T \gg 1$ %
, leading to %
\begin{equation} 
    (\phi_3^\perp|\phi_3^\perp)
    \simeq \frac{3150}{\pi^6}\,
    \frac{T}{\Delta t\, t^2_{\mathrm{rms}}}
    \left(f_\star T\right)^{-\min(\gamma,5)}.
\end{equation}

\subsection{Dynamic Limit}

In the dynamic limit, flybys have an encounter timescale short compared to the observing baseline, $\tau\ll T$; however, even after the distributional pieces are removed, their residuals are not compactly supported in time and leave low-frequency power that overlaps with the timing model. Working with bulk events --- those whose closest approach lies well inside the window, $t_0\gg\tau$ and $T-t_0\gg\tau$ ---  deep in the dynamic limit with $f_\star \tau \ll 1$, we show that this degeneracy is sharply suppressed for $\gamma > 3$ and $\gamma > 2$ in the Doppler and Shapiro cases, respectively. For the opposite limits, we show that the leftover degeneracy does not change the parametric scaling of the noise-weighted norm with $f_\star T$ and only contributes an order-unity prefactor.

\subsubsection{Doppler Signal}
\label{app:proj_dyn_dop}

Starting with the Fourier transform $\delta \tilde{t}_\mathcal{D}(f)$ from Sec.~\ref{sec:analytics_dop_dyn_pulsar}, Eq.~\eqref{eqn:dop_dyn_FT},
we first evaluate the unprojected noise-weighted norm of $\delta t_\mathcal{D}$, 
\begin{equation}   \label{eqn:unproj_norm_dop_dyn_1}
(\delta t_\mathcal D|\delta t_\mathcal D)
=
\frac{2G^2M^2}{\pi^2 v^4\Delta t\,t_{\rm rms}^2}
\int_{1/T}^{\infty}df\,
\frac{f^{\gamma-2}}{f_\star^\gamma+f^\gamma}
\left|
K_1(2\pi f\tau_P)(\unit b_P\cdot\unit n)
-
iK_0(2\pi f\tau_P)(\unit v\cdot\unit n)
\right|^2.
\end{equation}
The prefactor ${f^{\gamma-2}}/{(f_\star^\gamma+f^\gamma)}$ falls off as $f^{-2}$ for frequencies $f \gg f_\star$ at all values of $\gamma$, so the integrand is only supported for frequencies $f\lesssim f_\star$; within this band its weight sits near $f\sim f_\star$ or $f\sim 1/T$ depending on $\gamma$, as in App.~\ref{sec:appendix_integral}. Since $f_\star \tau_P \ll 1$, this means that the inner product probes $K_0$ and $K_1$ at small arguments. At small argument, $K_1(x)\sim 1/x$ dominates the logarithmically softer $K_0(x) \sim \ln(1/x)$, so the term proportional to $\unit{v} \cdot \unit{n}$ can be dropped in Eq.~\eqref{eqn:unproj_norm_dop_dyn_1}. We then have
\begin{equation}   \label{eqn:unproj_norm_dop_dyn_2}
    \begin{split}
        (\delta t_\mathcal D|\delta t_\mathcal D)
         &\approx 
        \frac{2G^2M^2}{\pi^2 v^4\Delta t\,t_{\rm rms}^2}(\unit b_P\cdot\unit n)^2
        \int_{1/T}^{\infty}df  \,
        \frac{f^{\gamma-2}}{f_\star^\gamma+f^\gamma}
        \left[K_1(2\pi f\tau_P)\right]^2 \\
        & \approx \frac{G^2M^2}{2\pi^4 v^4 \tau_P^2 \Delta t\,t_{\rm rms}^2}(\unit b_P\cdot\unit n)^2
        \int_{1/T}^{\infty}df  \,
        \frac{f^{\gamma-4}}{f_\star^\gamma+f^\gamma} \\
        &= \frac{G^2M^2 T^3}{2\pi^4 v^4 \tau^2_P \Delta t\,t_{\rm rms}^2}(\unit b_P\cdot\unit n)^2 \, I_{4, \gamma}.
    \end{split}
\end{equation}
We now compute the projected signal norm using Eq.~\eqref{eqn:projected_snr_pulsar},
\begin{equation} \label{eqn:explicit_proj_dyn_dop}
    (\delta t^\perp_\mathcal{D}|\delta t^\perp_\mathcal{D}) = (\delta t_\mathcal{D}|\delta t_\mathcal{D})  - \frac{(\phi_1|\delta t_\mathcal{D})^2}{(\phi_1|\phi_1)} - \frac{(\phi_2|\delta t_\mathcal{D})^2}{(\phi_2|\phi_2)} \, .
\end{equation}
Likewise, $(\phi_1|\delta t_\mathcal{D})$ and $(\phi_2|\delta t_\mathcal{D})$ are supported at $f \lesssim f_\star$ and we approximate the integrals as %
\begin{equation} \label{eqn:degen_dop_dyn}
        (\phi_1|\delta t_\mathcal{D}) \propto \int_{1/T}^{\infty}df  \,
        \frac{f^{\gamma-3}}{f_\star^\gamma+f^\gamma} \sin(2\pi ft_{\mathcal{D},0}), \
        (\phi_2|\delta t_\mathcal{D}) \propto \int_{1/T}^{\infty}df  \,
        \frac{f^{\gamma-4}}{f_\star^\gamma+f^\gamma} \cos(2\pi ft_{\mathcal{D},0}).
\end{equation}

$(\phi_2|\delta t_\mathcal{D})$ decoheres for $\gamma>3$: there the non-oscillatory prefactor is broadly peaked near $f \sim f_\star$, and since bulk events have $f_\star t_{\mathcal{D},0} \gg 1$, the oscillation $\cos(2\pi ft_{\mathcal{D},0})$ is rapid across that peak, decohering the integral and sharply suppressing the degeneracy. For $\gamma < 3$, the prefactor instead peaks at low frequencies $f \sim 1/T$, where $ft_{\mathcal{D},0}\sim t_{\mathcal{D},0}/T \lesssim 1$; the oscillation is now slow across the support, so the overlap no longer decoheres. Even so, because $\phi_2$ and $\delta t_\mathcal{D}$ share the same $1/f^2$ envelope, both $(\phi_2|\delta t_\mathcal{D})$ and $(\phi_2|\phi_2)$ scale like $I_{4,\gamma}$, so $(\phi_2|\delta t_\mathcal{D})^2/(\phi_2|\phi_2)$ has the same $\sim (f_\star T)^{-\min(\gamma, 3)}$ scaling as $(\delta t_\mathcal{D}|\delta t_\mathcal{D})$.

By the same argument, $(\phi_1|\delta t_\mathcal{D})$ decoheres for $\gamma>2$ and only requires examination at $\gamma<2$. Here, the penalty $(\phi_1|\delta t_\mathcal{D})^2/(\phi_1|\phi_1) \sim (f_\star T)^{-2\min(\gamma, 2)+\min(\gamma, 1)}$ is parametrically subdominant for $1 < \gamma < 2$, as $2\min(\gamma, 2) > \min(\gamma, 1)+\min(\gamma, 3)$, and for $\gamma < 1$ it shares the norm's scaling.

The scaling argument above shows that $(\delta t^\perp_\mathcal{D}|\delta t^\perp_\mathcal{D})$ scales the same as $(\delta t_\mathcal{D}|\delta t_\mathcal{D})$ with $f_\star T$. We now check numerically that their prefactors agree to within an order-unity factor. %
On the discrete modes $f_k=k/T$ up to the Nyquist frequency $f_{\rm Nyq}=1/(2\Delta t)$, we compute the raw norm $(\delta t_\mathcal{D}|\delta t_\mathcal{D})$ and the timing-model-projected norm $(\delta t^\perp_\mathcal{D}|\delta t^\perp_\mathcal{D})$ of Eq.~\eqref{eqn:explicit_proj_dyn_dop}, retaining the full Bessel envelope $K_1(2\pi f\tau_P)$ (and keeping only the leading $\unit b_P\cdot\unit n$ term) rather than the small-argument form $1/(2\pi f\tau_P)$ used to reduce Eq.~\eqref{eqn:unproj_norm_dop_dyn_1} to Eq.~\eqref{eqn:unproj_norm_dop_dyn_2}. We then report the fractional SNR loss $1-\sqrt{(\delta t^\perp_\mathcal{D}|\delta t^\perp_\mathcal{D})/(\delta t_\mathcal{D}|\delta t_\mathcal{D})}$, taken as the median over bulk events, $t_{\mathcal{D}, 0}/T\in[0.2,0.8]$, across the dynamic range $\tau_P/T \in \left[10^{-3}, 10^{-1}\right]$. For the extremal, best-fit, and theory benchmarks of Sec.~\ref{sec:results_and_discussion}, we find that the loss from timing-model projection is $\lesssim 9\%$, $\lesssim 2.5\%$, and $\lesssim 0.25\%$ respectively. %
These benchmarks span $\gamma\simeq2$--$4$; at lower $\gamma$ the loss grows monotonically but stays bounded, peaking at $\lesssim 41\%$ in the white-noise limit $\gamma\to0$. %
At the level of approximation of our analytic estimates, the projection is therefore 
not a concern. This means that we can ensemble average over $\left\langle(\unit b_P\cdot\unit n)^2\right\rangle = 1/3$ and from Eq.~\eqref{eqn:unproj_norm_dop_dyn_2} obtain 
\begin{equation}
    (\delta t^\perp_\mathcal D|\delta t^\perp_\mathcal D) \sim \frac{G^2M^2 T^3}{6\pi^4 v^4 \tau_P^2 \Delta t\,t_{\rm rms}^2} (f_\star T)^{-\min(\gamma, 3)},
\end{equation}
matching Eq.~\eqref{eqn:snr_dop_dyn}.

\subsubsection{Shapiro Signal}
\label{app:proj_dyn_shap}

We start from Sec.~\ref{sec:analytics_shap_dyn_pulsar}, Eq.~\eqref{eqn:shap_dyn_FT} and evaluate the unprojected noise-weighted norm of $\delta t_\mathcal{S}$, 
\begin{equation}   
\label{eqn:unproj_norm_shap_dyn_1}
    (\delta t_\mathcal S|\delta t_\mathcal S)
    =
    \frac{8G^2M^2}{\Delta t\,t_{\rm rms}^2}
    \int_{1/T}^{\infty}df\,
    \frac{f^{\gamma-2}}{f_\star^\gamma+f^\gamma}\,
    e^{-4\pi f\tau_\perp}.
\end{equation}
Once again, the integrand is supported only at $f\lesssim f_\star$, and since $f_\star \tau_\perp \ll 1$ the exponential is probed at small argument, so $e^{-4\pi f\tau_\perp}\approx 1$ over the support, leaving the bare $1/f$ envelope,
\begin{equation}   \label{eqn:unproj_norm_shap_dyn_2}
    (\delta t_\mathcal S|\delta t_\mathcal S)
    \approx
    \frac{8G^2M^2 T}{\Delta t\,t_{\rm rms}^2}\,I_{2,\gamma}.
\end{equation}
The projection penalty preserves the norm's scaling for the same reasons as in the Doppler case. Because $\delta t_\mathcal{S}$ has a $1/f$ envelope, $(\phi_1|\delta t_\mathcal{S})$ and $(\phi_2|\delta t_\mathcal{S})$ decohere for $\gamma > 1$ and $\gamma > 2$ respectively, and below these thresholds they either share the norm's scaling or remain parametrically subdominant for $f_\star T \gg 1$. 

The same numerical check applies to the Shapiro signal, now retaining the full exponential envelope $e^{-2\pi f\tau_\perp}$ rather than setting it to unity as in the reduction of Eq.~\eqref{eqn:unproj_norm_shap_dyn_1} to Eq.~\eqref{eqn:unproj_norm_shap_dyn_2}, and with no angular factor to average. Median over bulk events, the fractional SNR loss $1-\sqrt{(\delta t^\perp_\mathcal S|\delta t^\perp_\mathcal S)/(\delta t_\mathcal S|\delta t_\mathcal S)}$ across $\tau_\perp/T\in[10^{-3},10^{-1}]$ is $\lesssim 6.5\%$, $\lesssim 1.5\%$, and $\lesssim 0.1\%$ at the extremal, best-fit, and theory benchmarks, and $\lesssim 37\%$ in the white-noise limit $\gamma\to0$. The projection is therefore again only an order-unity effect, giving
\begin{equation}\label{eqn:proj_norm_shap_dyn}
    (\delta t^\perp_\mathcal S|\delta t^\perp_\mathcal S) \sim \frac{8G^2M^2 T}{\Delta t\,t_{\rm rms}^2}\,(f_\star T)^{-\min(\gamma, 1)},
\end{equation}
which reproduces Eq.~\eqref{eqn:snr_shap_dyn}.

\section{Details of Numerics}
\label{app:numerics}

This appendix describes how the projected reach of Sec.~\ref{sec:results_and_discussion} is computed. The deterministic reach is set by the expected number of detectable events $\bar N_{\rm th}$ of Eq.~\eqref{eqn:N_ev}, an integral over the subhalo phase space $\theta=(\vec r_0,\vec v)$. We evaluate it by Monte Carlo: we draw a population of worldlines as described in App.~\ref{app:sampling_geometries}, and for each we evaluate the integrand --- the detectability $\Theta[\text{SNR}_{\rm ev}(\theta)-\text{SNR}_{\rm th}]$ --- through the single-event SNR defined above Eq.~\eqref{eqn:N_ev}, which we compute using the timing-marginalized noise operator $\mathbf{N}^{-1}_\perp$ as described in App.~\ref{app:time_domain}. The integral is then the sampling volume times the mean integrand over the draws. The stochastic reach reuses the same Monte Carlo machinery, but the sampled integral is now the population covariance $\mathbf{S}_{(10)}$ of Eq.~\eqref{eq:stoch:Sperp} rather than an event count; the detection significance then follows from the trace formula of Eq.~\eqref{eq:DM_stoch_SNR}. This covariance carries a subtlety: it is dominated by the rare loudest events --- and, for the Doppler signal, divergent without a cutoff --- so it requires a percentile cut, a larger sampling region, and a check of its Gaussian validity, all treated in App.~\ref{app:stochastic_limit_cutoff}.

\subsection{Sampling Geometries}
\label{app:sampling_geometries}

\begin{figure}
    \centering
    \includegraphics[width=\linewidth]{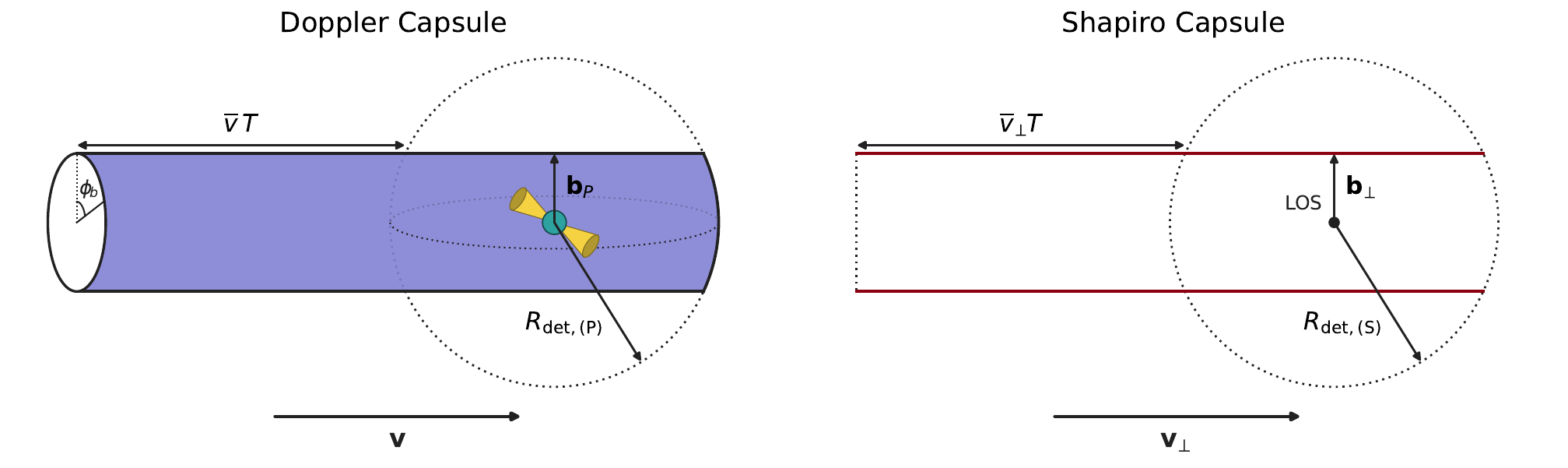}
    \caption{The Doppler-pulsar (left) and Shapiro line-of-sight (right) sampling capsules. A subhalo is detected if it passes within $R_{\text{det},(c)}(M)$ of the target during the observation window; because it moves during that window, it can start outside $R_{\text{det}}$ and still cross in, so the sampled region is not a ball but a capsule --- elongated along the flyby by the distance traveled, $\bar v T$ (Doppler) or $\bar v_\perp T$ (Shapiro). The impact parameter is $b_P$ to the pulsar (azimuth $\phi_b$ about $\mathbf v$) or $b_\perp$ to the line of sight (signed).}
    \label{fig:geometry_capsules}
\end{figure}

We first describe how the subhalo worldlines are sampled. Sampling uniformly over all of space would waste nearly every draw on undetectable worldlines, where the integrand is too small, so we restrict the draws to the region that contributes. The size of the sampled region is set by a quantity we call the \textit{detection distance}, $R_{\text{det},(c)}(M)$ for $c\in\{P,S\}$, which is defined such that a subhalo with mass $M$ that stays farther than $R_{\text{det},(c)}$ throughout the observing window $0\leq t\leq T$ from the pulsar ($P$) or its line of sight ($S$) is undetectable, \textit{i.e.} $\text{SNR}_{\rm ev}<\text{SNR}_{\rm th}=4$. We will describe how this is computed at the end of this subsection. %
Because a subhalo travels a distance $\bar v T$ during the observation, the worldlines that pass within $R_{\text{det}, (c)}$ (i.e. all detectable worldlines) for a given velocity direction $\unit{v}$ do not fill a ball but a capsule, which we parametrize in impact-parameter coordinates. Fig.~\ref{fig:geometry_capsules} shows this capsule geometry for both Doppler and Shapiro cases. For the pulsar sphere, we write $d^3 r_0 = b_P \, d b_{P} \, d \phi_b \, \bar{v}\, dt_{\mathcal{D}, 0}$, and each coordinate is sampled in the intervals
\begin{equation}
    t_{\mathcal{D}, 0} \in \left[-{\sqrt{R_{\text{det},(P)}^2-b_P^2}}/{\bar{v}}, \, {T+\sqrt{R_{\text{det},(P)}^2-b_P^2}}/{\bar{v}} \right], \, b_P \leq R_{\text{det},(P)}(M),\, d \phi_b \in [0, 2\pi ],
\end{equation}
where $d\phi_b$ is parametrized in the plane perpendicular to $\unit{v}$, and for the line-of-sight cylinder, we write $d^3 r_0 = db_\parallel \, d b_\perp \, \bar{v}_\perp \, dt_{\mathcal{S}, 0}$, and the sampling intervals are
\begin{equation}
    t_{\mathcal{S}, 0} \in \left[-{\sqrt{R_{\text{det},(S)}^2-b_\perp^2}}/{\bar{v}_\perp}, \, T+{\sqrt{R_{\text{det},(S)}^2-b_\perp^2}}/{\bar{v}_\perp}\right], \, |b_\perp| \leq  R_{\text{det},(S)}(M), \, |b_\parallel| \leq L/2.
\end{equation}
Each worldline is drawn in a fixed order: first the velocity --- speed $\bar v$ with direction $\unit v$ sampled isotropically --- which orients the capsule; then the transverse impact parameters (and azimuth $\phi_b$ for the pulsar sphere); and finally the closest-approach time, whose range is set by the impact parameter through the capsule end-caps above. We estimate each integral as the sampling volume times the mean integrand over the draws (its sum divided by the number of sampled worldlines), and we sample $b_P^2$ uniformly for the pulsar sphere (rather than $b_P$) to reproduce its radial measure $b_P\,\mathrm{d}b_P$.

\begin{figure}
    \centering
    \includegraphics[width=0.67\linewidth]{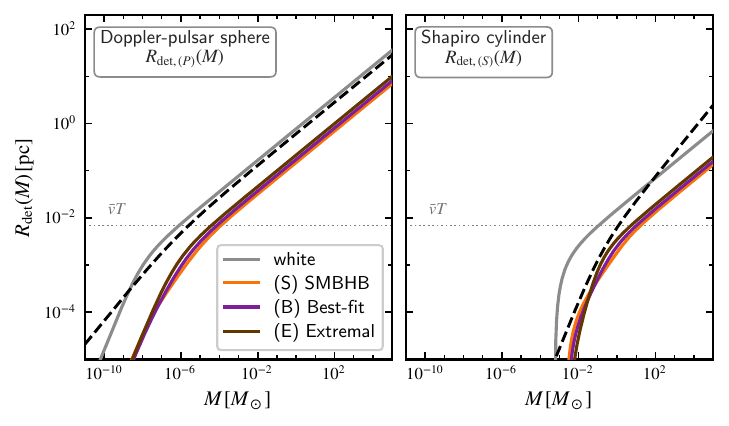}
    \caption{Detection distance $R_{\text{det},(c)}(M)$ (solid) for the Doppler-pulsar sphere ($P$) and the Shapiro line-of-sight cylinder ($S$) versus subhalo mass, for the white-noise benchmark and the three GWB posterior points (S/B/E). $R_{\text{det}}$ is the largest closest-approach distance --- within the observation window, to the pulsar ($P$) or its line of sight ($S$) --- at which a mass-$M$ subhalo still produces $\text{SNR}_{\rm ev}=\text{SNR}_{\rm th}=4$. It is the \emph{transverse} radius of the sampled region, not its full extent: because the subhalo travels $\sim\bar vT$ during the window, the region is an elongated capsule of transverse radius $R_{\text{det}}$ and length $\sim\bar vT$ (Fig.~\ref{fig:geometry_capsules}). The dotted line marks $\bar vT$, the static--dynamic boundary. The dashed line is the $90^{\rm th}$-percentile closest-approach distance of the subhalo population at $f_{\rm DM}=1$; where $R_{\text{det}}(M)$ lies below it, that region is not sensitive to substructure with $f_{\rm DM}<1$.}
    \label{fig:rdet_detection_radius}
\end{figure}

It remains to specify how the detection distance $R_{\text{det},(c)}(M)$ is computed. For a given subhalo worldline, we define $d_{\text{min}}$ as the smallest distance between the subhalo and the pulsar ($P$) or line of sight ($S$) during the observation window $[0,T]$. Then $R_{\text{det},(c)}(M)$ is then the largest $d_{\text{min}}$ for which there exists a subhalo worldline with $\text{SNR}_{\rm ev}(\theta)\ge\text{SNR}_{\rm th}=4$. To compute this, we will need to maximize $\text{SNR}_{\rm ev}(\theta)$ over $\theta$ under the constraint that this worldline has the closest distance $d_\text{min}$, which is done as follows. We split the maximization over worldlines into two cases depending on $t_0$, which is taken to be $t_{\mathcal{D}, 0}$ ($P$) or $t_{\mathcal{S}, 0}$ ($S$). In Case~1, $t_0$ falls inside the observing window $[0,T]$, and $d_\text{min}$ is simply the impact parameter. We then maximize over $t_0\in[0,T]$, the azimuth $\phi_b$ or the sign of $b_\perp$ for ($P$) and ($S$) respectively, and the direction $\unit v$. In Case~2 it $t_0$ falls outside the window, and can be arbitrarily far before or after the data, so it cannot be scanned over any finite range. The closest the worldline comes during the observation is then at a window edge, $t=0$ or $t=T$; we place the subhalo there and instead maximize over the approach angle $\mu = \unit d\cdot\unit v$, with $\mu\in[-1,0)$ or $(0,1]$, where $\unit d$ points from the target to the subhalo. In Fig.~\ref{fig:rdet_detection_radius}, we show $R_{\mathrm{det},(P)}(M)$ and $R_{\mathrm{det},(S)}(M)$ for the white-noise only benchmark, as well as the three GWB posterior points from Sec.~\ref{sec:results_and_discussion}.

\subsection{Integrand Evaluation in the Time Domain}
\label{app:time_domain}

After each subhalo worldline is drawn randomly, we have to evaluate the single-event SNR defined above Eq.~\eqref{eqn:N_ev}, which is the noise-weighted norm of the signal after marginalizing the timing model. In App.~\ref{app:pulsar_SNRs}, and in the analytic estimates of Sec.~\ref{sec:analytical_estimates}, we imposed this by projecting the signal, $(\delta\vec t^\perp\,|\,\delta\vec t^\perp)$. That is not the only route, and for a Monte Carlo over many signals it is not the cheapest: rather than projecting every sampled $\delta\vec t$, we absorb the projection into the inner product and evaluate it on the \emph{un}projected signal, %
\begin{equation}\label{eqn:snr_ev_marginalized}
    (\delta\vec t\,|\,\delta\vec t)_\perp \equiv \delta\vec t^\top\,\mathbf N^{-1}_\perp\,\delta\vec t = \text{SNR}^2_{\rm ev}. %
\end{equation}
Here $\mathbf N^{-1}_\perp$ is a timing-marginalized inverse-noise operator that folds the noise weighting and the timing-model projection into a single matrix acting directly on the raw signal $\delta\vec t$, which we define in the next paragraph. This norm computed using $\mathbf N^{-1}_\perp$ gives the same number as the projected norm, \textit{i.e.} $(\delta\vec{t}^{\perp}|\delta\vec{t}^{\perp})=(\delta\vec{t}|\delta\vec{t})_{\perp}$, but depends only on the noise, so we build it once and reuse it for every event. As shown in Ref.~\cite{vanHaasteren:2012hj}, this is done cleanly in the time domain with a singular value decomposition of the design matrix; we sketch the construction here.

We work on the grid of $N_t=\lfloor T/\Delta t\rfloor$ arrival times $t_i=i\,\Delta t$ (cadence $\Delta t$ of Sec.~\ref{sec:SNR}), on which the full gauge-invariant observable of Eq.~\eqref{eqn:delta_t_t} gives the signal vector $\delta\vec t(\theta)\in\mathbb{R}^{N_t}$. The quadratic timing model is the design matrix $\mathbf{M}=\big[1,\ t_i,\ t_i^{\,2}\big]\in\mathbb{R}^{N_t\times 3}$; its singular value decomposition supplies an orthonormal basis $\mathbf{G}\in\mathbb{R}^{N_t\times(N_t-3)}$ for the subspace orthogonal to the timing model ($\mathbf{G}^\top\mathbf{G}=\mathbf{I}$, $\mathbf{G}^\top\mathbf{M}=0$). The noise covariance combines white and GWB red noise, $\mathbf{N}=t_{\rm rms}^2\,\mathbf{I}+\mathbf{F}\,\mathbf{\Phi}\,\mathbf{F}^\top$ (following Ref.~\cite{NANOGrav:2023gor}), where $t_{\rm rms}^2$ is the per-sample white variance set by the PSD $2\Delta t\,t_{\rm rms}^2$ of Eq.~\eqref{eqn:explicit_Sn}, and $\mathbf{F}\,\mathbf{\Phi}\,\mathbf{F}^\top$ realizes the GWB PSD of the same equation on the Fourier modes $f_k=k/T$: the columns of $\mathbf{F}=[\sin(2\pi f_k t_i),\,\cos(2\pi f_k t_i)]$ are the sine/cosine pairs and
\begin{equation}
    \mathbf{\Phi}=\mathrm{diag}(\Phi_k), \quad \Phi_k= \frac{1}{T}\frac{A_{\rm GWB}^2}{12\pi^2}\left(\frac{1\,\mathrm{yr}^{-1}}{f_k}\right)^{\gamma}\,\mathrm{yr}^3.
\end{equation}
The marginalized operator is $\mathbf{N}^{-1}_\perp = \mathbf{G}(\mathbf{G}^\top\mathbf{N}\,\mathbf{G})^{-1}\mathbf{G}^\top$; we evaluate it with the Woodbury identity,
\begin{equation}\label{eqn:Nperp}
\mathbf{N}^{-1}_{\perp}=t_{\rm rms}^{-2}\mathbf{G}\mathbf{G}^\top-t_{\rm rms}^{-4}\,\mathbf{G}\mathbf{G}^\top\mathbf{F}\big(\mathbf{\Phi}^{-1}+t_{\rm rms}^{-2}\,\mathbf{F}^\top\mathbf{G}\mathbf{G}^\top\mathbf{F}\big)^{-1}\mathbf{F}^\top\mathbf{G}\mathbf{G}^\top,
\end{equation}
avoiding a full inversion of $\mathbf{N}$. Computed once for the noise benchmark, the same $\mathbf{N}^{-1}_\perp$ also carries the projection in the stochastic trace formula of Eq.~\eqref{eq:DM_stoch_SNR}, so the covariance $\mathbf{S}_{(10)}$ is left unprojected.

\subsection{Cutoff Insensitivity of the Stochastic Limit Reach}
\label{app:stochastic_limit_cutoff}

The stochastic covariance is dominated by the closest encounters: a handful of loud events can outweigh the entire population of weak ones. For the Doppler signal the ensemble covariance diverges, though only logarithmically (Sec.~\ref{sec:analytics_stochastic_regime}). We therefore regulate it with a percentile cut, retaining only events below a threshold $\text{SNR}_{10}$ chosen so that the loudest event exceeds it in $90\%$ of universes. The cut is strictly necessary only for the Doppler divergence, but we apply the same cut to the Shapiro signal so that the validity boundary $N_Q$ is defined uniformly for both. The regulated population covariance is
\begin{equation}\label{eq:stoch:Sperp}
  \mathbf{S}_{(10)}(M, f_\text{DM}) = \frac{\rho_{\mathrm{DM}}f_{\mathrm{DM}}}{M} \int d^3\vec{r}_0\int d^3\vec{v}\,f_{\vec{v}}(\vec{v}) \,\Theta\left[\text{SNR}_{10} - \text{SNR}_\text{ev}(\theta)\right] \, \delta \vec{t}({\theta})\,  \delta \vec{t}({\theta})^\top \, .
\end{equation}
Unlike the deterministic reach, which is confined to within $R_\text{det}$ of the target, the covariance also collects weak, distant events, so it is sampled over the same capsule but enlarged: $R_\text{det}$ is replaced by $3\max\!\left(\bar v T,\,b_{90}(M,f_\text{DM})\right)$ so that the trace is comfortably converged, with $b_{90}$ the $90^{\rm th}$-percentile impact parameter. Sampling proceeds as in App.~\ref{app:sampling_geometries}, except that, because the covariance receives comparable contributions from each decade of impact parameter, we sample $b$ logarithmically rather than uniformly. We split the covariance by region for the comparison in Fig.~\ref{fig:analytic_overlay}.

The same close-event domination limits the validity of the Gaussian trace formula of Eq.~\eqref{eq:DM_stoch_SNR}, which holds only when many events contribute comparably. To quantify this heuristically, we note that each event contributes the weight
\begin{equation}
    q_i \equiv q(\theta_i) =
    \delta \vec{t}(\theta_i)^\top
    \mathbf N^{-1}_\perp\,\mathbf S_{(10)}\,
    \mathbf N^{-1}_\perp
    \delta \vec{t}(\theta_i),
\end{equation}
to the sensitivity, and count how many contribute comparably through
\begin{equation}
\label{eqn:NQ_definition}
N_Q(M,f_{\rm DM}) = \frac{\left(\sum_i q_i\right)^2}{\sum_i q_i^2},
\end{equation}
the sum running over the retained population. When many events contribute, $N_Q$ is large and the Gaussian approximation holds; when a single event dominates, $N_Q\sim 1$ and the trace formula no longer captures the true statistics. We take $N_Q<10$ as the few-event regime, shown dotted in Fig.~\ref{fig:analytic_overlay}.

\begin{figure}
    \centering
    \includegraphics[width=\linewidth]{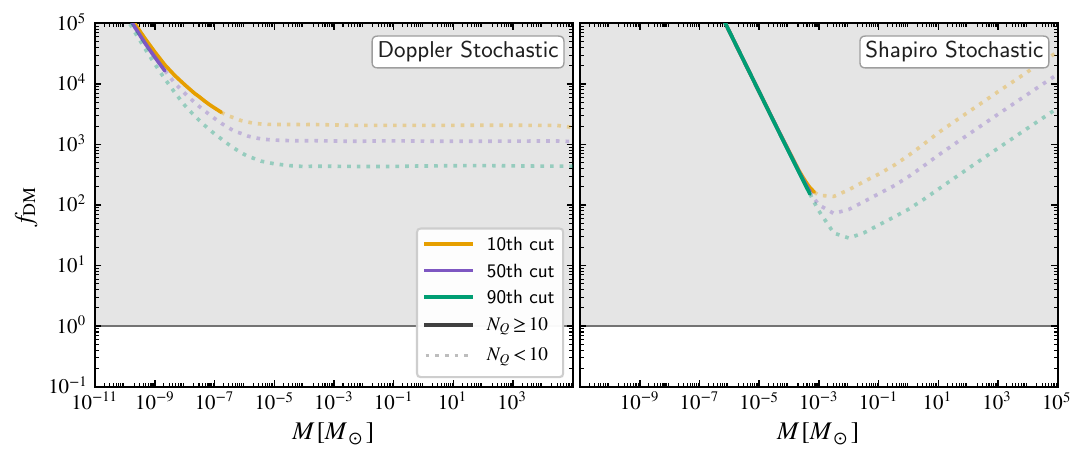}
    \caption{
    Stochastic projected reach $f_\text{DM}(M)$ at the SMBHB-baseline GWB for the Doppler (left) and Shapiro (right) signals, each computed with three percentile cuts on the retained covariance ($10^{\rm th}$, $50^{\rm th}$, $90^{\rm th}$). Solid segments ($N_Q\ge10$, Eq.~\eqref{eqn:NQ_definition}) mark where the Gaussian trace formula is valid; there the three cuts overlap, so the reach is insensitive to the choice of cut. Faded dotted segments ($N_Q<10$) are extrapolations and should not be read quantitatively. The $N_Q=10$ boundary is fixed across cuts for Shapiro but moves for Doppler, whose logarithmically divergent covariance makes $N_Q$ itself cut-dependent.
    }
 \label{fig:percentile_cut_insensitivity}
\end{figure}
In Fig.~\ref{fig:percentile_cut_insensitivity} we show the stochastic reach computed with three percentile cuts on the retained covariance. Where the Gaussian approximation holds ($N_Q>10$, solid curves), the three cuts overlap, so the reach is independent of the cut: the cut is only a device for defining a representative covariance. The figure also exposes the close-event sensitivity directly: for the Shapiro signal, whose covariance is not heavy-tailed, the $N_Q=10$ boundary sits at the same mass for every cut; for the Doppler signal it does not: its covariance is only logarithmically sensitive to the closest events --- enough to make $N_Q$ depend on the cut, so the boundary moves with it.

\end{document}